\documentclass[11pt,a4paper]{article}
\usepackage[left=1in, right=1in, top=1in, bottom=1in]{geometry}

\usepackage{xcolor}

\usepackage{path}
\usepackage{graphicx,floatflt} 
\usepackage{lineno} 

\newcommand{\Emph}[1]{\emph{#1}}

\usepackage{path}
\usepackage{graphicx} 
\usepackage{xspace}

\usepackage{amssymb}

\usepackage[ocgcolorlinks,citecolor=blue,colorlinks=true,linkcolor=black,urlcolor=blue]{hyperref}

\usepackage{amsthm}
\theoremstyle{plain}

\newtheorem{theorem}{Theorem}[section]
\newtheorem{lemma}[theorem]{Lemma}
\newtheorem{corollary}[theorem]{Corollary}

\newtheorem{definition}[theorem]{Definition}
\newtheorem*{convention*}{Convention}
\newtheorem{observation}[theorem]{Observation}
\newtheorem{conjecture}[theorem]{Conjecture}

\newenvironment{clm}
{\smallskip

\noindent
{\bf Claim.}
\em
}
{
\smallskip

\noindent\ignorespacesafterend
}

\newcommand{\ColorCyan}{}
\newcommand{\Thm}[1]{\ColorCyan{Thm.}\,\ref{#1}} 
\newcommand{\Thms}[1]{\ColorCyan{Theorems}~\ref{#1}} 
\newcommand{\Lm}[1]{\ColorCyan{Lemma}~\ref{#1}}  
\newcommand{\Obs}[1]{\ColorCyan{Obs.}\,\ref{#1}} 
\newcommand{\Cor}[1]{\ColorCyan{Cor.}\,\ref{#1}} 
\newcommand{\Def}[1]{\ColorCyan{Def.}\,\ref{#1}} 
\newcommand{\Defs}[1]{\ColorCyan{Definitions}~\ref{#1}} 
\newcommand{\Fig}[1]{\ColorCyan{Fig.}\,\ref{#1}} 
\newcommand{\Figs}[1]{\ColorCyan{Figures}~\ref{#1}} 
\newcommand{\FigLR}[2]{\ColorCyan{Fig.}\,\ref{#1}\,(#2)} 
\newcommand{\Sec}[1]{\ColorCyan{Sec.}\,\ref{#1}} 
\newcommand{\Secs}[1]{\ColorCyan{Sections}~\ref{#1}} 
\newcommand{\EqRef}[1]{\ColorCyan{(\ref{#1})}}
\newcommand{\ItemRef}[1]{\ColorCyan{\ref{#1}}} 

\newsavebox{\smallProofsym}
\savebox{\smallProofsym}  
{%
\begin{picture}(3.5,6)                                   
\put(0,0){\framebox(6,6){}}
\put(0,2){\framebox(4,4){}}
\end{picture}
}%
\newcommand{\smalleop}{\mbox{} \hfill \usebox{\smallProofsym}}

\newenvironment{MyProof}[1]
[Proof]
{\noindent
\emph{#1.}
}
{
\smalleop
\smallskip

\noindent\ignorespacesafterend
}

\newcommand{\ie}{i.e., }
\newcommand{\eg}{e.g., }
\newcommand{\etal}{et\,al.}
\newcommand{\etc}{etc.\ }

\newcommand{\Case}[2]{
\smallskip 

\noindent 
{\sc Case}~#1. \emph{#2}
}

\usepackage{enumitem}

\newenvironment{EnumRom}
{
\begin{enumerate}
[leftmargin=2em, 
topsep=0.3em, 
parsep=0.2em, 
itemsep=0em, 
labelsep=0.2em, 
label={(\roman*)}]
}
{
\end{enumerate}
}

\newenvironment{EnumCapRom}
{
\begin{enumerate}
[leftmargin=2em, 
topsep=0.3em, 
parsep=0.2em, 
itemsep=0em, 
labelsep=0.2em, 
label={(\Roman*)}]
}
{
\end{enumerate}
}

\newenvironment{EnumAlph}
{
\begin{enumerate}
[leftmargin=2em, 
topsep=0.3em, 
parsep=0.2em, 
itemsep=0em, 
labelsep=0.2em, 
label={(\alph*)}]
}
{
\end{enumerate}
}

\newenvironment{EnumCapAlph}
{
\begin{enumerate}
[leftmargin=2em, 
topsep=0.3em, 
parsep=0.2em, 
itemsep=0em, 
labelsep=0.2em, 
label={(\Alph*)}]
}
{
\end{enumerate}
}

\newenvironment{EnumNo}
{
\begin{enumerate}
[leftmargin=2em, 
topsep=0.3em, 
parsep=0.2em, 
itemsep=0em, 
labelsep=0.2em, 
label={--~}]
}
{
\end{enumerate}
}

\newcommand{\placefig}[2]
        {\includegraphics[width=#2]{FIGS/#1.pdf}}

\newcommand{\RR}{\mathbb{R}}
\newcommand{\NN}{\mathbb{N}}
\newcommand{\NNnull}{\NN_0}

\usepackage{xspace}
\newcommand{\one}{$1$\xspace}
\newcommand{\two}{$2$\xspace}
\newcommand{\three}{$3$\xspace}
\newcommand{\four}{$4$\xspace}
\newcommand{\five}{$5$\xspace}
\newcommand{\six}{$6$\xspace}

\newcommand{\eight}{$8$\xspace}
\newcommand{\Eight}{$8$\xspace}

\newcommand{\ten}{$10$\xspace}

\newcommand{\restr}[2]{{#1\hspace{-0.06em}\raisebox{-0.185ex}{\large $|$}}\raisebox{-0.44ex}{\hspace{-0.06em}$_{#2}$}}
\newcommand{\ind}[2]{\restr{#1}{#2}}

\usepackage{units}
\newcommand{\Nicefrac}[2]{\nicefrac{#1}{#2}}

\newcommand{\cl}[1]{\overline{#1}} 

\newcommand{\bfg}{\ColorCyan{bistellar flip graph}}
\newcommand{\efg}{\ColorCyan{edge flip graph}}

\newcommand{\ptriangulation}{triangulation\xspace}
\newcommand{\ptriangulations}{triangulations\xspace}
\newcommand{\psubdivision}{subdivision\xspace}
\newcommand{\psubdivisions}{subdivisions\xspace}

\newcommand{\region}{region\xspace}
\newcommand{\regions}{regions\xspace}

\newcommand{\coarsener}{coarsener\xspace}
\newcommand{\coarseners}{coarseners\xspace}

\newcommand{\fT}{{\cal T}_{\mathsf{full}}} 
\newcommand{\rT}{{\cal T}_{\mathsf{reg}}} 
\newcommand{\pT}{{\cal T}_{\mathsf{part}}} 
\newcommand{\fTref}[1]{{\cal T}_{\mathsf{full}}{\langle #1 \rangle}} 
\newcommand{\pTref}[1]{{\cal T}_{\mathsf{part}}{\langle #1 \rangle}} 

\newcommand{\ext}[1]{{\sf xtr}#1} 
\newcommand{\inn}[1]{#1^\circ} 
\newcommand{\innNb}{n^{\!\circ}} 
\newcommand{\extNb}{h} 
\newcommand{\pts}{\mathrm{\sf V}} 
\newcommand{\Pts}[1]{\mathrm{\sf V}\hspace{-0.05em}#1} 
\newcommand{\ptsInn}{\mathrm{\sf \inn{V}}}
\newcommand{\PtsInn}[1]{\mathrm{\sf \inn{V}}\hspace{-0.05em}#1}
\newcommand{\PtsBy}[1]{\mathrm{\sf V^{by}\hspace{-0.4ex}}#1} 
\newcommand{\PtsInv}[1]{\mathrm{\sf V^{inv}\hspace{-0.4ex}}#1} 

\newcommand{\EdsHull}{{\sf E_\mathsf{hull}}} 
\newcommand{\Eds}[1]{\mathrm{\sf E}#1} 
\newcommand{\EdsInn}[1]{\mathrm{\sf \inn{E}}\hspace{-0.05em}#1} 
\newcommand{\ElmtsFlip}[1]{\mathrm{\sf F}#1} 

\newcommand{\Reg}[1]{\mathrm{\sf R}#1} 

\newcommand{\Striv}{\mathsf{S}_{\mathsf{triv}}} 

\newcommand{\SecPoly}[1]{\Sigma\mbox{-}\mathsf{poly}(#1)}

\newcommand{\slack}{{\sf sl}}
\newcommand{\Slack}[1]{\slack\hspace{0.1em}#1}

\newcommand{\refslack}{{\sf sl^*}}
\newcommand{\Refslack}[1]{\refslack#1}

\newcommand{\height}{{\sf h}}
\newcommand{\Height}[1]{\height #1}
\newcommand{\HeightMax}{\height_\mathsf{max}}

\newcommand{\terr}{{\sf terr}}
\newcommand{\Terr}[1]{\terr\hspace{0.1em}#1}
\newcommand{\TerrTwo}[2]{\terr_{#1}\hspace{0.1em}#2}

\newcommand{\precDir}{\prec_{\mathrm{dir}}}
\newcommand{\precPerf}{\prec_1}
\newcommand{\precPerfStar}{\prec^*_1}

\newcommand{\meet}{\wedge}

\newcommand{\incr}{{\sf inc}}
\newcommand{\Incr}[1]{\incr\hspace{0.1em}#1}

\newcommand{\fLink}[1]{{\mathsf{Lk}}_{\mathsf{full}}#1}
\newcommand{\pLink}[1]{{\mathsf{Lk}}_{\mathsf{part}}#1}

\newcommand{\comp}{\diamond}
\newcommand{\notcomp}{\mbox{$\not\!\diamond\,\,$}}

\newsavebox{\Ridge}
\savebox{\Ridge}  
{\begin{picture}(4,4)  \put(0,1){\framebox(4,4){}} \put(4,1){\framebox(4,4){}} \end{picture} }

\newcommand{\orient}[1]{\vec{#1}}

\newcommand{\FBound}{\max\{\lceil \frac{n}{2}-2\rceil, h-3\}}
\newcommand{\FBoundMinus}{\max\{\lceil \frac{n}{2}-3\rceil, h-4\}}

\newcommand{\pFlip}[2]{#1_{\pm #2}}

\title{Connectivity of Triangulation Flip Graphs in the Plane%
\thanks{This is a full and revised version of \cite{WWSoCG20} (on partial triangulations) in \emph{Proceedings of the 36th Annual International Symposium on Computational Geometry (SoCG`20)} and of some of the results in  \cite{WW20} (on full triangulations) in \emph{Proceedings of the 31st Annual ACM-SIAM Symposium on Discrete Algorithms (SODA`20)}. }
\thanks{This research started at the $11^{\mathrm{th}}$ Gremo's Workshop on Open Problems (GWOP), Alp Sellamatt, Switzerland, June 24-28, 2013, motivated by a question posed by Filip Mori{\'c} on full triangulations. Research was supported by the Swiss National Science Foundation within the collaborative DACH project Arrangements and Drawings as SNSF Project 200021E-171681, and by IST Austria and Berlin Free University during a sabbatical stay of the second author. We thank Michael Joswig, Jes\'us De Loera, and Francisco Santos for helpful discussions on the topics of this paper, and Daniel Bertschinger for carefully reading an earlier version and for many helpful comments.}
}

\author{
Uli Wagner, IST Austria\\
Am Campus 1\\
A-3400 Klosterneuburg, Austria\\
{\tt uli@ist.ac.at}
\and Emo Welzl, ETH Z\"urich\\
Department of Computer Science\\
CH-8092 Z\"urich, Switzerland 
\\ 
{\tt emo@inf.ethz.ch}
}

\date{}

\begin{document}

\maketitle

\begin{abstract}
Given a finite point set $P$ in \emph{general position} in the plane, a \emph{full triangulation} of $P$ is a maximal straight-line embedded plane graph on $P$. A \emph{partial triangulation} of $P$ is a full triangulation of some subset $P'$ of $P$ containing all extreme points in $P$. A \emph{bistellar flip} on a partial triangulation either flips an edge (called \emph{edge flip}), removes a non-extreme point of degree 3, or adds a point in $P \setminus P'$ as vertex of degree 3. The \emph{bistellar flip graph} has all partial triangulations as vertices, and a pair of partial triangulations is adjacent if they can be obtained from one another by a bistellar flip. The \emph{edge flip graph} is defined with full triangulations as vertices, and edge flips  determining the adjacencies. Lawson showed in the early seventies that these graphs are connected. The goal of this paper is to investigate the structure of these graphs, with emphasis on their vertex connectivity. 

For sets $P$ of $n$ points in the plane in general position, we show that the edge flip graph is $\lceil \frac{n}{2}-2\rceil$-vertex connected, and the bistellar flip graph is $(n-3)$-vertex connected; both results are tight. The latter bound matches the situation for the subfamily of regular triangulations (\ie partial triangulations obtained by lifting the points to 3-space and projecting back the lower convex hull), where $(n-3)$-vertex connectivity has been known since the late eighties through the secondary polytope due to Gelfand, Kapranov \& Zelevinsky and Balinski's Theorem. For the edge flip-graph, we additionally show that the vertex connectivity is as least as large as (and hence equal to) the minimum degree (i.e., the minimum number of flippable edges in any full triangulation), provided that $n$ is large enough.

Our methods also yield several other results: (i) The edge flip graph can be covered by graphs of polytopes of dimension $\lceil \frac{n}{2} -2\rceil$ (products of associahedra) and the bistellar flip graph can be covered by graphs of polytopes of dimension $n-3$  (products of secondary polytopes). (ii) A partial triangulation is regular, if it has distance $n-3$ in the Hasse diagram of the partial order of partial subdivisions from the trivial subdivision. (iii) All partial triangulations of a point set are regular iff the partial order of partial subdivisions has height $n-3$. (iv) There are arbitrarily large sets $P$ with non-regular partial triangulations and such that every proper subset has only regular triangulations, \ie there are no small certificates for the existence of non-regular triangulations.
\end{abstract}

\section{Introduction}
\label{se:Introduction}
Triangulations of point sets play a role in many areas including mathematics, numerics, computer science, and processing of geographic data, \cite{Ede01,LRS10,Ber17,LS17}. A natural way to provide structure to the \emph{set of all triangulations} is to consider a graph, called \emph{flip graph}, with the triangulations as vertices and with pairs of triangulations adjacent if they can be obtained from each other by a minimal local change, called a \emph{flip} (see below for the precise definition). One of the first and most prominent results on flip graphs of planar points sets (for edge flips), proved by Lawson \cite{L72} in 1972, is that they are connected. The corresponding question of connectedness of flip graphs in higher dimensions remained a mystery until Santos \cite{San00} showed, in 2000, that in dimension \five and higher, there exist point sets for which the graph (for bistellar flips) is not connected.  The question is still open in dimension \three and \four. 

Here, we concentrate on point sets in general position in the plane and investigate ``how'' connected the flip graphs are, \ie determine the largest $k$ (in terms of $n$, the size of the underlying point set) such that $k$-vertex connectivity holds. Moreover, we supply some structural results for the flip graph.

The preceding discussion swept under the rug that there are several types of triangulations, most prominently \emph{full triangulations} (as mostly considered in computational geometry), \emph{partial triangulations} (as primarily considered in discrete geometry), and \emph{regular triangulations} (also called coherent triangulations or weighted Delaunay triangulations). We will address the vertex-connectivity for full and partial triangulations, and we will discuss some implications for regular triangulations, for which the vertex-connectivity of the flip graph has been known since the late eighties via the secondary polytope introduced by Gelfand \etal\ \cite{GKZ90} (see \cite[Cor.\,5.3.2]{LRS10}).

\begin{figure}[htb]
\centerline{
\placefig{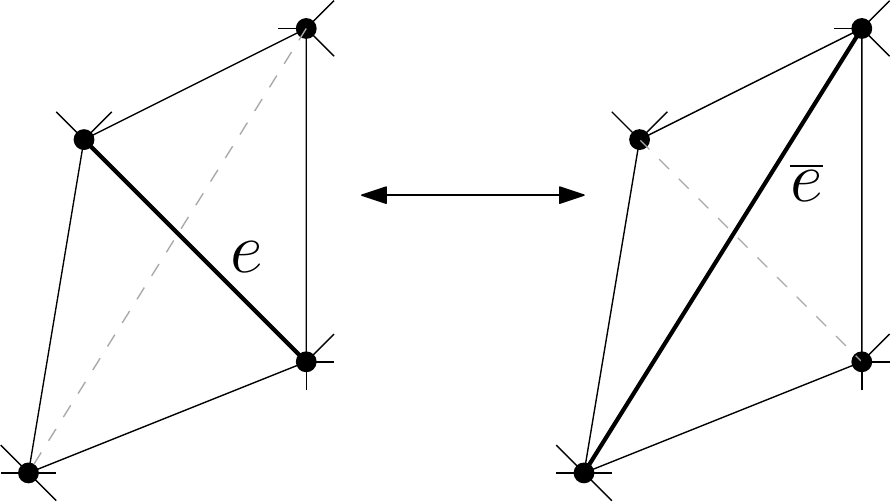}{0.22\textwidth}
\hspace{0.2\textwidth}
\placefig{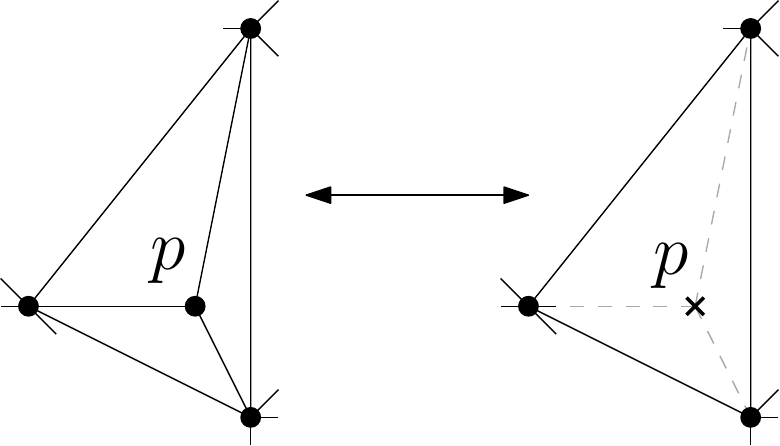}{0.2\textwidth}
}
\caption{Edge flips and point flips (point removal, left to right; point insertion, right to left).}
\label{fi:Flips}
\end{figure}

\begin{figure}[htb]
\centerline{
\begin{minipage}[c]{0.28\textwidth}
\placefig{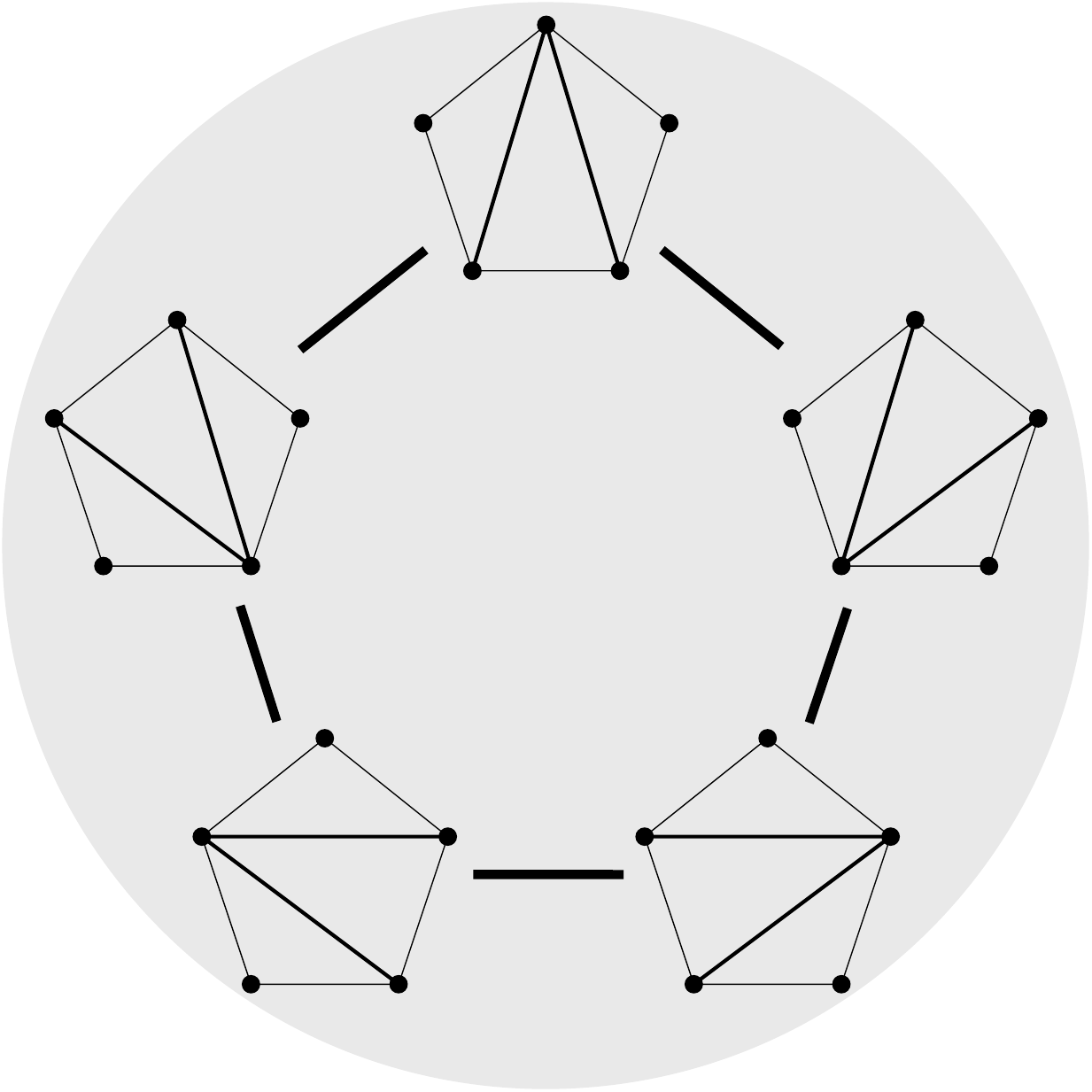}{\textwidth}
\end{minipage}
~~~
\begin{minipage}[c]{0.27\textwidth}
\placefig{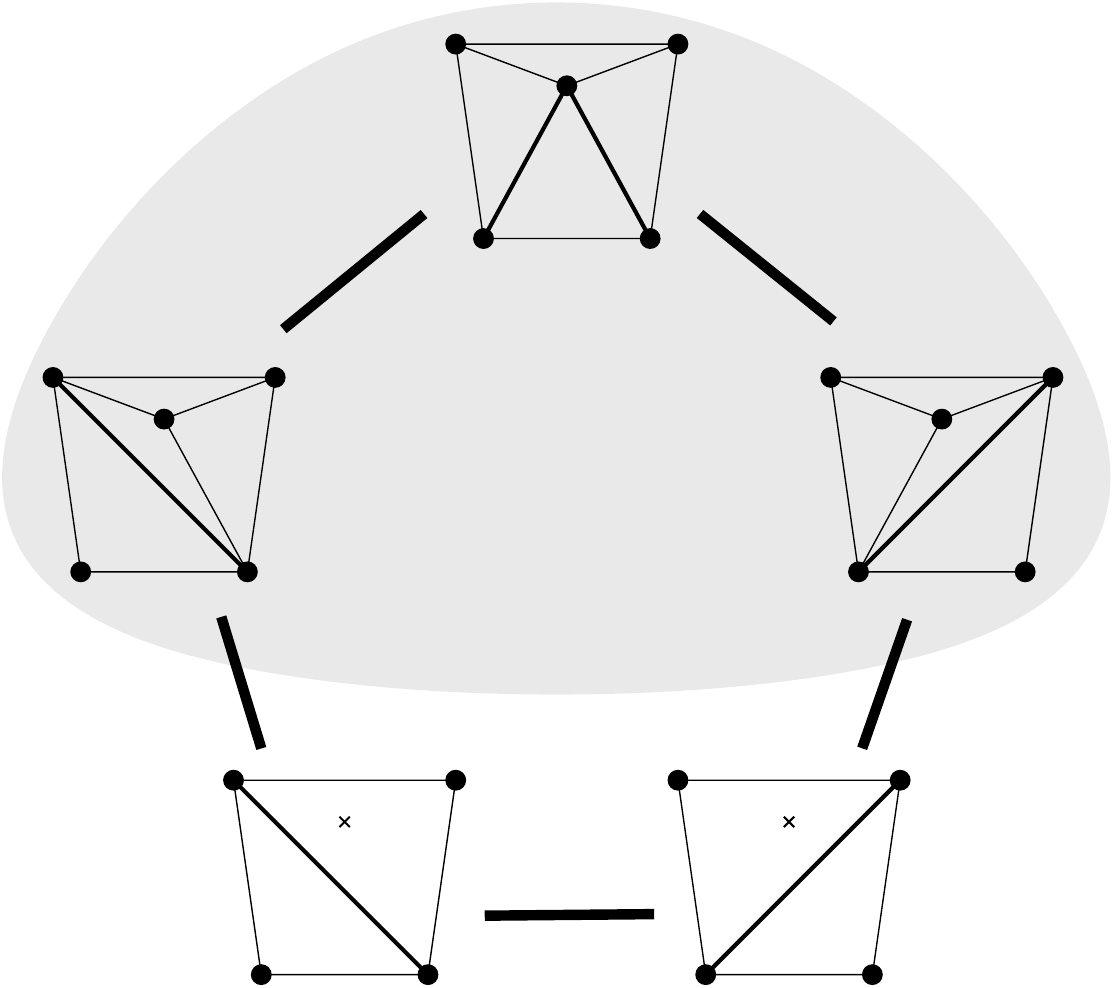}{\textwidth}
\end{minipage}
~~~
\begin{minipage}[c]{0.25\textwidth}
\placefig{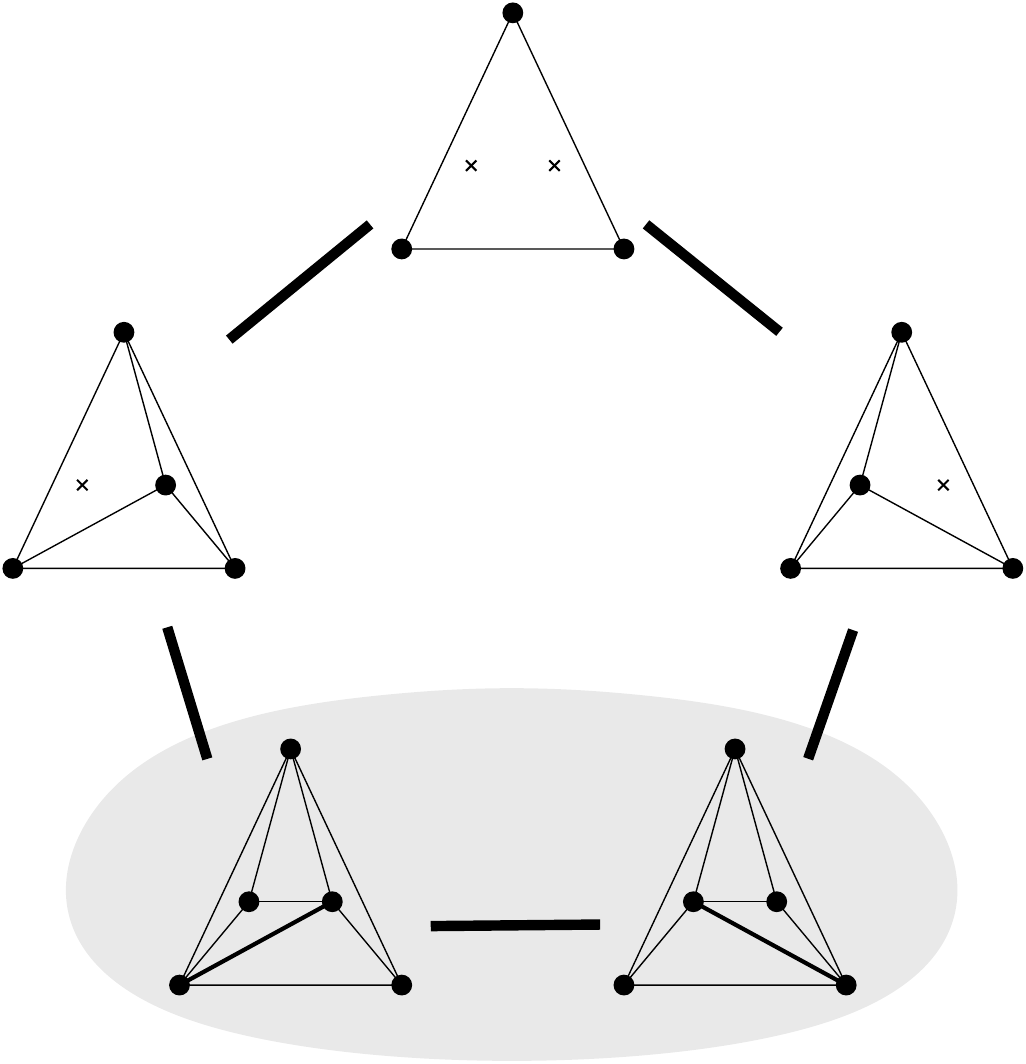}{\textwidth}
\end{minipage}
}
\caption{Bistellar flip graphs for \five points. Small crosses indicate skipped points in $P$. The shaded areas contain the edge flip graph, the subgraph induced by full triangulations.}
\label{fi:TheFiveExamples}
\end{figure}

\begin{figure}[htb]
\centerline{
\begin{minipage}[b]{0.43\textwidth}
\placefig{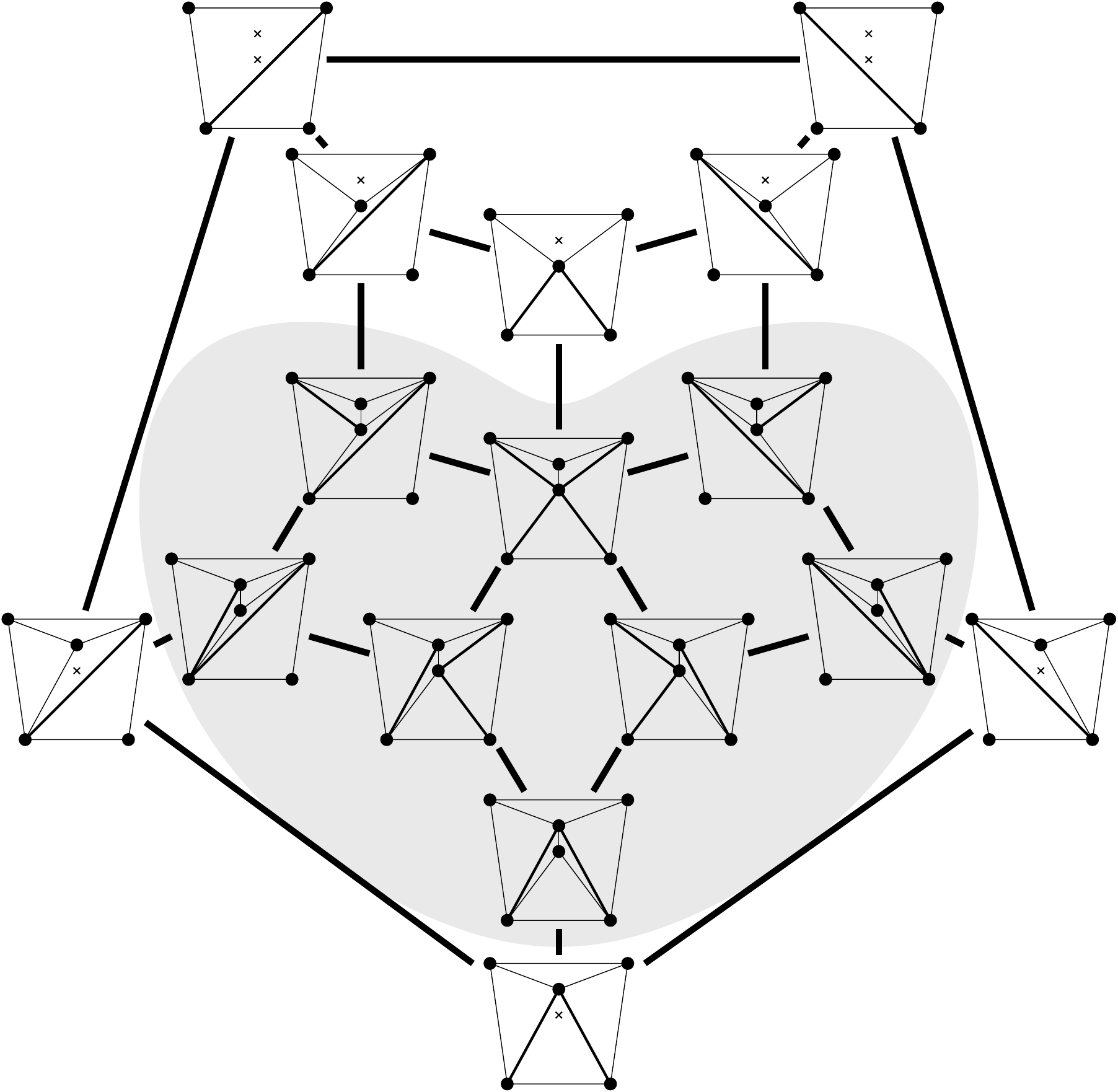}{0.9\textwidth}
\end{minipage}
\hspace{-2.7em}
\begin{minipage}[b]{0.13\textwidth}
\placefig{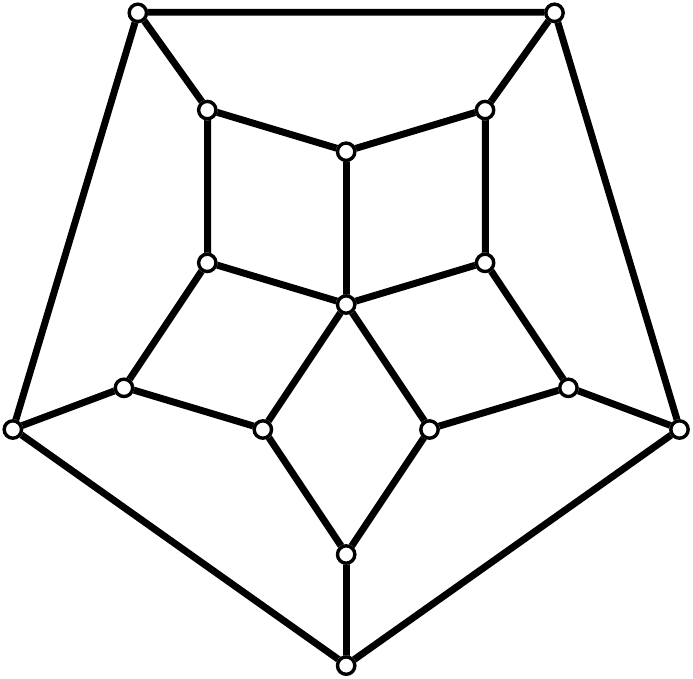}{\textwidth}
\end{minipage}
\hspace{-1.3em}
\begin{minipage}[b]{0.43\textwidth}
\placefig{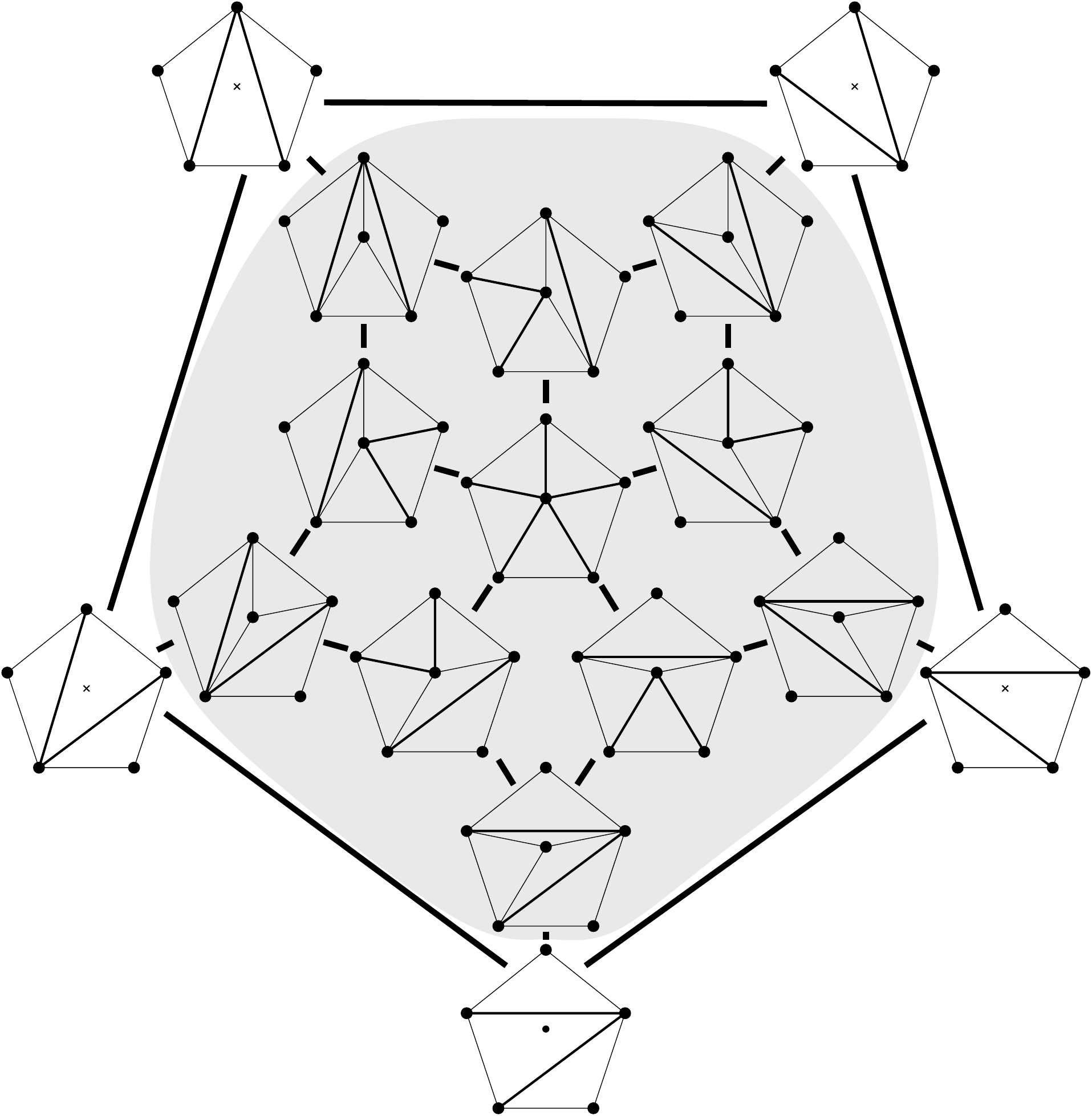}{0.9\textwidth}
\end{minipage}
}
\vspace{-0.36cm}
\centerline{
\,
\begin{minipage}[b]{0.43\textwidth}
\placefig{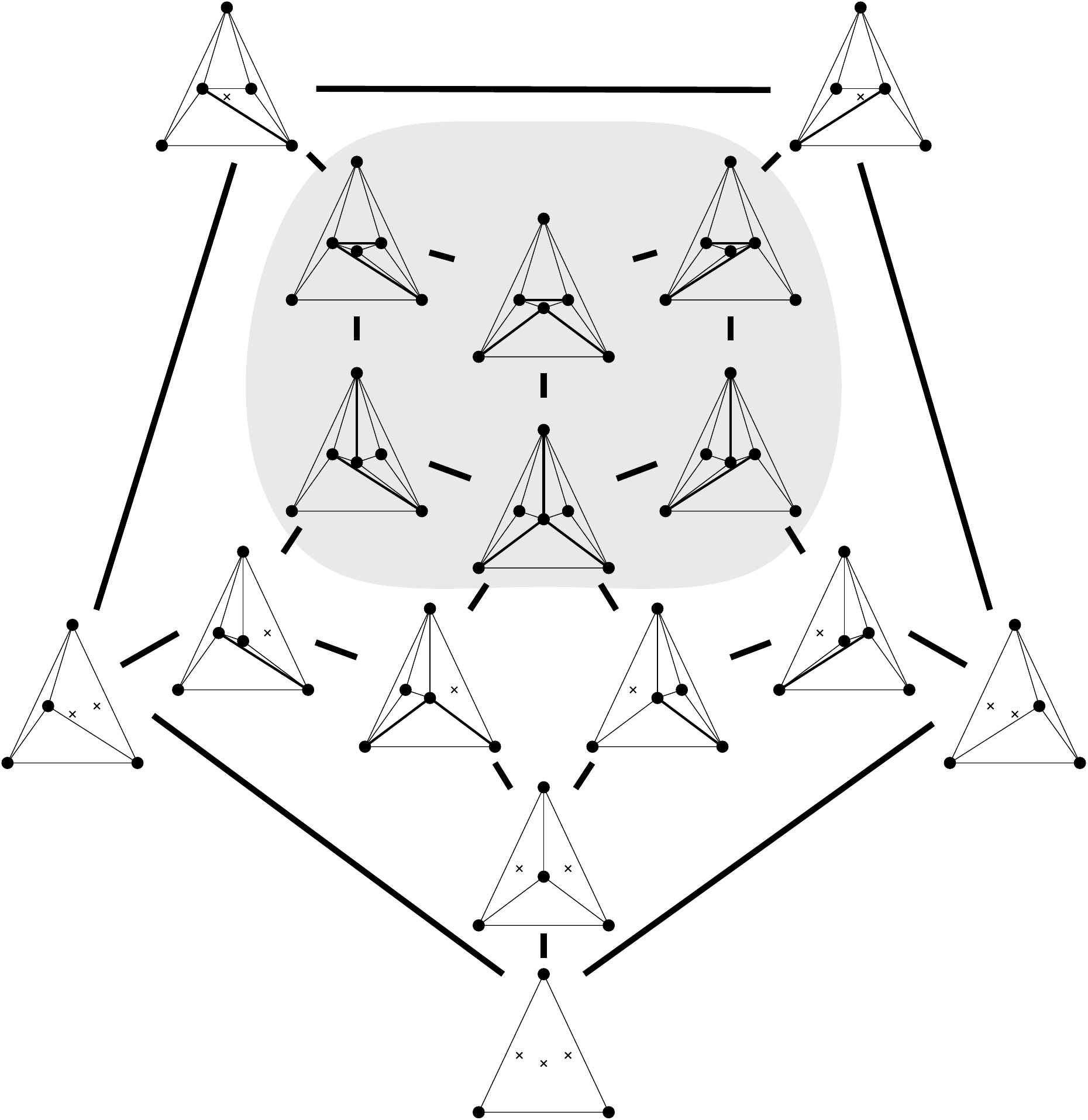}{0.9\textwidth}
\end{minipage}
}
\caption{Sets of \six points with isomorphic bistellar flip graphs of partial triangulations. The shaded areas cover the edge flip graphs of full triangulations. (Crosses indicate skipped points.)}
\label{fi:ExampleTwoInWheel}
\end{figure}

Let us first supply the basic definitions and then discuss the results in more detail.
\begin{definition}
[point set]
\label{d:PointSet}
Throughout this paper we let $P$ denote a finite planar point set in \Emph{general position} (\ie no three points on a line) with $n\ge 3$ points. The set of \Emph{extreme points} of $P$ (\ie the vertices of the convex hull of $P$) is denoted by $\ext{P}$, and we let $\inn{P}:= P \setminus \ext{P}$ denote the  set of \Emph{inner} (\ie non-extreme) \Emph{points} in $P$. We consistently use the notation $\extNb=\extNb(P):= |\ext{P}|$ and $\innNb = \innNb(P) := |\inn{P}| = n-\extNb$, and we let $\EdsHull = \EdsHull(P) \subseteq {P \choose 2}$ denote the set of edges of the convex hull of $P$, $|\EdsHull|=\extNb$.
\end{definition}
\begin{definition}
[plane graph]
\label{d:Plane}
For graphs $G=(P',E)$, $P' \subseteq P$, $E \subseteq {P' \choose 2}$, on $P'$ we often identify edges $\{p,q\}\in E$ with their corresponding straight line segments $pq$. We let $\Pts{G} := P'$ and $\Eds{G}:=E$.

A graph $G$ on $P$ is \Emph{plane} if no two straight line segments corresponding to edges in $\Eds{G}$ cross (\ie they are disjoint except for possibly sharing an endpoint). For $G$ plane, the bounded connected components of the complement of the union of the edges are called \Emph{regions}, the set of regions of $G$ is denoted by $\Reg{G}$.
\end{definition}
Note that regions are \emph{open sets}. Note also that isolated points in a plane graph are ignored in the definition of regions.
\begin{definition}
[full, partial, regular triangulation]
\label{d:Triang}
\begin{EnumAlph}
\item
A \Emph{full triangulation} of $P$ is a maximal plane graph $T=(P,E)$.
\item
A \Emph{partial triangulation} of $P$ is a full triangulation $T=(P',E)$ with $\ext{P} \subseteq P' \subseteq P$ (hence $\EdsHull \subseteq \Eds{T}$). 
\item
A \Emph{regular triangulation} of $P$ is a triangulation obtained by projecting (back to $\RR^2$) the edges of the lower convex hull of a generic lifting of $P$ to $\RR^3$ (\ie we add third coordinates to the points in $P$ so that no four lifted points are coplanar).
\end{EnumAlph}

\noindent
Points in $\PtsInn{T} := \inn{P} \cap \Pts{T}$ are called \Emph{inner points} of $T$ and points in $\inn{P}\setminus\PtsInn{T}$ are called \Emph{skipped} in $T$ (clearly, for a full triangulation $\PtsInn{T}=\inn{P}$ and no points are skipped).  Edges in $\EdsInn{T} := \Eds{T}\setminus\EdsHull$ are called \emph{inner edges} and edges in $\EdsHull$ are called \Emph{boundary edges}. 

$\fT(P)$, $\pT(P)$, and $\rT(P)$ will denote the set of all full, partial, and regular triangulations of $P$, respectively.
\end{definition}
Every full and every regular triangulation of $P$ is also a partial triangulation of $P$. If $P$ is in convex position (\ie $\ext{P}=P$), all three notions coincide. It is well-known, that there are point sets with non-regular triangulations, \cite{LRS10}, see \Sec{s:Mother} and \Fig{f:Mother}.
\begin{definition}
[edge flip, point insertion flip, point removal flip]
\label{d:BistFl}
Let $T \in \pT(P)$. 

An edge $e \in \EdsInn{T}$ is called \Emph{flippable in} $T$ if removing $e$ from $T$ creates a convex quadrilateral region $Q$. In this case, we denote by $T[e]$ the triangulation with the other diagonal $\overline{e}$ of $Q$ added instead of $e$, \ie $\Pts{T[e]} := \Pts{T}$ and $\Eds{T[e]} := \Eds{T} \setminus \{e\} \cup \{\overline{e}\} = \Eds{T} \oplus \{e,\overline{e}\}$;\footnote{Here and throughout this paper, we use the notation $\oplus$ for the symmetric difference between sets.} we call this an \Emph{edge flip}. Occasionally, when we want to emphasize the new edge $\overline{e}$, we will also use the alternative notation $T[\Nicefrac{e}{\overline{e}}]$ instead of $T[e]$, or write $T[e]=T[\Nicefrac{e}{\overline{e}}]$. 
\smallskip

A point $p \in \inn{P}$ is called \Emph{flippable in} $T$ if $p\in \inn{P} \setminus \PtsInn{T}$, or if $p \in \PtsInn{T}$ is of  degree \three in $T$.
(a)~If $p\in \inn{P} \setminus \PtsInn{T}$ then $T[p]$ is the \ptriangulation with $p$ added as a point of degree \three (there is a unique way to do so); we call this a \Emph{point insertion flip}.
(b)
If $p \in \PtsInn{T}$ is of degree \three in $T$ then $T[p]$ is obtained by removing $p$ and its incident edges; we call this a \Emph{point removal flip}. See \Fig{fi:Flips}.

A \Emph{bistellar flip} is one of the three: an edge flip, a point insertion flip, or a point removal flip.
\end{definition}
Hence, whenever we write $T[x]$ for a partial triangulation $T$, then $x$ is either a flippable point in $\inn{P}$ or a flippable edge in $\EdsInn{T}$. We will use $T[x,y]$ short for $(T[x])[y]$, \etc
\begin{definition}[bistellar flip graph, edge flip graph]
The \Emph{bistellar flip graph} of $P$ is the graph with vertex set $\pT(P)$ and edge set $\{\{T,T[x]\} \,|\, T \in \pT(P)\mbox{, $x\in \inn{P} \cup \EdsInn{T}$ flippable in~} T \}$. 

The \Emph{edge flip graph} of $P$ is the graph with vertex set $\fT(P)$ and edge set $\{\{T,T[e]\} \,|\, T \in \fT(P)\mbox{, $e\in  \EdsInn{T}$ flippable in~} T \}$.
See \Figs{fi:TheFiveExamples} and \ref{fi:ExampleTwoInWheel} for examples.
\end{definition}
%
%
\subsection{Results -- full triangulations}
The \Emph{edge flip graph} is the subgraph of the bistellar flip graph induced by $\fT(P)$ (see \Figs{fi:TheFiveExamples} and \ref{fi:ExampleTwoInWheel}). Lawson \cite{L72} showed that the edge flip graph is connected. A trivial upper bound for the vertex connectivity (see \Def{de:Connectivity}) of any graph is the minimum vertex degree of the graph; for the edge flip graph this is the minimum number of flippable edges in any triangulation of $P$. This number has been investigated by Hurtado \etal\ in 1999, \cite[Thm.\,4.1]{HNU99}.
\begin{theorem}[\cite{HNU99}]
\label{t:FullNbFlips}
Any $T \in \fT(P)$ has at least $\lceil\frac{n}{2}-2\rceil$ flippable edges. This bound is tight for all $3 \le n \in \NN$\,\footnote{We let $\NN$ denote the set of positive integers and we let $\NNnull := \NN \cup \{0\}$.} (\ie for every $n$ there is a set $P$, $|P|=n$, with a triangulation with exactly $\lceil\frac{n}{2}-2\rceil$ flippable edges).
\end{theorem}
We provide a lower bound on the vertex connectivity of the edge flip graph that matches this lower bound of $\lceil\frac{n}{2}-2\rceil$ for the minimum vertex degree, actually in the refined form $\FBound$, see Hoffmann \etal\ \cite{HSSTWbook13}. Note, however, that there are sets of points where all full triangulations allow more than $\FBound$ edge flips.\footnote{Consider, \eg the top left point set in \Fig{fi:ExampleTwoInWheel}: $n=6$, $h=4$, thus $\FBound = 1$, but the edge flip graph has minimum vertex degree \two.} We show that, for $P$ large enough, the minimum vertex degree always determines the vertex connectivity of the edge flip graph.
\begin{theorem}
\label{t:MainFull}
\begin{EnumRom}
\item
\label{it:iMainFull}
There exists $n_0 \in \NN$, such that the edge flip graph of any set of $n \ge n_0$ points in general position in the plane is $\delta$-vertex connected, where $\delta$ is the minimum vertex degree in the edge flip graph.
\item
\label{it:iiMainFull}
For $5 \le n \in \NN$, the edge flip graph of any set of $n$ points with $h$ extreme points in general position is $\FBound$-vertex connected. This is tight: For every $n \in \NN$ there is a triangulation of some set of $n$ points with no more than $\FBound$ flippable edges.
\end{EnumRom}
\end{theorem}
Obviously, for $n\ge n_0$, \ItemRef{it:iMainFull} implies \ItemRef{it:iiMainFull}. In fact, we do not know whether the restriction ``$n$ large enough'' is required in \ItemRef{it:iMainFull}. Still, apart from covering the range to $n_0$, we consider the proof for \ItemRef{it:iiMainFull} of independent interest, since it provides some extra insight to the structure of the edge flip graph via so-called subdivisions, and it is an introduction to the proof for the bistellar flip graph.
%
%
\subsection{Results -- partial triangulations}
The bistellar flip graph is connected, as it follows easily from the connectedness of the edge flip graph, see \cite[Sec.\,3.4.1]{LRS10}. Here is the counterpart of \Thm{t:FullNbFlips} addressing the minimum vertex degree in the bistellar flip graph, shown by De Loera \etal\ in 1999, \cite[Thm.\,2.1]{LSU99}.
\begin{theorem}[\cite{LSU99}]
\label{t:PartNbFlips}
Any $T \in \pT(P)$ allows at least $n-3$ flips. This bound is tight for all $P$.
\end{theorem}
Again, we show that the vertex connectivity equals the minimum degree. 
\begin{theorem}
\label{t:MainPart}
Let $4 \le n \in \NN$. The bistellar flip graph of any set of $n$ points in general position in the plane is $(n-3)$-vertex connected. This is tight: Any triangulation of a point set that skips all inner points has degree $n-3$ in the bistellar flip graph.\footnote{There are exactly $h-3$ edge flips and exactly $n-h$ point insertion flips.}
\end{theorem}
This answers (for points in general position) a question mentioned by De Loera, Rambau \& Santos in 2010, \cite[Exercise\,3.23]{LRS10}, and by Lee \& Santos in 2017, \cite[pg.\,442]{LS17}. 
\bigskip

Before we mention further results, we provide some context. Along the way, we encounter some tools and provide intuition relevant later in the paper.
%
%
\subsection{Context -- convex position, associahedron, and Balinski's Theorem}
\label{s:Associahedron}
\begin{figure}[htb]
\centerline{
\placefig{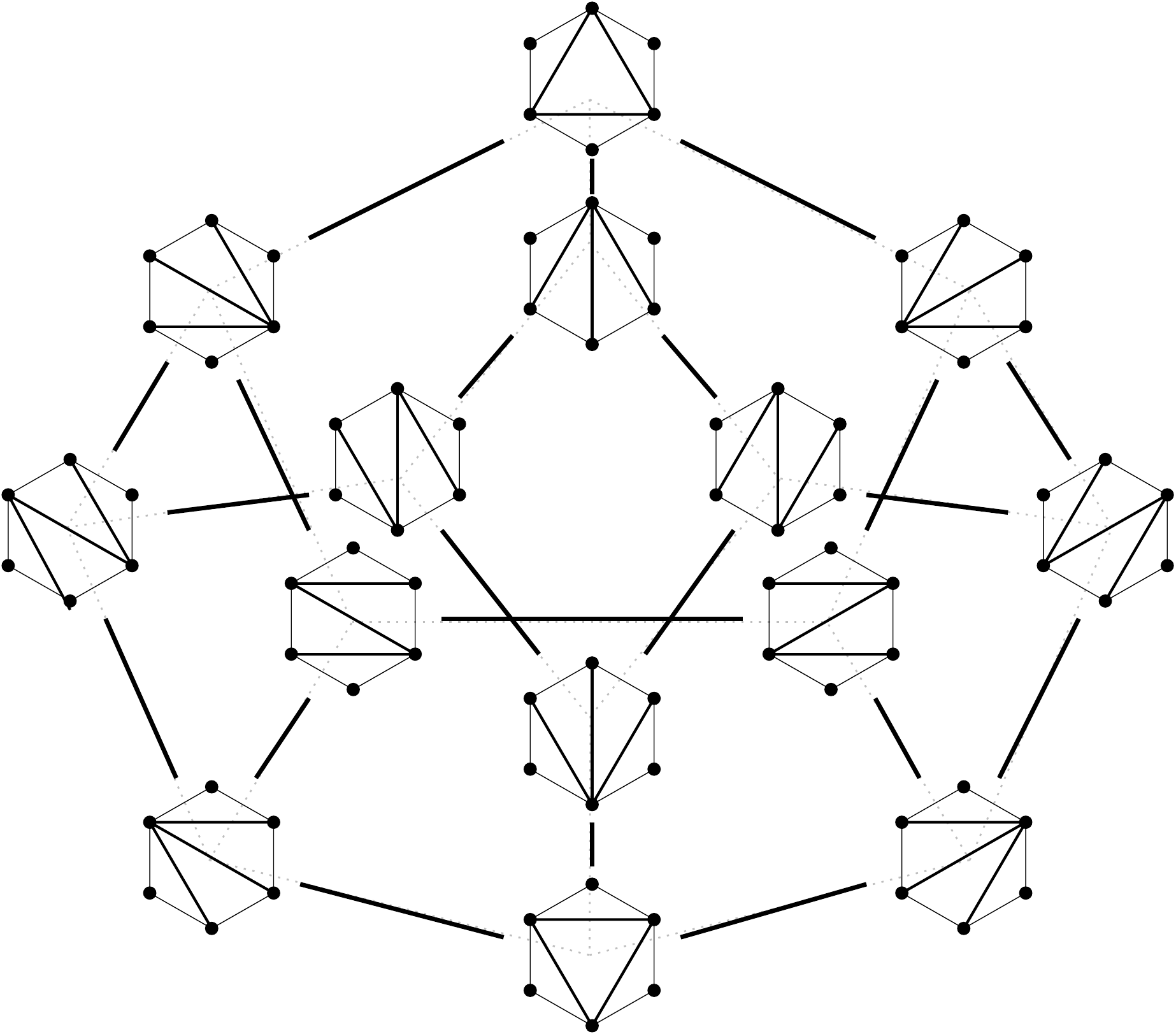}{0.3\textwidth}
\hspace{3em}
\placefig{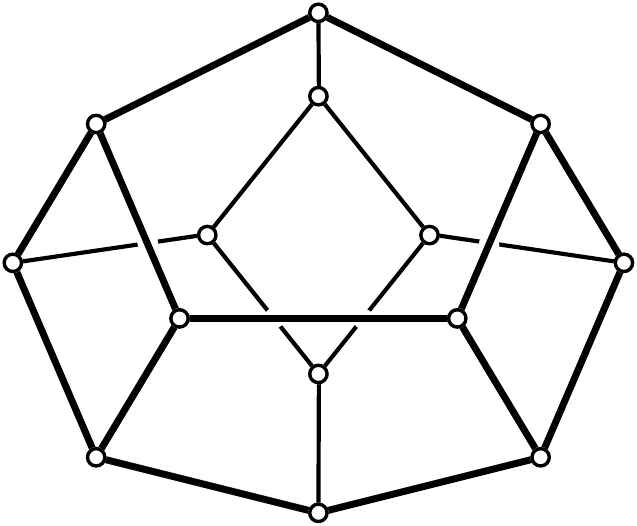}{0.15\textwidth}
}
\caption{The flip graph of a convex hexagon, the graph of the 3-dimensional associahedron.}
\label{fi:SixExampleConvex}
\end{figure}

\begin{figure}[htb]
\centerline{
\placefig{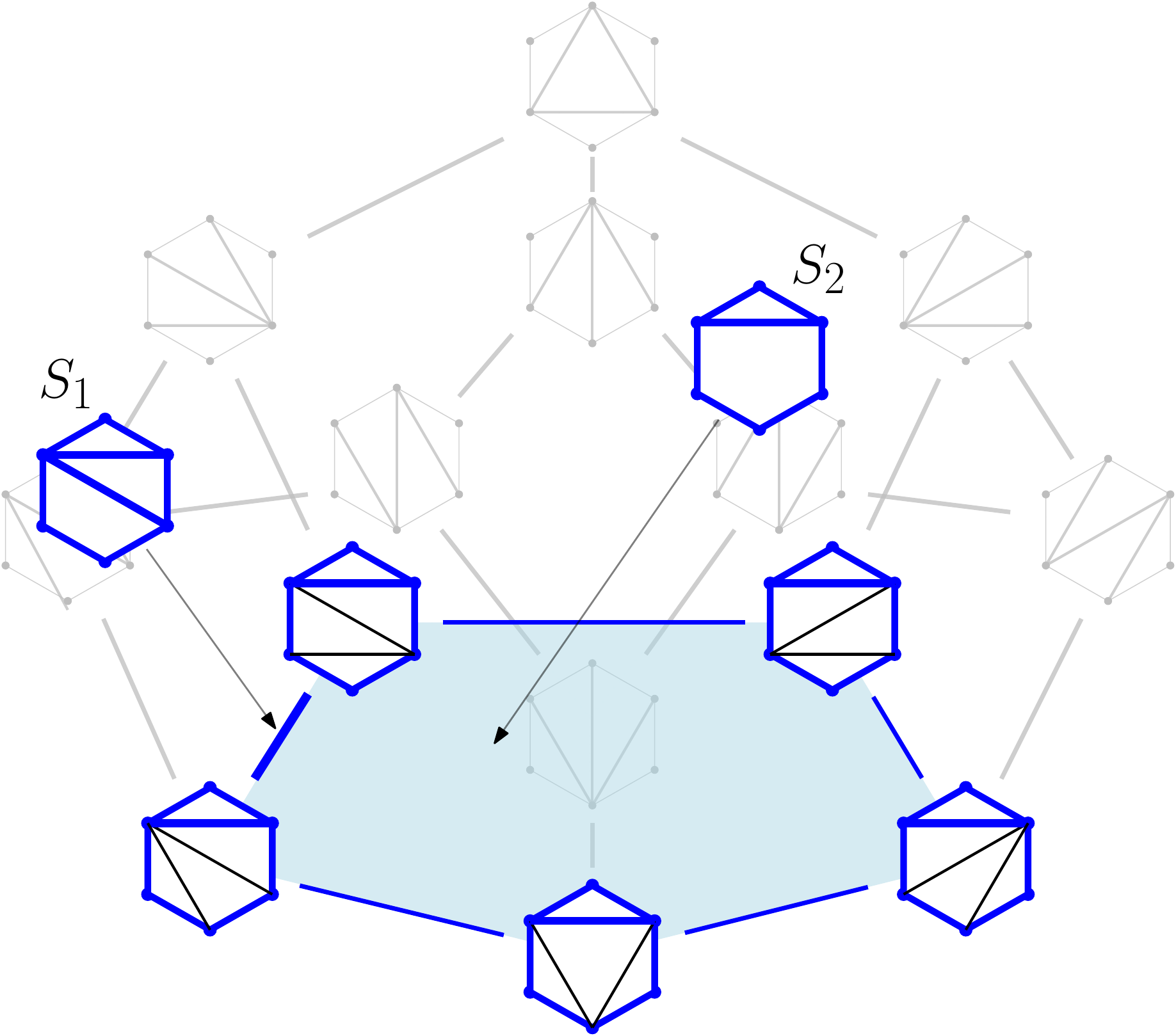}{0.3\textwidth}
\hspace{3em}
\placefig{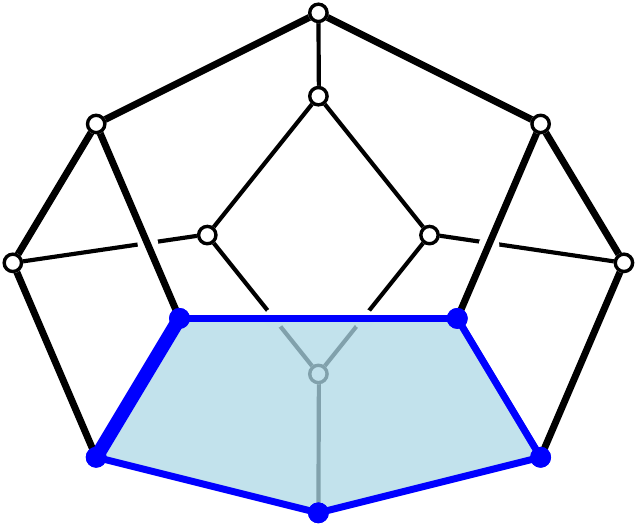}{0.15\textwidth}
}
\caption{A subdivision $S_1$ with one edge missing representing an edge (1-face) of the 3-dimensional associahedron, and a subdivisions $S_2$ with two edges missing representing a facet (2-face).}
\label{fi:SubdVsFace}
\end{figure}

Suppose $P$ is in convex position. Then $\fT(P)=\pT(P)=\rT(P)$ is the set of triangulations of a convex $n$-gon whose study goes back to Euler, with one of the first appearances of the Catalan Numbers. There is an $(n-3)$-dimensional convex polytope, the \emph{associahedron}, whose vertices correspond to the triangulations of a convex $n$-gon, and whose edges correspond to edge flips between these triangulations, see \cite{CSZ15} for a historical account. That is, the 1-skeleton (graph) of this polytope is isomorphic to the flip graph of $P$, see \Fig{fi:TheFiveExamples}(left) for $n=5$ and \Fig{fi:SixExampleConvex} for $n=6$. Here  Balinski's Theorem from 1961 comes into play.
\begin{theorem}[Balinski's Theorem, \cite{Balinski61}]
\label{t:Balinski}
The 1-skeleton of a convex $d$-dimensional polytope is at least $d$-vertex connected.
\end{theorem}
Thus we can conclude that the flip graph of $n$ points in convex position is $(n-3)$-vertex connected. The face structure of the associahedron is easily explained via so-called subdivisions (also known as polyhedral subdivisions or convex decompositions), which will feature prominently in our arguments. Specialized to the setting of convex position, a subdivision $S$ is a plane graph on $P$ with all boundary edges present, and some diagonals. We identify every such subdivision with the set of all its possible completions to a triangulation (which we will call \Emph{refinements}). \Fig{fi:SubdVsFace} illustrates on two examples for $n=6$, that if $S$ misses $i$ edges towards a triangulation, then its completions to triangulations correspond to the vertices of an $i$-face of the associahedron; in particular, no edge missing (a triangulation) corresponds to a vertex, and one edge missing corresponds to an edge (a flip). This correspondence will guide our intuition, even in more general settings in which there is no polytope in the background.
%
%
\subsection{Context -- regular triangulations and secondary polytope}
\label{s:SecondaryPolytope}
\begin{figure}[htb]
\centerline{
\begin{minipage}[c]{0.45\textwidth}
\placefig{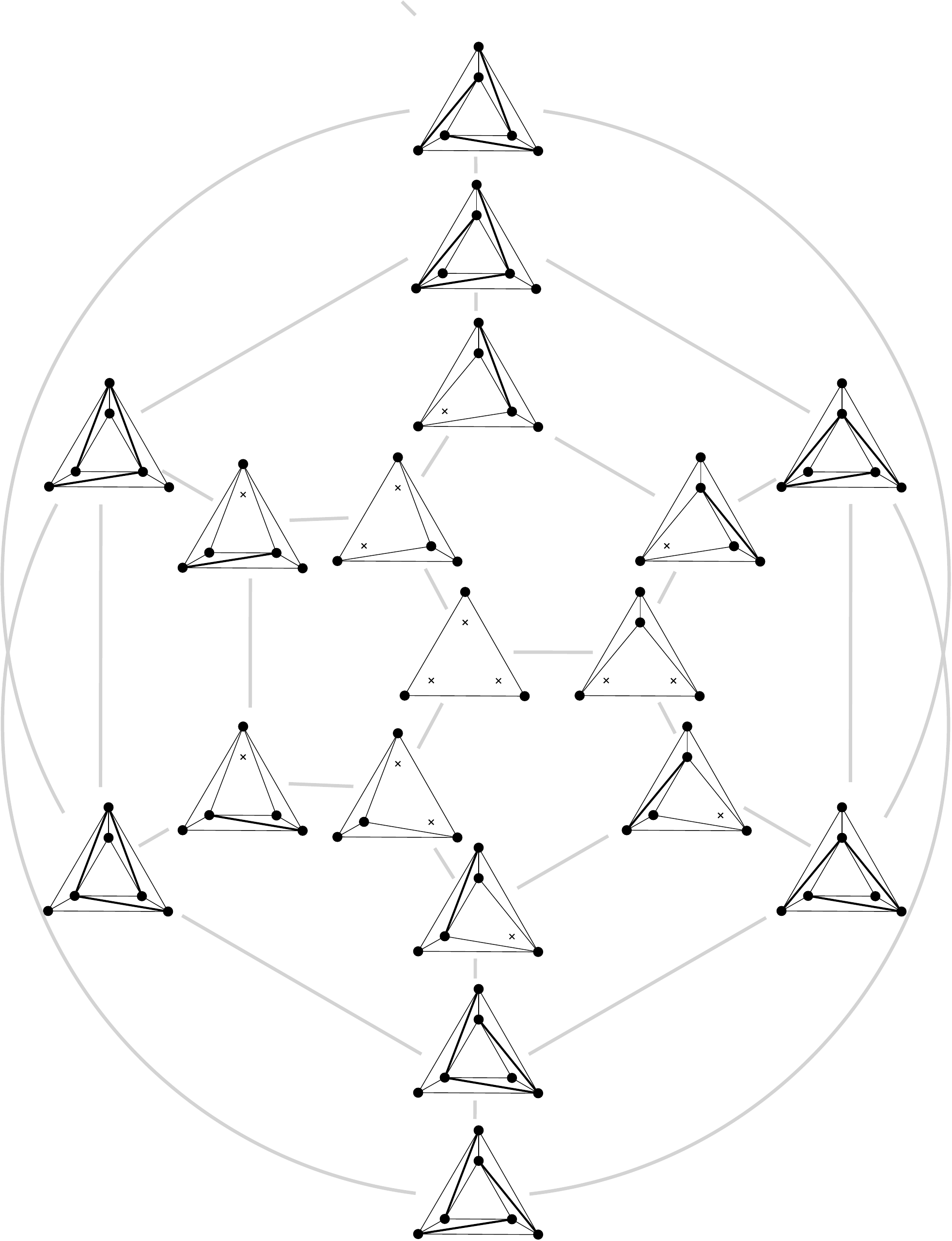}{\textwidth}
\end{minipage}
\hspace{0.5em}
\begin{minipage}[c]{0.25\textwidth}
\centerline{\placefig{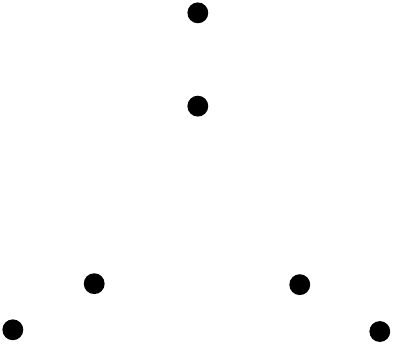}{0.63\textwidth}}
\ \\[1ex]
\mbox{~} \hfill \ \\[2ex]
\centerline{\placefig{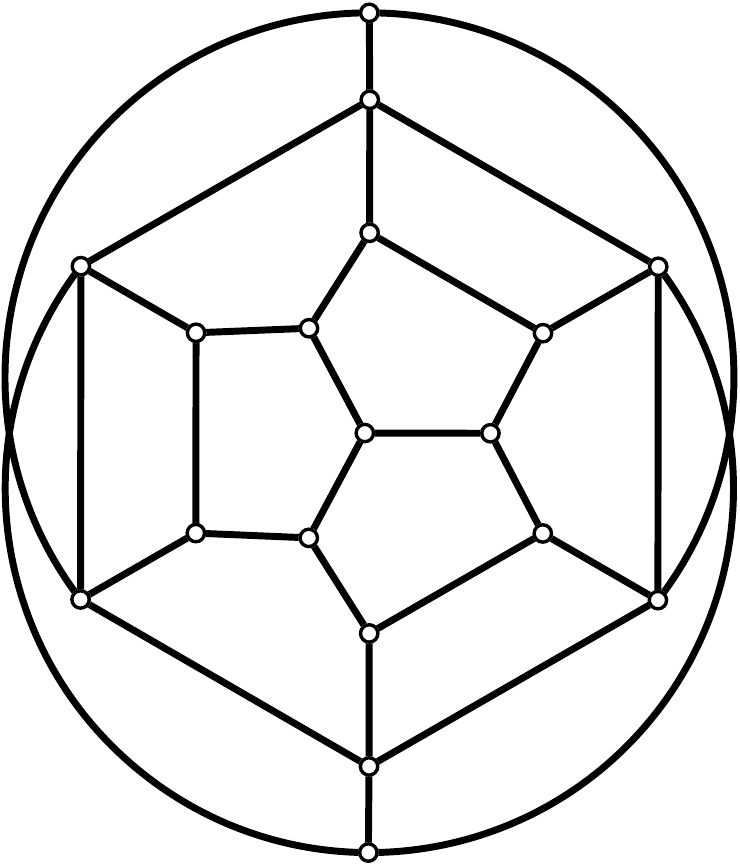}{0.8\textwidth}}\\
\mbox{~} \hfill 
\end{minipage}
}
\caption{The \bfg\ of the mother-of-examples configuration.}
\label{f:MotherFlipGraph}
\end{figure}

All examples of bistellar flip graphs we have seen so far -- \Figs{fi:TheFiveExamples} and \ref{fi:ExampleTwoInWheel}, and sets in convex position -- are graphs of polytopes, which is not true in general.\footnote{Up to \six points, there is only one order type which has partial triangulations that are not regular, see \Fig{f:MotherFlipGraph} in \Sec{s:ImplRegularityPreserv}.} The reason is simply that for the underlying point sets in these examples all partial triangulations are regular. Specifically, there is the following generalization of the associahedron, a highlight of the topic of triangulations from the late eighties due to Gelfand \etal\ \cite{GKZ90} (see \cite[Thm.\,5.1.9]{LRS10}) (here specialized to points in the plane).
\begin{theorem}[secondary polytope, \cite{GKZ90}]
\label{t:SecondaryPolytope}
For every set $P$ of $n$ points in general position in the plane, there is an $(n-3)$-dimensional polytope $\SecPoly{P}$, the \emph{secondary polytope} of $P$, whose 1-skeleton is isomorphic to the bistellar flip graph of \emph{regular triangulations }of $P$.
\end{theorem}
Again, with the help of Balinski's Theorem (\Thm{t:Balinski}), we immediately get: The bistellar flip graph of \emph{regular} triangulations of $P$ is $(n - 3)$-vertex connected. Our \Thm{t:MainPart} is the corresponding result for the bistellar flip graph of \emph{all partial} triangulations. Note, however, that it is not a generalization, since it does not imply the result for regular triangulations.
%
%
\subsection{Context -- simplicial complex of plane graphs and its dual flip complex}
\label{s:FlipComplex}
A polytopal representation of the edge flip graph or the bistellar flip graph of all partial triangulations is not known. In fact, a clean result as with the secondary polytope is not possible (see \cite{LRS10} for details). However, there is a construction, due to Orden \& Santos \cite{OS05}, of a simple high-dimensional polytope whose $1$-skeleton represents all so-called \emph{pseudo-triangulations} of a planar point set and flips between them, following an earlier construction by Rote \etal\ \cite{RSS03} of a corresponding polytope for \emph{pointed pseudo-triangulations}.

Moreover, the edge flip graph of full triangulations of a planar point set does form the $1$-skeleton of a closely related higher-dimensional structure, the \Emph{flip complex}, first described by Orden \& Santos \cite{OS05} and later rediscovered by Lubiw \etal\ \cite{LMW19}. The flip complex is not a polytope, but it is a polytopal complex (informally, a collection of convex polytopes that intersect only in common faces) with a particularly simple topology (it is homotopy equivalent to a ball). For the reader familiar with these notions, we briefly review how our findings fit into this context (for a quick reference for the relevant terminology from topology theory, see, \eg \cite{B95}). This is not essential as a tool for our proofs and the rest of the paper can be followed without this context. Rather, our results shed some extra light on these structures considered in the literature.

Following \cite{LMW19}, let us consider
\[
\mathbb{T} := \mathbb{T}(P) = \{\mbox{$E \subseteq {P \choose 2}$} \mid \mbox{~no two edges in $E$ cross}\},
\]
\ie these are the edge sets of plane graphs on $P$. Since this family of sets is closed under taking subsets, it is a simplicial complex, and we call its elements \emph{faces} of $\mathbb{T}$. By definition, the \emph{dimension} of a face $F \in \mathbb{T}$ is $\dim F := |F| - 1$. The inclusion-maximal faces (called \emph{facets} of $\mathbb{T}$) are exactly the edge sets of full triangulations of $P$. All facets have equal dimension\footnote{A simplicial complex all of whose facets have the same dimension is called \emph{pure}.} $m-1$ (where $m:=3n-h-3$, see \Lm{le:EdgeCount}).

In \cite{LMW19}, it is shown that $\mathbb{T}$ is shellable and homeomorphic to an $(m-1)$-dimensional ball (this also follows also from the results in \cite{OS05}). This fact was instrumental in the proof of the so-called \emph{orbit conjecture} regarding flips in edge-labeled full triangulations formulated in \cite{Bose:Flipping-edge-labelled-triangulations-2018} and proved in \cite{LMW19}.

A face $F$ of dimension $m-2$ is a plane graph on $P$ with exactly one edge missing towards a full triangulation. If $F$ is obtained from a full triangulation by removal of a flippable edge, there are exactly two ways to complete it to a triangulation -- we call $F$ an \Emph{interior $(m-2)$-face}; otherwise, there is exactly one way to do so -- we call $F$ a \Emph{boundary $(m-2)$-face}. A face of any dimension is called a \Emph{boundary face} if it is contained in a boundary $(m-2)$-face, and it is called an \emph{interior face}, otherwise (hence, all facets are interior faces). We have the following, \cite[Prop.\,3.7]{LMW19}.
\begin{theorem}[\cite{LMW19}]
\label{t:InteriorFace}
$F \in \mathbb{T}$ is an interior face iff $\EdsHull \subseteq F$, and all bounded regions of $F$ are convex; (in other words, iff $(P,F)$ is what we will call a subdivision, see \Def{d:FullSubdivision} and also discussion in \Sec{s:Associahedron}).
\end{theorem}
From the Coarsening Lemma~\ref{l:FullCoarsening} in \Sec{s:FullSubdivisions}, the following property of $\mathbb{T}$ is implied.
\begin{theorem}
\label{t:CoarseningSimplicialComplex}
Every interior face $F$ of  $\mathbb{T}$ contains an interior face of dimension $m +1 - \lceil \frac{n}{2} \rceil$.
\end{theorem}
The flip graph can be seen as a structure dual to $\mathbb{T}$, with the vertices of the flip graph corresponding to the facets of $\mathbb{T}$, and two vertices adjacent in the flip graph, if the corresponding facets of $\mathbb{T}$ share (\ie contain) a common $(m-2)$-face (which is clearly interior). This can be generalized to the flip complex $\mathbb{X}= \mathbb{T}^*$, see \cite{LMW19}, with each face of $\mathbb{X}$ (which is dual to an interior face of $\mathbb{T}$) corresponding to a subdivision as  we define it below (see \Thm{t:InteriorFace} above). Each such face is a product of associahedra, and the vertices of such a face of the flip complex are the triangulations refining this subdivision (\Def{d:FullCoarsening}). In terms of the flip complex, the Coarsening Lemma~\ref{l:FullCoarsening} and \Thm{t:CoarseningSimplicialComplex} can be restated as saying that every inclusion-maximal face of the flip complex is of dimension at least $\max\{\frac{n}{2} - 2, h-3\}$ (see also \Thm{th:EdgeInProdAssoc}).
%
%
\subsection{Results -- regular triangulations}
We study the (well-known, see \cite{LRS10}) partially ordered sets of \emph{full and partial subdivisions} of $P$, respectively (see \Defs{d:FullSubdivision} and \ref{d:PartSubdivision}), in which triangulations are the minimal elements. We introduce the notions of \Emph{slack} of a subdivision (\Defs{d:FullSlack} and \ref{d:PartSlack}), \Emph{perfect coarsening} (\Def{d:DirectPerfectCoarsening}) and \Emph{perfect coarsener} (\Def{d:Coarsener}), and we prove the so-called Coarsening Lemmas~\ref{l:FullCoarsening} and \ref{le:CoarsenPartSubd} (these can be considered extensions of \Thms{t:FullNbFlips} and \ref{t:PartNbFlips}). We consider these notions and lemmas our main contributions besides \Thms{t:MainFull} and \ref{t:MainPart}. Together with a sufficient condition for the regularity of partial triangulations and subdivisions (\Thm{t:LowClosureRegular} and Regularity Preservation Lemma~\ref{le:RegularToRegular}), these yield several other results on the structure of flip graphs. In particular, they allow us to settle, in the unexpected direction, a question by F.\,Santos \cite{SanPersCom} regarding the size of certificates for the existence of non-regular triangulations of a given point set in the plane.
\begin{theorem}
\label{t:MinNonRegExample}
For all $n \in \NN$ there is a set of at least $n$ points in general position in the plane, which has non-regular triangulations, and for which any proper subset has only regular triangulations (in other words, only full triangulations of the set can be non-regular).
\end{theorem}
This should be seen in contrast with the situation in higher dimensions. In large enough dimension, every point set with non-regular triangulations has a subset of bounded size (in the dimension) with non-regular triangulations.  This holds since,  (a) the vertices of any realization of the cyclic polytope with 12 vertices in dimension 8 has non-regular triangulations, \cite{LHSS96} (see \cite[Sec.\,5.5.2]{LRS10}), and (b) every large enough set of points in general position in $\RR^8$ has a subset of 12 points which are vertices of a cyclic polytope (this follows from Ramsey's Theorem, \cite{GRS90, EM13,Suk14}).
%
%
\subsection{Approach}
\label{s:Approach}
All our vertex connectivity bounds rely on a local variant of Menger's Theorem, the Local Menger Lemma~\ref{le:LocalMenger}. This lemma says that, assuming connectedness, in order to show $k$-vertex connectivity, it is enough to show $k$ internally{} vertex-disjoint paths between any two vertices \emph{at distance \two}. Then, in order to establish the min-degree bound of \Thm{t:MainFull}\ref{it:iMainFull}, we explicitly construct the necessary paths between vertices at distance \two, \ie triangulations $T[e]$ and $T[f]$.

\begin{figure}[htb]
\centerline{\placefig{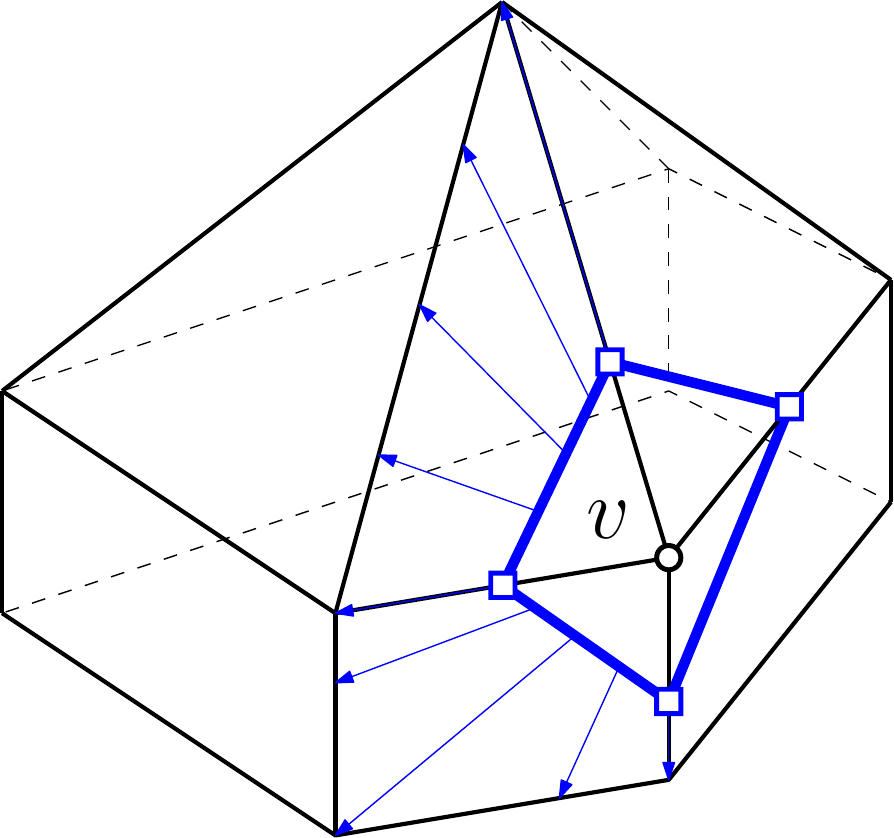}{0.2\textwidth}}
\caption{The vertex figure of a polytop, with the mapping of paths in the vertex figure to the 1-skeleton of the polytope indicated.}
\label{f:VertexFigure}
\end{figure}

For the bounds in \Thms{t:MainFull}\ref{it:iiMainFull} and \ref{t:MainPart}, we look at the neighborhood of a triangulation $T$ (which is in one-to-one correspondence with the flippable elements in $T$), supplied with a compatibility relation between the flippable elements (for the edge flip graph, two flippable edges $e$ and $f$ are compatible, if $e$ remains flippable after flipping $f$, \Def{d:FlipEdgeRel}). We call this the link\footnote{Links are a basic notion in the theory of simplicial and polytopal complexes. In terms of the flip complex $\mathbb{X}$ discussed above, $T$ corresponds to a vertex of $\mathbb{X}$, and what we call link here is the $1$-skeleton of the link (in the topological sense) of the vertex $T$ in $\mathbb{X}$.} of $T$, a structure motivated by the \Emph{vertex figure} of a vertex in a polytope, see \cite[pg.\,54]{Zie95}: Recall that for a vertex $v$ in a $d$-polytope ${\cal P}$, its vertex figure is the $(d-1)$-polytope ${\cal P}'$ obtained by intersecting ${\cal P}$ with a hyperplane that separates $v$ from the remaining vertices of the polytope. Vertices of ${\cal P}'$ correspond to edges of ${\cal P}$ incident to $v$, edges in the graph of ${\cal P}'$ correspond to $2$-faces of ${\cal P}$ incident to $v$. As indicated in  \Fig{f:VertexFigure}, there is a natural way of mapping paths in the graph of ${\cal P}'$ to paths in the graph of ${\cal P}$. 
Using the Local Menger Lemma~\ref{le:LocalMenger}, this can be easily made an inductive proof of Balinski's Theorem (\Thm{t:Balinski}). We follow exactly this line of thought for flip graphs (where 4- and 5-cycles will play the role of 2-faces), except that we will not need induction: The link of a triangulation avoids 4-cycles \emph{in its complement}, which turns out to directly yield sufficient vertex connectivity (\Lm{l:C4FreeConn}).

That is, we borrow intuition from polytope theory, although we know that the edge flip graph and the bistellar flip graph are in general not graphs of polytopes. However, as we will see in \Secs{s:fCoveringFlipGraph} and \ref{s:Covering}, the edge flip graph and the bistellar flip graph can be covered by graphs of $\lceil \frac{n}{2}-2\rceil$- and $(n-3)$-polytopes, respectively.

It is perhaps worthwhile to mention, that interestingly our proofs never supply insight about the flip graphs to be connected, this is assumed and never proved here. That is, the techniques will probably \emph{not} be able to say something about the connectedness of flip graphs in higher dimension (3 and 4, where this question is open). If at all, it might help analyzing the vertex connectivity of the connected components of flip graphs.
%
%
\subsection{Paper organization}
\label{s:Organization}
In \Sec{se:GraphConnectivity} we show the two lemmas on graph connectivity we mentioned already in \Sec{s:Approach}: The Local Menger Lemma and a lemma about the vertex connectivity of graphs with no 4-cycle in their complement. These may be of independent interest, if only as exercises after teaching Menger's Theorem in class.

Then the paper splits in three parts, where the second and third part on partial triangulations is largely independent of the first part about full triangulations.
\begin{EnumNo}
\item \Sec{se:MinDegreeBound} shows the min-degree bound of \Thm{t:MainFull}\ref{it:iMainFull}. \Sec{s:FullSubdivisions} prepares the proof of \Thm{t:MainFull}\ref{it:iiMainFull}, which is presented in \Sec{s:FullBound2}. We prove the Unoriented Edges Lemma~\ref{le:Unorient}, which captures the essence of and entails \Thms{t:FullNbFlips} (from \cite{HNU99}) and \ref{t:PartNbFlips} (from \cite{LSU99}) above. It allows us to extend them further for our purposes. Moreover, along the way, we give a short proof of the bound for so-called simultaneously flippable edges by Souvaine \etal\ \cite{STW11}, and we indicate, how it may help getting in insight on so-called $k$-holes in point sets, \cite{Scheu20}. In \Sec{s:fCoveringFlipGraph} we briefly indicate polytopal substructures in the \efg. 
\item
\Secs{se:PartTriangulations}-\ref{se:PartLink} show the $(n-3)$-bound in \Thm{t:MainPart}.
\item
Building on the tools developed in \Secs{se:PartTriangulations}-\ref{se:PartLink}, in particular the Coarsening Lemma~\ref{le:CoarsenPartSubd}, \Sec{s:SuccPerfRef} proves a nontrivial new sufficient condition for the regularity of partial triangulations (\Thm{t:LowClosureRegular}), which can be considered as a generalization of the regularity of stacked triangulations (\ie triangulations obtained by successively adding points of degree \three to a triangulation of $\ext{P}$). This will give us a number of implications to be presented in \Sec{s:ImplRegularityPreserv}: Covering the bistellar flip graph by graphs of $(n-3)$-polytopes, a characterization of point sets for which all partial triangulations are regular, and, finally, \Thm{t:MinNonRegExample} on the size of certificates for the existence of non-regular triangulations in $\pT(P)$.
\end{EnumNo}
We conclude by discussing  open problems in \Sec{s:Discuss}.
%
%
\section{Graph Connectivity}
\label{se:GraphConnectivity}
We follow the graph theory books of Bollob\'as, \cite{Bollobas98}, and Diestel,  \cite{Diestel97}.
\begin{definition}
\label{de:Connectivity}
For $k \in \NN$, a simple undirected graph $G$ is \emph{$k$-vertex connected} if $G$ is connected, has at least $k+1$ vertices, and removing any set of at most $k-1$ vertices (and their incident edges) leaves the graph connected.
\end{definition}
Note that a graph is $1$-vertex connected iff it is connected and has at least two vertices. Here is a classical result due to Menger from 1927, \cite{Menger27}, see \cite[Theorem~III.5]{Bollobas98}.
\begin{theorem}[Menger's Theorem, \cite{Menger27}]
\label{t:Menger}
Let $u$ and $v$ be distinct nonadjacent vertices of a graph $G$. Then the minimal number of vertices separating $u$ from $v$ is equal to the maximal number of internally{} vertex-disjoint $u$-$v$ paths.
\end{theorem}
\begin{convention*} 
From now on, we we will use \emph{vertex-disjoint} short for ``internally{} vertex-disjoint.''
\end{convention*}
We will need a local variant of Menger's Theorem.
\begin{lemma}[Local Menger]
\label{le:LocalMenger}
Let $k\in\NN$ and let $G$ be a simple undirected graph. Assume that $G$ is connected. Then $G$ is $k$-vertex connected iff $G$ has at least $k+1$ vertices and for any pair of vertices $u$ and $v$ \emph{\underline{at distance \two}} there are $k$ pairwise vertex-disjoint $u$-$v$-paths.
\end{lemma}
\begin{MyProof} 
The direction $(\Rightarrow)$ follows from Menger's Theorem~\ref{t:Menger}. For $(\Leftarrow)$, suppose that for any pair $u$ and $v$ of vertices at distance 2 there are $k$ pairwise vertex-disjoint $u$-$v$-paths. We show that no pair of vertices $x$ and $y$ can be separated by removal of a set $V'$ of at most $k-1$ vertices (different from $x$ and $y$). This is true for $x$ and $y$ adjacent. If $x$ and $y$ are not adjacent, let $(x=x_0,x_1,\ldots,x_\ell = y)$ be an $x$-$y$-path which uses the minimal number of vertices in $V'$, and among those, a shortest such path (hence $\{x_{i-1},x_{i+1}\} \notin \Eds{G}$ for $0 < i  <\ell$). If no vertex on this path is in $V'$ we are done. Otherwise, consider $x_i \in V'$, $0 < i < \ell$. We can replace the subpath $(x_{i-1},x_i,x_{i+1})$ by an $x_{i-1}$-$x_{i+1}$-path using none of the vertices in $V'$ as internal vertices ($x_{i-1}$ and $x_{i+1}$ have distance \two and hence such a path exists, since there are $k$ vertex-disjoint $x_{i-1}$-$x_{i+1}$-paths and $|V'| < k$). We obtain an $x$-$y$-walk\footnote{A walk is a path with repetitions of vertices allowed.} with one less overlap with $V'$, and we can turn this into an $x$-$y$-path with less overlap with $V'$; contradiction. 
\end{MyProof}
For the proofs of \Thms{t:MainFull}\ref{it:iiMainFull} and \ref{t:MainPart}, here is a special property of a graph that guarantees that the minimum vertex degree determines exactly the vertex connectivity. As briefly discussed in \Sec{s:Approach}, we will apply this lemma not directly to the flip graph, but rather to the link, a neighborhood structure of a triangulation.
\begin{lemma} 
\label{l:C4FreeConn}
Let $G$ be a graph with its complement having no cycle of length \four, \ie for any sequence $(x_0, x_1, x_2, x_3)$ of four distinct vertices in $G$ there exists $i\in \{0,1,2,3\}$ with $\{x_i, x_{i+1\!\bmod\!4}\} \in \Eds{G}$.  
Then $G$ is $\delta$-vertex connected, where $\delta$ is the minimum vertex-degree in $G$.
\end{lemma}
\begin{MyProof} 
It suffices to show that for any two distinct nonadjacent vertices $u$ and $v$ there are $\delta$ vertex-disjoint $u$-$v$-paths. Let $z_1, z_2, \ldots, z_\ell$ be the set of vertices adjacent both to $u$ and to $v$. This gives $u$-$v$-paths $(u,z_i,v)$, $1 \le i \le \ell$. If $\ell \ge \delta$ we are done. Otherwise, there are vertices $x_{\ell+1}, \ldots, x_\delta$ adjacent to $u$ and not to $y$, and there are vertices $y_{\ell+1}, \ldots, y_\delta$ adjacent to $v$ and not to $u$. For $\ell < j \le \delta$ consider the sequence $(u,y_j,x_j,v)$: None of $\{u,y_j\}$, $\{x_j,v\}$, and $\{u,v\}$ are edges in $G$, hence $\{y_j,x_j\} \in \Eds{G}$ and $(u,x_j,y_j,v)$  is a $u$-$v$-path. That is, we have found another $\delta-\ell$ paths connecting $u$ to $v$. All paths constructed are easily seen to be vertex-disjoint.
\end{MyProof}
%
%
\section{Min-Degree Bound for Full Triangulations}
\label{se:MinDegreeBound}
%
\begin{convention*} 
In \Secs{se:MinDegreeBound}-\ref{s:fCoveringFlipGraph} we will mostly use \emph{triangulation} short for ``full triangulation.''
\end{convention*}
In this section we prove \Thm{t:MainFull}\ref{it:iMainFull} (\Sec{s:ProofiMainFull}) via the following lemma (and the Local Menger Lemma \ref{le:LocalMenger}).
\begin{lemma} 
\label{le:ManyPaths}
There exists $n_0 \in \NN$, such that any set $P$ with $|P| \ge n_0$ has the following property:
If $T[e]$ and $T[f]$ are distinct triangulations obtained from $T \in \fT(P)$ by flipping edges $e$ and $f$, respectively, then there are $\delta'$ vertex-disjoint $T[e]$-$T[f]$-paths, with $\delta'$ the minimum degree of the two vertices $T[e]$ and $T[f]$ in the \efg\ of $P$. 
\end{lemma}
For the proof of \Lm{le:ManyPaths} (see \Sec{se:ManyPathsProof}), we need a better understanding of how flippable edges interact. This will exhibit short cycles, more concretely, 4- or 5-cycles in the \efg\ (called elementary cycles in \cite{LMW19}). Subpaths of these short cycles will be the building blocks for the $T[e]$-$T[f]$-paths as claimed in \Lm{le:ManyPaths}.
%
%
\subsection{Basic terminology}
\begin{definition}[territory of an edge]
\label{d:Territory}
For $T \in \fT(P)$ and $e \in \Eds{T}$, we define the \emph{territory of $e$}, $\Terr{e} = \TerrTwo{T}{e}$, as the \underline{interior} of the closure of the union of its one or two incident regions in $T$. (Recall that the unbounded face of $T$ is not a region, see \Def{d:Plane}.)
\end{definition}
If $e$ is a boundary edge, $\Terr{e} $ is is an open triangle (one of the regions of $T$). Otherwise, for $e$ an inner edge, it is an open quadrilateral. Obviously, an inner edge $e$ is flippable in $T$ iff $\Terr{e}$ is convex. 

We can observe right away that if $\Terr{e}$ and $\Terr{f}$ are disjoint for flippable edges in $T$, then we can flip $e$ and $f$ in any order leading to the same triangulation, \ie $T[e,f] = T[f,e]$ and $(T,T[e],T[e,f],T[f])$ is a 4-cycle in the \efg.
%
%
\subsection{Two consecutive flips}
\begin{lemma} 
\label{l:TwoFlips}
Let $T \in \fT(P)$ and let $e$ be a flippable edge in $T$ with $T[e] = T[\Nicefrac{e}{\overline{e}}]$ (notation as introduced in \Def{d:BistFl}). Then
\begin{EnumRom}
\item
\label{i:iTwoFlips}
$\Eds{T} \oplus \Eds{T[e]} = \{e,\overline{e}\}$.
\item 
\label{i:iiTwoFlips}
For $f$ an edge flippable in $T[e]$ with $T[e,f] = \left(T[e]\right)[\Nicefrac{f}{\overline{f}}]$, we have $T[e,f]=T$, if $f = \overline{e}$, or 
\begin{eqnarray}
\Eds{T} \oplus \Eds{T[e,f]} &=& \{e,\overline{e},f,\overline{f}\} \mbox{~,~and}
\label{eq:1TwoFlips}\\
\Eds{T} \setminus \Eds{T[e,f]} &=& \{e,f\} ~,
\label{eq:2TwoFlips}
\end{eqnarray}
otherwise.
\end{EnumRom}
\end{lemma}

\begin{MyProof} \ItemRef{i:iTwoFlips} is immediate by definition. For \ItemRef{i:iiTwoFlips} we observe that 
$\Eds{T} \oplus \Eds{T[e,f]} = \Eds{T} \oplus \Eds{T[e]} \oplus \Eds{T[e]} \oplus \Eds{T[e,f]} 
= \{e,\overline{e}\} \oplus \{f,\overline{f}\}$. If $f=\overline{e}$, then $\overline{f} = e$ and we are done. Otherwise, for \EqRef{eq:1TwoFlips} it is left to show $\{e,\overline{e}\} \cap \{f,\overline{f}\} = \emptyset$. We have $f \neq e$ since $e \not\in \Eds{T[e]}$, we have $f\neq \overline{e}$ by assumption, we have $\overline{f} \neq \overline{e}$ since $\overline{e} \in \Eds{T[e]}$, and we have $\overline{f} \neq e$ since $e$ crosses $\overline{e}$ which is present in $T[e,f]$ (by assumption $f\neq \overline{e}$). Finally, \EqRef{eq:2TwoFlips} follows from \EqRef{eq:1TwoFlips}, from $e,f \in \Eds{T}$ (by assumption $f\neq \overline{e}$), and from $|\Eds{T}|= |\Eds{T[e,f]}|$.
\end{MyProof}

Thus, $\left|\Eds{T} \oplus \Eds{T[e]}\right| = 2$ and $\left|\Eds{T} \oplus \Eds{T[e,f]}\right|=4$, unless $T[e,f]=T$. This directly implies:
\begin{corollary}
\label{c:TriangleFree}
The \efg\ of $P$ is triangle-free.
\end{corollary}
%
%
\subsection{Interplay of two flippable edges}
\label{s:Interplay}
\begin{figure}[htb]
\centerline{
\placefig{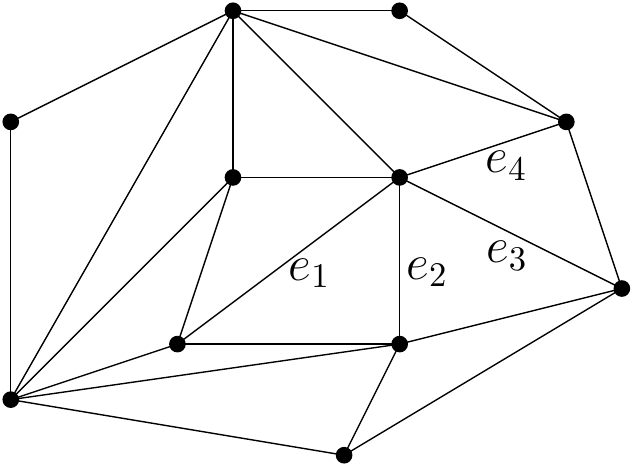}{0.26\textwidth}
}
\caption{Edges $e_1$ and $e_3$ are independently flippable, edges $e_2$ and $e_3$ are weakly independently flippable, and edges $e_3$ and $e_4$ are dependently flippable.}
\label{f:FlipEdgeRel}
\end{figure}

Here comes an essential lemma about the interplay of two flippable edges in a triangulation.
\begin{lemma} 
\label{fi:SymmetryInFlippability}
Let $e$ and $f$ be two distinct edges both of which are \underline{flippable} in $T \in \fT(P)$. Then
$e$ is flippable in $T[f]$ iff $f$ is flippable in $T[e]$.
\end{lemma}
\begin{MyProof} We distinguish three cases, depending on the shape of $\Terr{e} \cup \Terr{f}$. This can be composed of two disjoint quadrilaterals (recall that territories are open sets), or it is a pentagon, if $e$ and $f$ are incident to a common \region of $T$.
\begin{EnumAlph} 
\item
\label{it:TwoDisjoint}
If $\Terr{e} \cap \Terr{f} = \emptyset$ then, as observed above, $e$ is flippable in $T[f]$ and $f$ is flippable in $T[e]$ (see, \eg $e_1$ and $e_3$, or $e_1$ and $e_4$ in \Fig{f:FlipEdgeRel}).
\item 
\label{it:ConvPentagon} 
If $\Terr{e} \cup \Terr{f}$ is a convex pentagon, then $e$ is flippable in $T[f]$ and $f$ is flippable in $T[e]$ (see, \eg $e_2$ and $e_3$ in \Fig{f:FlipEdgeRel}).
\item
\label{it:NonconvPentagon}
We are left with the case of $\Terr{e} \cup \Terr{f}$ a nonconvex pentagon  (see, \eg $e_3$ and $e_4$ in \Fig{f:FlipEdgeRel}). Let $p$ be a reflex vertex in this pentagon. $e$ or $f$ have to be incident to $p$, otherwise $T$ has a reflex vertex in one of its regions. If only one of $e$ and $f$ is incident to $p$, say $e$, then $\TerrTwo{T}{e}$ is not convex and $e$ is not  flippable. Hence,  \emph{both} $e$ and $f$ are incident to $p$. But after flipping $e$, the other edge $f$ is left ``alone'' at this vertex $p$, \ie  $\TerrTwo{T[e]}{f}$ is not convex and thus not flippable in $T[e]$; similarly, after flipping $f$, edge $e$ is not flippable. That is, $e$ is not flippable in $T[f]$ and $f$ is not flippable in $T[e]$.
\end{EnumAlph}
\end{MyProof}
In short, the lemma states that, provided $e$ and $f$ are flippable in $T$, $T[e,f]$ is defined iff $T[f,e]$ is defined. If $T[e,f]$ and $T[f,e]$ are defined, they may be equal or not. With this in mind, we give the following definition (\Fig{f:FlipEdgeRel}).

\begin{definition} 
\label{d:FlipEdgeRel}
Let  $T\in\fT(P)$ and let $e$ and $f$ be flippable edges in $\Eds{T}$ with $e \neq f$.
\begin{EnumRom} 
\item
$e$ and $f$ are called \Emph{independently flippable in} $T$ if 
$e$ is flippable in $T[f]$, and $T[e,f] = T[f,e]$.
\item
$e$ and $f$ are called \Emph{weakly independently flippable in} $T$ if $e$ is flippable in $T[f]$, and $T[e,f] \neq T[f,e]$.
\item 
$e$ and $f$ are called \Emph{compatible in} $T$ if $e$ and $f$ are either independently or weakly independently flippable in $T$.
\item
$e$ and $f$ are called \Emph{dependently flippable in} $T$ if $e$ is not flippable in $T[f]$.
\end{EnumRom}
\end{definition}
\Lm{fi:SymmetryInFlippability} shows that all relations in \Def{d:FlipEdgeRel} are symmetric.
\begin{lemma}
\label{le:Cycles}
Let $T\in\fT(P)$ and let $e$ and $f\neq e$ be flippable edges in $T$. Then
\begin{EnumRom} 
\item
\label{it:Cycles1}
\begin{itemize}
\setlength\itemsep{-0.25em}
\item[] $e$ and $f$ are independently flippable  
\item[$\Leftrightarrow$] $\Terr{e} \cap \Terr{f} = \emptyset$
\item[$\Leftrightarrow$] $(T, T[e], T[e,f], T[f], T)$ is an induced 4-cycle in the \efg\ of $P$, see \Fig{fi:Cycles} (left)
\end{itemize}
\item
\label{it:Cycles2}
\begin{itemize}
\setlength\itemsep{-0.25em}
\item[] $e$ and $f$ are weakly independently flippable  
\item[$\Leftrightarrow$] $\Terr{e} \cup \Terr{f}$ is a convex pentagon
\item[$\Leftrightarrow$] $(T, T[e], T[e,f], T[f,e], T[f], T)$ is an induced 5-cycle in the \efg\ of $P$, see \Fig{fi:Cycles} (right)
\end{itemize}
\end{EnumRom}
\end{lemma}
\begin{MyProof} The proof of \Lm{fi:SymmetryInFlippability} has identified three disjoint cases for $e$ and $f$ flippable in $T$. \ItemRef{it:TwoDisjoint} ``$\Terr{e} \cap \Terr{f} = \emptyset$'' immediately yields the three properties listed under \ItemRef{it:Cycles1}. \ItemRef{it:ConvPentagon} ``$\Terr{e} \cup \Terr{f}$ is a convex pentagon'' implies the three properties listed under \ItemRef{it:Cycles2}, and \ItemRef{it:NonconvPentagon} ``$\Terr{e} \cup \Terr{f}$ is a nonconvex pentagon'' contradicts all conditions in \ItemRef{it:Cycles1} and \ItemRef{it:Cycles2}, since we have shown that $e$ is not flippable in $T[f]$ and $f$ is not flippable in $T[e]$ in that case.

The 4- and 5-cycles have to be induced since the \efg\ is triangle-free (\Cor{c:TriangleFree}).
\end{MyProof}

\begin{figure}[htb]
\centerline{\placefig{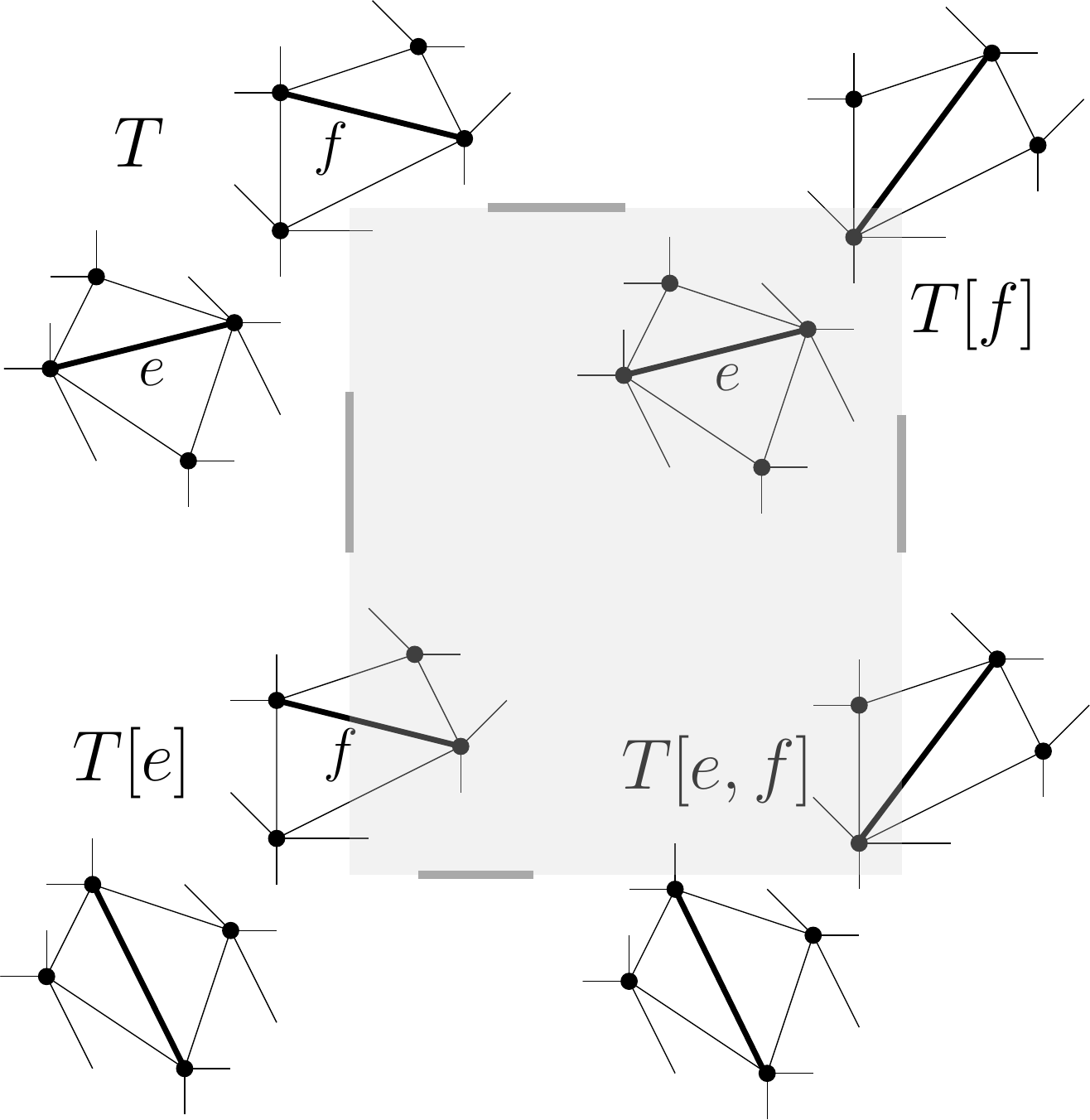}{0.32\textwidth} \hspace{3em} \placefig{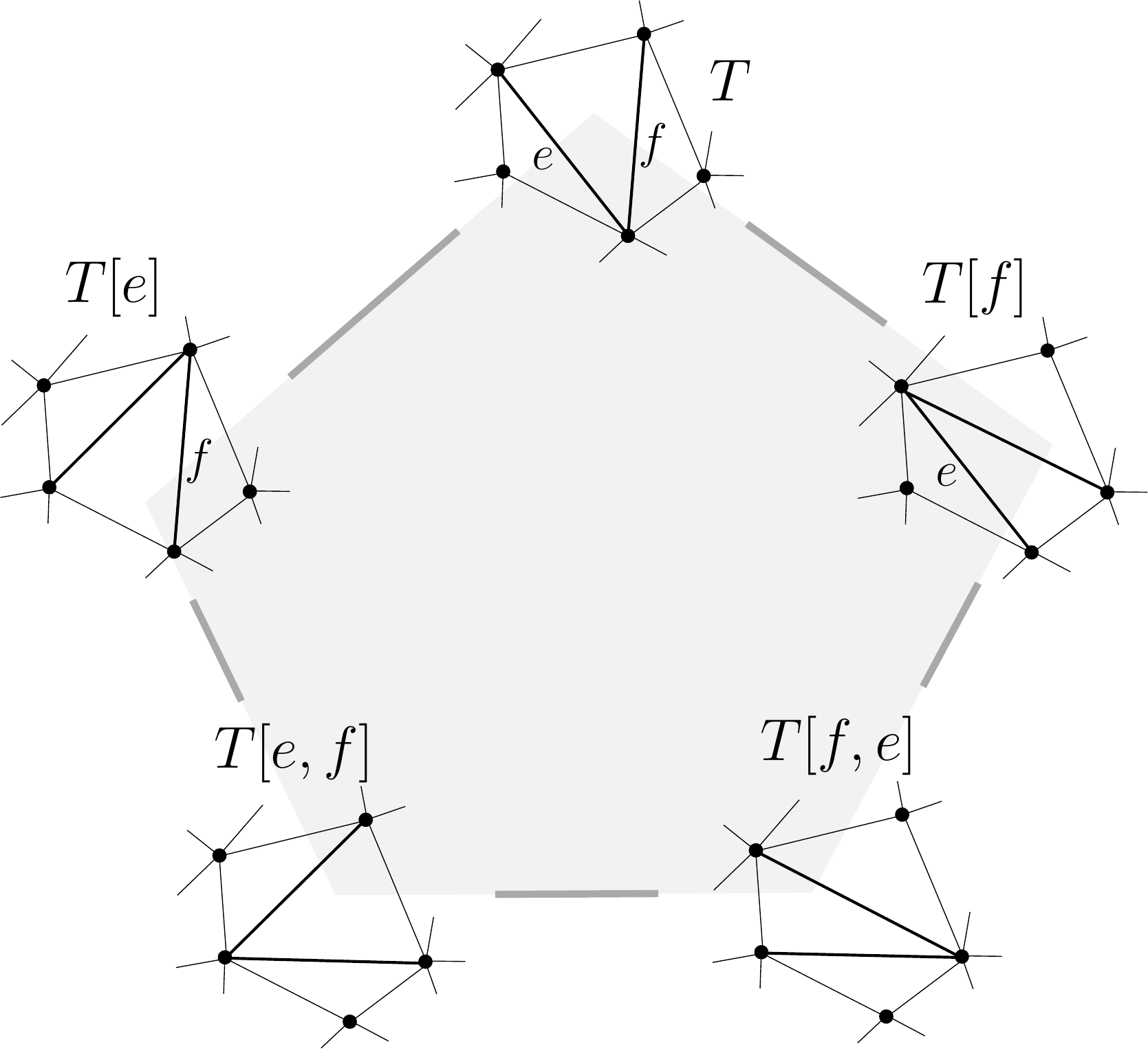}{0.405\textwidth}}
\caption{4-Cycle $(T, T[e], T[e,f], T[f], T)$ and 5-cycle $(T, T[e], T[e,f], T[f,e], T[f], T)$ for $e$ and $f$ independently flippable and weakly independently flippable, respectively.}
\label{fi:Cycles}
\end{figure}
%
%
\subsection{Proof of \Lm{le:ManyPaths}}
\label{se:ManyPathsProof}
\begin{figure}[htb]
\centerline{
\placefig{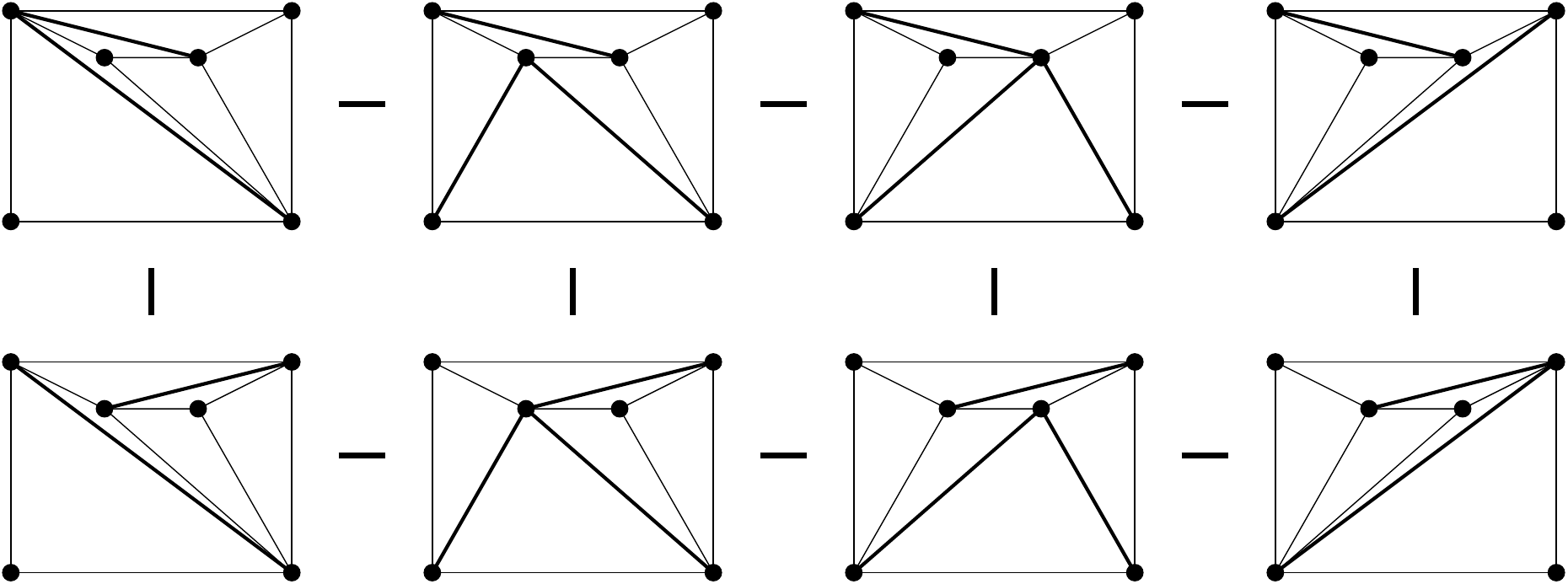}{0.5\textwidth}
\hspace{3em}
\placefig{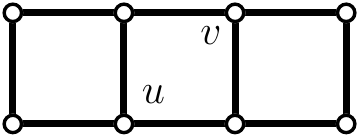}{0.15\textwidth}
}
\caption{A \efg\ with two vertices $u$ and $v$ of degree \three at distance \two, which do not allow \three vertex-disjoint connecting paths.}
\label{fi:6Example3a}
\end{figure}

\begin{MyProof} Let $e$ and $f\neq e$ be flippable edges in $T$, \ie $T[e] = T[\Nicefrac{e}{\overline{e}}]$ and $T[f] = T[\Nicefrac{f}{\overline{f}}]$ are defined for appropriate edges $\overline{e}$ and $\overline{f}$. Suppose the degree of $T[e]$ is at most the degree of $T[f]$. Our task is to identify, for each edge $g$ flippable in $T[e]$, a path $(T[e],T[e,g], \ldots, T[f,g^*],T[f])$, with these paths required to be vertex-disjoint. For this we distinguish four cases, depending on $g$.
\begin{EnumAlph}
\item 
$g=\overline{e}$: $(T[e],T,T[f])$ (note $T = T[e,\overline{e}]$).
\item 
$g=f$, if $f$ is flippable in $T[e]$: $(T[e], T[e,f],T[f])$ or $(T[e], T[e,f],T[f,e], T[f])$.
\item 
$g$ is flippable in $T[e]$, $T$, and $T[f]$:
$
(T[e], \!\!
\begin{array}{c}
T[e,g] \mbox{~or}\\ 
T[e,g], T[g,e] 
\end{array} \!\!,
T[g], \!\!
\begin{array}{c}
T[g,f] \mbox{~or}\\ 
T[g,f], T[f,g] 
\end{array} \!\!,
T[f]) ~.
$
\item
\label{i:LongPaths}
We still miss the paths $(T[e], T[e,g], \ldots)$ for edges $g \not\in \{\overline{e},f\}$ flippable in $T[e]$ but not flippable in $T$ or not flippable in $T[f]$. There are at most \eight such edges $g$, since for this to happen, $g$ must be an edge of $\TerrTwo{T}{e}=\TerrTwo{T[e]}{\overline{e}}$ or of $\TerrTwo{T}{f}$. For such an edge $g$ choose an edge $g^* \not\in \{\overline{f},e\}$ flippable in $T[f]$ but not flippable in $T$ or $T[e]$. Because of our degree condition, $g^*$ must exist. Note that $g$ may equal $g^*$, if $g$ is flippable in $T[f]$ but not in $T$. Now we choose edges $x$ and $y$ flippable in $T$ with the three sets
\[
\TerrTwo{T}{x},~ \TerrTwo{T}{y}, \mbox{~and~} \TerrTwo{T}{e} \cup \TerrTwo{T}{f} \cup \TerrTwo{T}{g} \cup \TerrTwo{T}{g^*}
\]
pairwise disjoint. These conditions allow for the following path
\begin{eqnarray*}
 &(T[e],T[e,g],T[e,g,x], T[e,g,x,y],T[e,x,y],T[x,y],\\
& T[f,x,y], T[f,g^*\!,x,y], T[f,g^*\!,x], T[f,g^*],T[f])~.
\end{eqnarray*}
Every $g$ in this final category is paired up with a different $g^*$ and a different pair $\{x,y\}$ is chosen for building such a $T[e]$-$T[f]$-path. $P$ large enough will allow us to do so, by \Thm{t:FullNbFlips} and since we have to deal only with at most a constant (at most \eight) such cases. 
\end{EnumAlph}
For vertex-disjointness, we define for $T'$ an internal vertex on such a $T[e]$-$T[f]$-path the \Emph{signature} $(\Eds{T[e]} \cap \Eds{T[f]}) \setminus \Eds{T'}$. The internal vertices of our long paths in \ref{i:LongPaths} have signatures
\begin{eqnarray*}
&(\{g\},\{g,x\},\{g,x,y\},\{x,y\},\{x,y\},&\\
&\{x,y\},\{g^*\!,x,y\},\{g^*\!,x\},\{g^*\})&
\end{eqnarray*}
while  previous cases gave signatures (again of internal vertices)
\[
\underbrace{(\emptyset)}_{\mathrm{(a)\,} g=\overline{e}},~ \underbrace{(\emptyset) \mbox{\,or\,}(\emptyset,\emptyset)}_{\mathrm{(b)\,} g=f}, \mbox{~or~}\underbrace{(\{g\}, \{g\}, \ldots, \{g\})}_{\mathrm{(c)\,} g \mathrm{\,compatible\,with\,} e \mathrm{\,and\,} f \mathrm{\,in\,} T}~.
\]
The vertex-disjointness of these paths can be readily concluded from these sequences and the proof is completed.
\end{MyProof}
The restriction ``$P$ large enough'' is essential in \Lm{le:ManyPaths}, see \Fig{fi:6Example3a}. 

%
%
\subsection{Proof of \Thm{t:MainFull}\ref{it:iMainFull}}
\label{s:ProofiMainFull}
\begin{MyProof}
The \efg\ is connected and triangulations $T'$ and $T''$ at distance \two in the \efg\ can be written as $T' = T[e]$ and $T'' = T[f]$, for some $T \in \fT(P)$ and $e,f \in \Eds{T}$. Therefore, \Lm{le:ManyPaths} indeed implies \Thm{t:MainFull}\ref{it:iMainFull} by the Local Menger Lemma~\ref{le:LocalMenger}.
\end{MyProof}
As mentioned before, while ``$P$ large enough'' is essential in \Lm{le:ManyPaths}, we do not know whether it can be removed in \Thm{t:MainFull}\ref{it:iMainFull}.
%
%
\section{Coarsening Full Subdivisions}
\label{s:FullSubdivisions}
In preparation of the proof of \Thm{t:MainFull}\ref{it:iiMainFull}, which will be presented in \Sec{s:FullBound2}, we introduce full subdivisions of a point set, which form a partially ordered set under a relation we call refinement,  and slack, a parameter of subdivisions.\footnote{As mentioned when discussing the flip complex in \Sec{s:FlipComplex} above, subdivisions correspond to the faces of the flip-complex. The refinement partial order corresponds to inclusion of one face in another, and the slack of a subdivision is equal to the dimension of the corresponding face of the flip complex.} This will allow us to prove a lower bound on how many edges are compatible with a given flippable edge in a triangulation.
%
%
\subsection{Full subdivisions}
\begin{definition}
[full subdivision] 
\label{d:FullSubdivision}
A \emph{full subdivision} $S$ of $P$ is a connected plane graph with $\Pts{S} = P$ and $\EdsHull \subseteq \Eds{S}$, and all regions of $S$ convex.
\end{definition}
\begin{convention*} In \Sec{s:FullSubdivisions} up to \Sec{s:fCoveringFlipGraph} we will use \emph{subdivision} short for ``full subdivision.''
\end{convention*}
\begin{definition}[coarsening, refinement]
\label{d:FullCoarsening}
\begin{EnumAlph}
\item
Given subdivisions $S_1$ and $S_2$ of $P$, we call $S_1$ a \emph{refinement of $S_2$}, in symbols $S_1 \preceq S_2$, if $\Eds{S_1} \supseteq \Eds{S_2}$ (or $S_2$ a \emph{coarsening of $S_1$}, in symbols $S_2 \succeq S_1$). 
\item
Given a subdivision $S$ of $P$, we let $\fTref{S} := \{ T \in \fT(P) \,|\, T \preceq S\}$, the set of triangulations of $P$ refining $S$.\footnote{Sometimes also called $S$-constrained triangulations of $P$.}
\end{EnumAlph}
\end{definition}
Obviously, $\preceq$ is a partial order on subdivisions of $P$ with the triangulations the minimal elements. Note that if $S_1 \preceq S_2$, then every region of $S_1$ is contained in some region of $S_2$, hence the name ``refinement.'' Here is a notion that allows us to identify edges in a subdivision that can be individually removed (not necessarily simultaneously removed) in order to obtain a coarsening.

\begin{figure}[htb]
\centerline{
\begin{minipage}[c]{0.26\textwidth}
\placefig{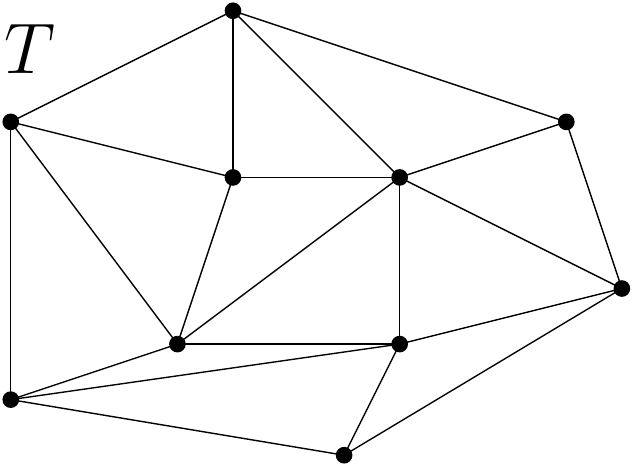}{\textwidth} 
\end{minipage}
~~$\preceq$~~
\begin{minipage}[c]{0.26\textwidth}
\placefig{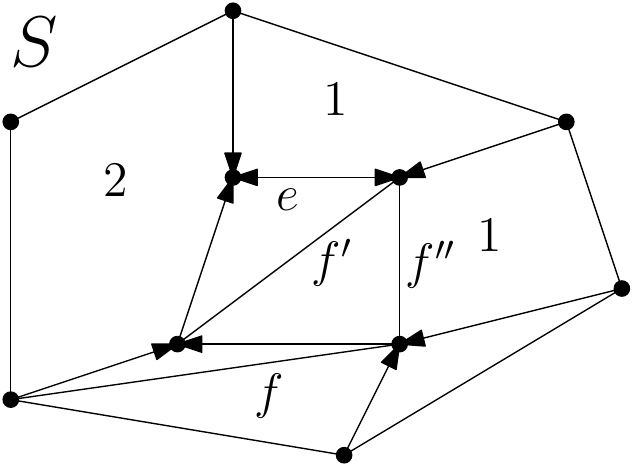}{\textwidth} 
\end{minipage}
~~$\preceq$~~
\begin{minipage}[c]{0.26\textwidth}
\placefig{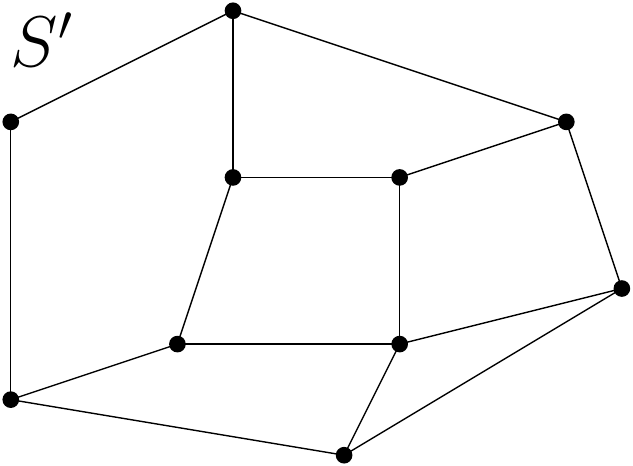}{\textwidth} 
\end{minipage}
}
\caption{A subdivision $S$, with edges directed to endpoints where they are locked. Edge $e$ is locked at both endpoints. Edges $f$, $f'$, and $f''$ are not locked, each one can be removed obtaining a coarsening subdivision. Slacks of non-triangular regions are indicated (triangular regions have slack $0$). $\Slack{S}=2+1+1=4$. Removal of $f$ and $f'$ yields a $\preceq$-maximal subdivision $S'$ coarsening $S$. $T$ is a triangulation refining $S$.}
\label{fi:Cycles}
\end{figure}
\begin{definition}[locked]
\label{d:Lock}
In a graph $G$ on $P$, an edge $e \in \Eds{G}$ is \Emph{locked at endpoint} $p$ if the angle obtained at $p$ (between the edges adjacent to $e$ in radial order around $p$) after removal of $e$ is at least $\pi$ (in particular, if $p$ has degree \one or \two in $G$).
\end{definition}
An edge in a triangulation is flippable iff it is not locked at any endpoint (in a triangulation, an edge can be locked at at most one of its endpoints). If $S$ is a subdivision, then removal of an edge $e$ in $S$ gives a subdivision iff $e$ is locked at none of its endpoints in $S$ (here, for a subdivision, an edge can be locked at both of its endpoints). Hence, the $\preceq$-maximal subdvisions are those with all of their edges locked.

Here are some simple fundamental properties of locking.
\begin{observation} Let $G$ be a graph on $P$ and $p \in \Pts{G}$.
\label{o:LockBasic}
\begin{EnumRom}
\item 
\label{i:LockBasicConsecutive}
Any two edges locked at $p$ must be consecutive in the radial order around $p$.
\item 
\label{i:LockBasicAtMost23}
There are at most \two edges locked at $p$, unless $p$ has degree \three in $G$.
\item 
\label{i:LockBasicTrglDeg3}
If $G$ is a subdivision and $p \in \inn{P}$ is of degree \three, then the three edges incident to $p$ are locked at $p$.
\end{EnumRom}
\end{observation}
\begin{definition}[slack, refined slack]
\label{d:FullSlack}
Let $r$ be a region of a subdivision $S$, a convex set bounded by a $k$-gon, $k \ge 3$. Then we define the \Emph{slack}, $\Slack{r}$,  of $r$ as $k-3$, and the \Emph{refined slack}, $\Refslack{r}$, of $r$ as $\lceil \frac{\Slack{r}}{2} \rceil$. 

The \Emph{slack}, $\Slack{S}$, of subdivision $S$ is the sum of the slacks of its regions. The \Emph{refined slack}, $\Refslack{S}$, is the sum of the refined slacks of its regions.
\end{definition}
Note that $k-3$ is the number of edges it takes to triangulate a $k$-gon. Hence, the slack of a subdivision is the number of edges which have to be added towards a triangulation. We need the following well-known facts on the number of edges and regions of triangulations.
\begin{lemma} 
\label{le:EdgeCount}
For $T \in \fT(P)$, the number of edges, $|\Eds{T}|$, equals $|\EdsInn{T}| + \extNb = 3n-3-\extNb = 3\innNb-3+2\extNb$ and the number of regions, $|\Reg{T}|$, equals $2n-2-\extNb = 2\innNb-2+\extNb$ (recall that the unbounded face is not a \region).
\end{lemma}
\begin{observation}
\label{o:SlackMissingEdges}
For a subdivision $S$, we have 
\begin{eqnarray*}
\Slack{S} &=& 3n-3-\extNb - |\Eds{S}| = 3\inn{n} - 3 + h - |\EdsInn{S}| \mbox{~,~and}\\
&=& 2n-2 -h -|\Reg{S}| = 2\inn{n}-2+h -|\Reg{S}| ~.
\end{eqnarray*}
\end{observation}
%
%
\subsection{Unoriented Edges Lemma}
\begin{definition}
Let $\orient{G}$ be a graph $G$ on $P$ with \underline{some} of its edges oriented to \underline{one} of its endpoints and some edges unoriented; we call this a \Emph{partially oriented graph}. $\orient{G}$ is \emph{well-oriented} if (a) no edge is directed to a point in $\ext{P}$, and (b) if edges $e$ and $f$ are directed to a common point $p$, then $e$ and $f$ have to be consecutive in radial order around $p$.
\end{definition}
Suppose, given a subdivision, we orient each locked \emph{inner} edge to one locking endpoint (and we leave all other edges, in particular the boundary edges, unoriented), then \Obs{o:LockBasic} shows that this is a well-oriented graph. This is how we will employ the following lemma.
\begin{lemma}
[Unoriented Edges Lemma]
\label{le:Unorient}
Let $\orient{S}$ be a partially oriented subdivsion of $P$. Let $C_i$, $i \in \NNnull$, be the number of inner points of $\orient{S}$ with indegree $i$ and let  $D:=\Slack{S}$ and $D^* := \Refslack{S}$.
\begin{EnumRom}
\item 
\label{it:UnorientOne}
If $C_i = 0$ for $i \ge 4$ then the number of unoriented inner edges equals
\[
n-3 - C_3 - D +(C_1+2C_0) \ge 
n-3 - C_3 - D \ge h-3-D
~.
\]
\item 
\label{it:UnorientTwo} Suppose $\orient{S}$ is well-oriented. Then the number of unoriented inner edges is at least
\begin{equation}
\frac{n}{2} - 2 - \frac{D+D^*}{2}
\ge  
\frac{n}{2} - 2 - D ~.
\label{eq:UnorientTwo}
\end{equation}
\end{EnumRom}
\end{lemma}
\begin{MyProof} 
\ItemRef{it:UnorientOne}
$\innNb = C_3 + C_2 +C_1 +C_0$ since $C_i = 0$ for $i \ge 4$. $|\EdsInn{S}|=3\innNb - 3 + h -D$ (\Obs{o:SlackMissingEdges}). Since $\orient{S}$ has $3C_3 +2 C_2 + C_1$ oriented edges, the number of unoriented inner edges is 
\begin{eqnarray}
\underbrace{3\innNb - 3 + h}_{=n-3+2(C_3 + C_2 +C_1 +C_0)}\hspace{-2em} -D - (3C_3 +2 C_2 + C_1)  
&=&  n-3 - C_3 - D +C_1+2C_0 \label{eq:UnorientedExact} \nonumber \\[-1em]
&\ge& n-3 - C_3 -D  \nonumber \\
&\ge& h- 3-D \mbox{\hspace{10ex} since $C_3 \le n-h$.} \nonumber
\end{eqnarray}
\smallskip

\noindent
\ItemRef{it:UnorientTwo} 
Let every inner point charge \one to each region incident to it that lies between two incoming edges. The overall charge made is exactly $3 C_3 + C_2$ (if the indegree is \three, the degree is \three; if the indegree is \two, the two incoming edges are consecutive). While each triangular \region can be charged at most once, other \regions can be multiply charged: A \region with slack $d$, a $(d+3)$-gon, is charged at most $\lfloor \frac{d+3}{2} \rfloor = 1 + \lfloor \frac{d+1}{2} \rfloor = 1 + \lceil \frac{d}{2} \rceil$ times (vertices charging a region $r$ cannot be consecutive along the boundary of $r$). Hence, if $\ell$ is the number of regions charged at all, the overall charge made to these $\ell$ regions is at most $\ell + D^*$. Thus  $3 C_3 + C_2 \le \ell + D^*$, \ie $\ell$ is at least $3 C_3 + C_2 - D^*$. Moreover, $\ell$ is at most $2 \innNb - 2 + h - D$, the overall number of regions (\Obs{o:SlackMissingEdges}). That is,
\begin{eqnarray*}
&&3 C_3 + C_2 - D^* \le 2 \innNb - 2 + h - D 
=\underbrace{\innNb + h}_{=n} -2 +\overbrace{C_3 + C_2 +C_1 +C_0}^{\innNb=} - D\\
&\Leftrightarrow& C_3 \le \frac{n}{2} - 1 - \frac{D-D^*}{2}  +\frac{C_0 + C_1}{2}
\end{eqnarray*}
We plug this bound on $C_3$ into the number of unoriented edges derived in \EqRef{eq:UnorientedExact} above:
\begin{eqnarray*}
&& n-3 - C_3 -D +C_1+2C_0\\
&\ge & n-3- \frac{n}{2} + 1 + \frac{D-D^*}{2} - \frac{C_0 + C_1}{2}-D +C_1+2C_0\\
&=& \frac{n}{2} - 2 - \frac{D+D^*}{2} + \frac{C_1 + 3 C_0}{2} 
\end{eqnarray*}
The left bound in \ItemRef{it:UnorientTwo} on the number of unoriented follows readily. Moreover, $\left\lceil \frac{d}{2} \right\rceil \le d$ for $d \in \NNnull$, hence $D^* \le D$, \ie $\frac{D+D^*}{2} \le D$ and the right bound in \ItemRef{it:UnorientTwo} is implied.
\end{MyProof}
\begin{lemma}[Coarsening Lemma for full subdivisions]
\label{l:FullCoarsening}
Any $\preceq$-maximal subdivision $S$ of $P$ (\ie all edges in  $\EdsInn{S}$ are locked) has slack at least $\max\{\frac{n}{2} - 2, h-3\}$.
\end{lemma}
\begin{MyProof}
Orient each edge in $S$ to a locking endpoint (ties broken arbitrarily). This gives a well-oriented graph $\orient{S}$ without unoriented inner edges. Since \Lm{le:Unorient}\ItemRef{it:UnorientOne} and \ItemRef{it:UnorientTwo} guarantees at least $\max\{h-3- \Slack{S},\frac{n}{2} - 2- \Slack{S}\} $ unoriented inner edges, we have $\max\{\frac{n}{2} - 2,h-3\} - \Slack{S} \le 0$.
\end{MyProof}
Here is the essential implication on the number of compatible edges.
\begin{corollary} 
\label{c:ManyFlipComp}
\begin{EnumRom}
\item 
\label{i:iManyFlipComp}
Every $T \in \fT(P)$ has at least $\max\{\frac{n}{2} - 2,h-3\}$ flippable edges.
\item
\label{i:iiManyFlipComp}
Given $T \in \fT(P)$ and $e$ flippable in $T$, there are at least $\max\{\frac{n}{2} - 3,h-4\}$ edges in $\Eds{T}$ compatible with $e$.
\end{EnumRom}
\end{corollary}
\begin{MyProof}
\ItemRef{i:iManyFlipComp}
Let $S$ be a $\preceq$-maximal subdivision with $S \succeq T$. Then all edges in $\Eds{T} \setminus \Eds{S}$ are flippable in $T$. Since $|\Eds{T} \setminus \Eds{S}| = \Slack{S}$, the claim follows by \Lm{l:FullCoarsening}.
\smallskip

\noindent
\ItemRef{i:iiManyFlipComp}
Let $T_{-e}$ be the graph obtained by removing the edge $e$ from $T$. Since $e$ is flippable, $T_{-e}$ is a subdivision. Let $S$ be a $\preceq$-maximal coarsening of $T_{-e}$. All edges in $\Eds{T_{-e}} \setminus \Eds{S}$ are compatible with $e$ in $T$. Since $|\Eds{T_{-e}} \setminus \Eds{S}| = \Slack{S}-1$, the claim follows again by \Lm{l:FullCoarsening}.
\end{MyProof}
\medskip

We see that \Cor{c:ManyFlipComp}\ItemRef{i:iManyFlipComp} is \Thm{t:FullNbFlips} by Hurtado \etal\ \cite{HNU99}. Actually, the set $\Eds{T} \setminus \Eds{S}$ in the argument for \Cor{c:ManyFlipComp}\ItemRef{i:iManyFlipComp} is what is called \emph{ps-flippable} (\emph{pseudo-simultaneously flippable}) by Hoffmann \etal\ \cite{HSSTWbook13}, where a lower bound of $\max\{\frac{n}{2} - 2,h-3\}$ is shown for such ps-flippable edges. The proof (not the statement) of \cite[Lemma~1.3]{HSSTWbook13} implies \Lm{l:FullCoarsening} above. We want to claim that the proof given above, including the proof of  the Unoriented Edges \Lm{le:Unorient} employed, is more concise. We emphasize, though, that our proof of the Unoriented Edges Lemma and its set-up is clearly inspired by the proof of the $\lceil \frac{n}{2}-2 \rceil$-bound for flippable edges in \cite{HNU99} (see also \cite{LSU99} for \Lm{l:FullCoarsening}\ref{it:UnorientOne}). 
\medskip

Here is another application of \Lm{l:FullCoarsening} which we mention here, although it will play no role in the rest of the paper. A set of pairwise independently flippable edges is often called \emph{simultaneously flippable} in the literature. Here is a streamlined proof of the known tight lower bound on their number by Souvaine \etal\ from 2011, \cite{STW11}, based on the Unoriented Edges Lemma~\ref{le:Unorient}.
\begin{theorem}[\cite{STW11}]
\label{t:SimFlipp}
Every triangulation $T$ of $P$ has a set of at least $\lceil\frac{n-4}{5}\rceil$ edges that are pairwise independently flippable (simultaneously flippable).
\end{theorem}
\begin{MyProof} Let $S$ be a $\preceq$-maximal coarsening of $T$. Orient edges in $S$ (all locked) to a locking endpoint (ties broken arbitrarily).  \Lm{le:Unorient}\ref{it:UnorientTwo} guarantees at least $\frac{n-4-(\Slack{S}+\Refslack{S})}{2}$ unoriented edges in $\orient{S}$. Since all inner edges are oriented, $\Slack{S}+\Refslack{S} \ge n-4$ holds. 

For each region $r$ of $S$ of slack $d$, $T$ induces a triangulation of this $(d+3)$-gon. Its $d$ diagonals can be 3-colored with no triangle incident to two edges of the same color (start with an edge and spread the coloring along the dual tree of the triangulation of $r$). Each color class offers a set of simultaneously flippable edges, one of size at least $\lceil \frac{d}{3} \rceil$. It is easy to verify that
$\left\lceil \frac{d}{3} \right\rceil \ge \frac{d + \lceil d/2\rceil}{5}$  (check for  $d=0,1,2$; for $d\ge 3$, $\lceil \frac{d}{3} \rceil \ge \frac{d}{3} \ge \frac{d + (d+1)/2}{5}\ge \frac{d + \lceil d/2\rceil}{5}$). We combine the simultaneously flippable edges collected for each region of $S$, which gives an overall set of at least $\frac{\Slack{S}+\Refslack{S}}{5} \ge \frac{n-4}{5}$ simultaneously flippable edges.
\end{MyProof}
While the lower bounds of $\frac{n-4}{2}$ for the number of flippable edges, and of $\frac{n-4}{5}$ for the size of a set of simultaneously flippable edges are tight (see \cite{HNU99} and \cite{GHNPU03}, respectively), it is interesting to observe from the proofs that both bounds cannot be simultaneously attained for the same point set. That is, if the number of flippable edges is close to its lower bound then this forces the existence of a set of more than $\lceil\frac{n-4}{5}\rceil$ simultaneously flippable edges, and vice versa. This can be quantified as follows.
\begin{theorem} If the number of flippable edges in a triangulation $T$ of $P$ is $\alpha$, and if the largest set of simultaneously flippable edges in $T$ has size $\beta$, then $\alpha + \beta \ge \frac{4(n-4)}{5}$.\footnote{The claim is still true if $\alpha$ is a the largest size of a set of pseudo-simultaneously  flippable edges (see \cite{HSSTWbook13}).}
\end{theorem}
\begin{MyProof} Let $S$ be a $\preceq$-maximal coarsening of $T$, with $d_1$, $d_2$, \ldots, $d_{|\Reg{S}|}$ the slacks of the regions of $S$. Then $\alpha \ge \sum_i d_i$ and, by the proof of \Thm{t:SimFlipp}, $\beta \ge \sum_i \lceil \frac{d_i}{3}\rceil$. Note that $d + \lceil \frac{d}{3} \rceil \ge \frac{4}{5}(d + \lceil d/2\rceil)$ for all $d \in \NNnull$ (for $d=0,1,2$ we have $d + \lceil \frac{d}{3} \rceil = d + \lceil d/2\rceil$ and for $d \ge 3$, $d + \lceil \frac{d}{3} \rceil  \ge \frac{4d}{3} \ge \frac{4(d + (d+1)/2)}{5} \ge \frac{4(d + \lceil d/2\rceil)}{5}$). Therefore, $\alpha + \beta \ge \frac{4(\Slack{S} + \Refslack{S})}{5} \ge \frac{4(n-4)}{5}$ (\Lm{le:Unorient}\ItemRef{it:UnorientTwo}).
\end{MyProof}
\medskip

Let us conclude this section with an observation about interior-disjoint $k$-holes, as it emerged from a discussion with M.\,Scheucher (see also \cite{Scheu20}). A \Emph{$k$-hole} of $P$ is a subset of $k$ points in convex position whose convex hull is disjoint from all other points in $P$. A $k$-hole and an $\ell$-hole are called interior-disjoint, if their respective convex hulls are interior-disjoint (they can share up to \two points). Harborth showed in 1978, \cite{Har78}, that every set of at least $10$ points has a $5$-hole. This was recently strengthened to the existence of another interior-disjoint $4$-hole, \cite[Thm.\,2]{HU20}. We show, how our framework allows an easy proof of this fact.
\begin{theorem}[\cite{HU20}]
Every set $P$ of at least $10$ points has a $5$-hole and a $4$-hole which are interior-disjoint.
\end{theorem}
\begin{MyProof} Since $P$ has a 5-hole, \cite{Har78}, there is a subdivision of $P$ with a region of slack $2$. Consider a $\preceq$-maximal coarsening $S$ of such a subdivision with $r$ a region of largest slack. We know that $\Slack{r} \ge 2$. If $\Slack{r} \ge 4$, \ie it is a $k$-gon with $k\ge 7$, then we can add a diagonal to $r$ which divides $r$ into a $5$-gon and a $(k-3)$-gon which readily gives the claimed interior-disjoint holes. Otherwise, if $k \le 6$, we have $\Slack{r} +\Refslack{r} \le 5$, while $\Slack{S} + \Refslack{S} \ge n-4 \ge 6$ (\Lm{le:Unorient}\ItemRef{it:UnorientTwo}). Hence, there must be another region $r'\in \Reg{S}$ with positive slack, \ie  a $k'$-gon with $k' \ge 4$.  
\end{MyProof} 
%
%
\section{$\FBound$-Bound for Full Triangulations}
\label{s:FullBound2}
The proof of \Thm{t:MainFull}\ref{it:iiMainFull} needs one more ingredient.
%
%
\subsection{Link of a full triangulation}
\label{s:fLink}
The \Emph{link} of a triangulation is the graph representing the compatibility relation among its flippable edges.  In \Sec{s:Approach}, the intuition for links as counterparts of vertex figures in polytopes was briefly explained.
\begin{definition}
[link of full triangulation]
\label{d:fLink}
For $T \in \fT(P)$, the \Emph{link of} $T$, denoted $\fLink{T}$, is the edge-weighted graph with vertices $\ElmtsFlip{T}:= \{e \in  \EdsInn{T}  \mid e \mbox{~flippable~in~} T\}$ and edge set $\{\{e,f\}\in{ \ElmtsFlip{T} \choose 2} \mid e \mbox{~and~} f \mbox{~compatible}\}$. The \Emph{weight of} an edge $\{e,f\}$ is $2$ if $e$ and $f$ are independently flippable, and $3$ if $e$ and $f$ are weakly independently flippable.
\end{definition}

\begin{figure}[htb]
\begin{center}
\begin{minipage}[c]{0.375\textwidth}
\hspace{5.7em} \placefig{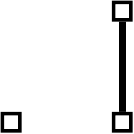}{0.18\textwidth}\\[2ex]
\placefig{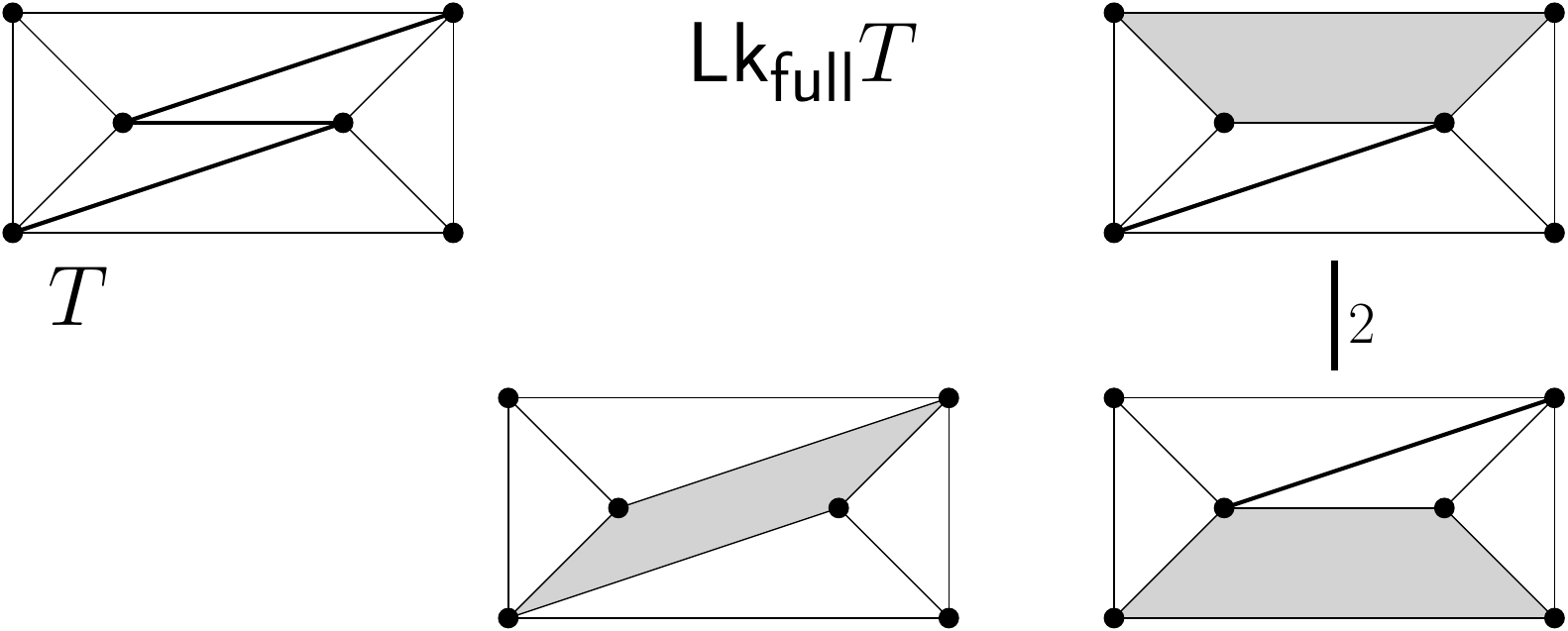}{0.95\textwidth}
\end{minipage}
\hspace{3em}
\begin{minipage}[c]{0.47\textwidth}
\hspace{4.05em} \placefig{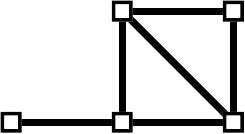}{0.25\textwidth}\\[2ex]
\placefig{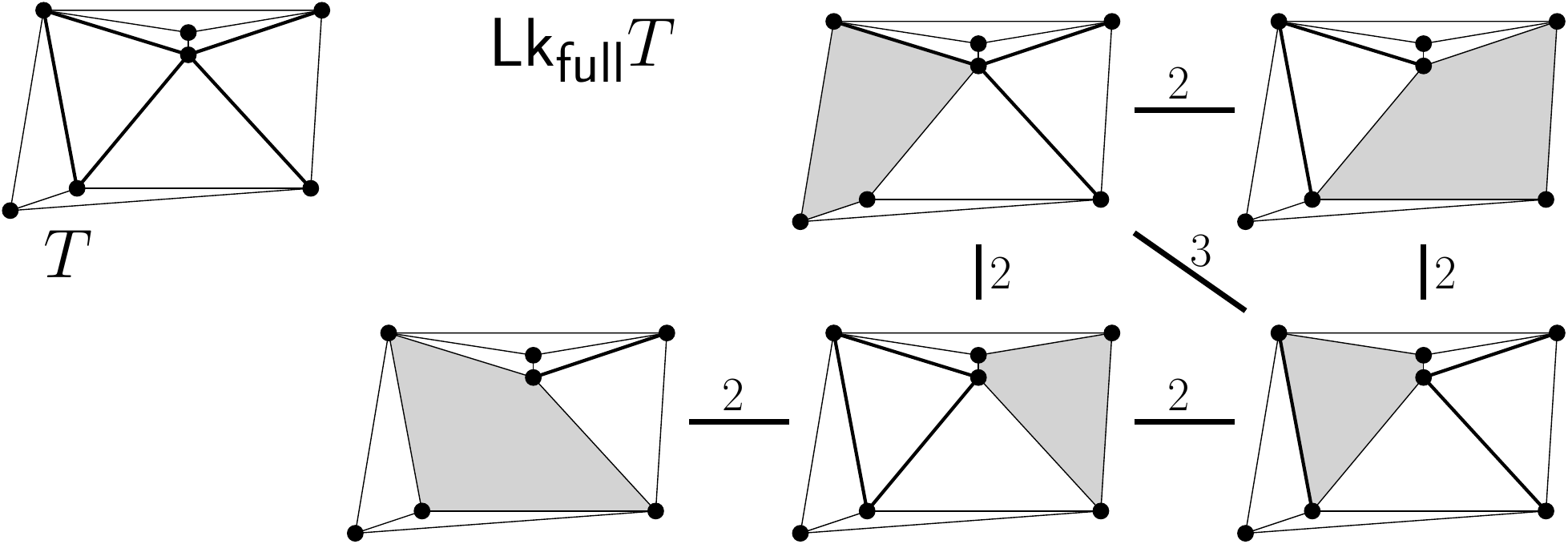}{0.95\textwidth}
\end{minipage}
\end{center}
\caption{Two examples of links, the left link has \three vertices, one of which is isolated. We indicate vertices of the link, \ie $e \in \ElmtsFlip{T}$, also as $T[e]=T[\Nicefrac{e}{\overline{e}}]$ with $\TerrTwo{T}{e}=\TerrTwo{T[e]}{\overline{e}}$ shaded. Small examples are not typical: For large sets, the link is a dense highly connected graph.}
\label{fi:EXPLINKsix-TwoInOpp}
\end{figure}
\begin{figure}[htb]
\centerline{
\placefig{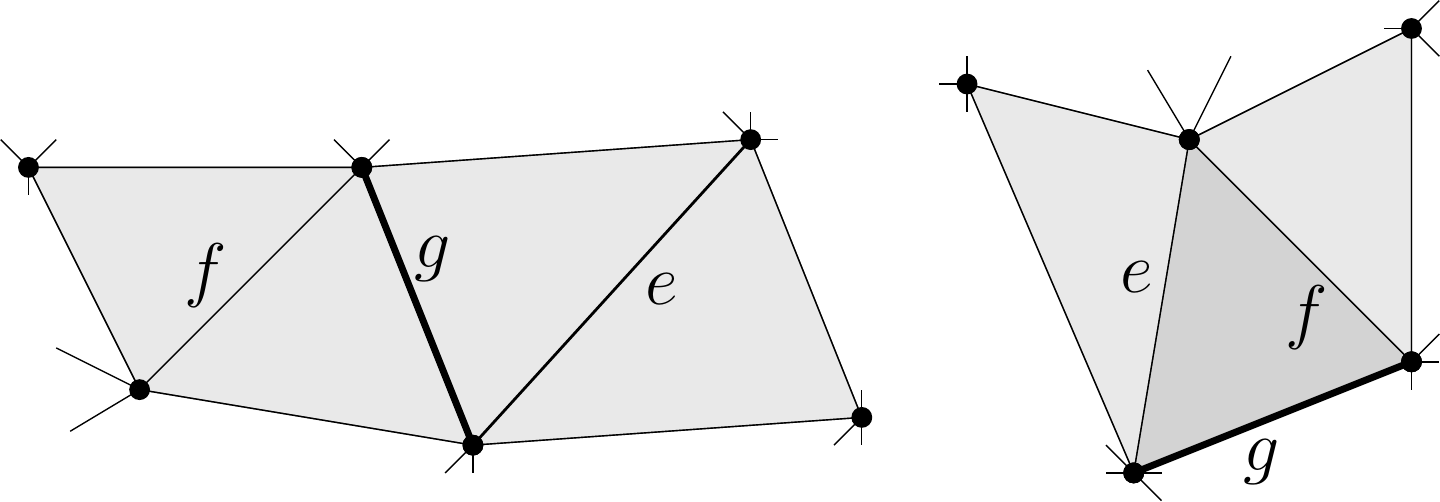}{0.4\textwidth}
}
\caption{Edge $g$ in the intersection of boundaries of territories of flippable edges $e$ and $f$.}
\label{fi:CommonBoundaryFull}
\end{figure}

We will see that for proving \Thm{t:MainFull}\ref{it:iiMainFull} (in \Sec{s:ProofiiMainFull} below) it is enough to prove $\lceil \frac{n}{2} - 3 \rceil$-vertex connectivity of links.  Here is the special property of links that will immediately show that the vertex connectivity is determined by the minimum vertex degree (via \Lm{l:C4FreeConn}).
\begin{lemma} 
\label{l:fLinkC4Free}
For $T \in \fT(P)$, the complement of $\fLink{T}$ has no cycle of length \four, \ie if $(e_0, e_1, e_2, e_3)$ are flippable edges in $T$, then there exists $i\in \{0,1,2,3\}$ such that $e_i $ is compatible with $e_{i+1\!\bmod\! 4}$.
\end{lemma}
\begin{MyProof}
For $e$ and $f$ flippable edges  in $T$ we show that there is at most one flippable edge $g$ that is compatible with neither $e$ nor $f$; this implies the lemma. Such a $g$ has to be an edge of both $\Terr{e}$ and $\Terr{f}$. If $\Terr{e}$ and $\Terr{f}$ are disjoint, they share at most one edge (since $\Terr{e}$ and $\Terr{f}$ are convex, see \Fig{fi:CommonBoundaryFull} (left). Otherwise, $\Terr{e}$ and $\Terr{f}$ overlap in a triangle $\Delta$, of which $e$ and $f$ are edges; the third edge $g$ of this triangle is the a common edge of $\Terr{e}$ and $\Terr{f}$, see \Fig{fi:CommonBoundaryFull} (right). No other edge of $\Terr{f}$ can appear on the boundary of $\Terr{e}$: Consider the line $\lambda$ through edge $f$. All edges of $\Terr{f}$ other than $e$ and $g$ lie on the side of $\lambda$ opposite to $\Delta$, and $\Terr{e}$ is on the same side of $\lambda$ as $\Delta$, since it contains $\Delta$ and $f$ is an edge of $\Terr{e}$; again, convexity of $\Terr{e}$ and of $\Terr{f}$ is essential here.
\end{MyProof}

\begin{lemma}
\label{le:fLink}
For $T \in \fT(P)$, the link $\fLink{T}$ is $\FBoundMinus$-vertex connected.
\end{lemma}
\begin{MyProof}
Every flippable edge in $T$ is compatible with at least $\FBoundMinus$ edges (\Cor{c:ManyFlipComp}\ItemRef{i:iiManyFlipComp}), \ie the minimum vertex degree in $\fLink{T}$ is at least $\FBoundMinus$. $\fLink{T}$ has no cycle of length \four in its complement (\Lm{l:fLinkC4Free}). The lemma follows (\Lm{l:C4FreeConn}).
\end{MyProof}
\begin{lemma} 
\label{l:fLinkToFlipGraph}
Given flippable edges $e$ and $f$, $e \neq f$, in $T \in \fT(P)$, every $e$-$f$-path of weight $w$ in $\fLink{T}$ induces a $T$-avoiding $T[e]$-$T[f]$-path of length $w$ in the \efg, in a way that vertex-disjoint $e$-$f$-paths in the link induce vertex-disjoint $T[e]$-$T[f]$-paths.
\end{lemma}

\begin{figure}[htb]
\begin{center}
\placefig{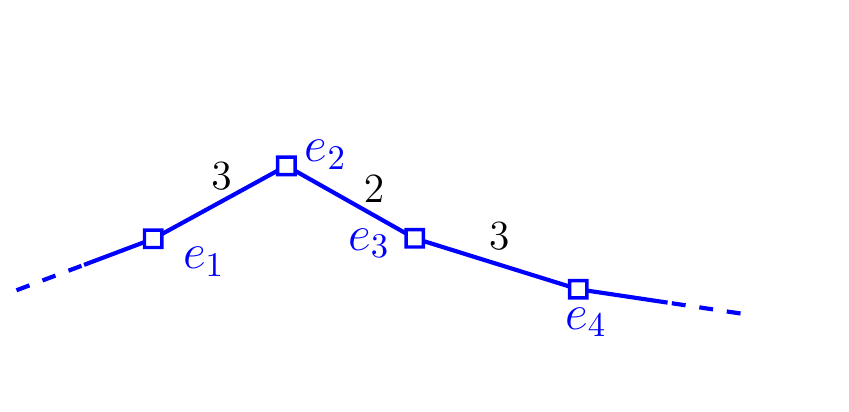}{0.3\textwidth}\hfill
\placefig{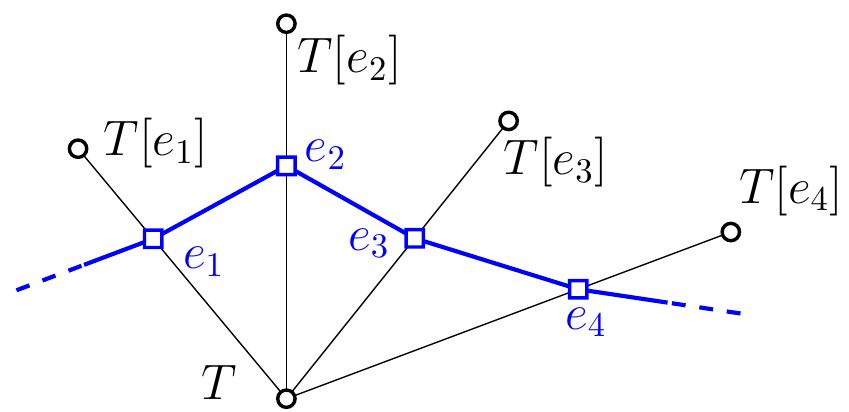}{0.3\textwidth}\hfill
\placefig{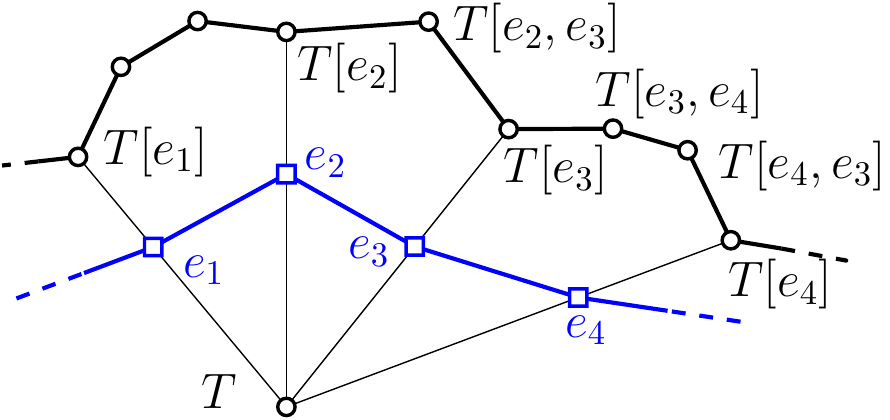}{0.3\textwidth}
\end{center}
\caption{From a path in the link $\fLink{T}$ to a path in the \efg.}
\label{fi:LinkPathToFGPath}
\end{figure}
\begin{MyProof} Given an $e$-$f$-path in $\fLink{T}$, we replace each edge $\{z',z''\}$ on this path (\ie $z'$ and $z''$ are compatible) by the path $(T[z'],T[z',z''],T[z''])$ or $(T[z'],T[z',z''],T[z'',z'],T[z''])$, depending on whether $z'$ and $z''$ are independently flippable (weight of $\{z',z''\}$ is \two) or weakly independently flippable (weight of $\{z',z''\}$ is \three), respectively, see \Lm{le:Cycles} (\Fig{fi:LinkPathToFGPath}). Note that the vertices $T[z',z'']$ and $T[z'',z']$ at distance \two from $T$ (recall triangle-freeness, \Cor{c:TriangleFree}) on these substitutes satisfy $\Eds{T} \setminus \Eds{T[z',z'']} = \Eds{T} \setminus \Eds{T[z'',z']} = \{z',z''\}$ (\Lm{l:TwoFlips}\ItemRef{i:iiTwoFlips}\EqRef{eq:2TwoFlips}). Therefore, these vertices cannot appear on any substitute for another edge on the given $e$-$f$-path, nor on substitutes for any other vertex-disjoint $e$-$f$-path. Clearly, also the internal vertices at distance \one from $T$  (\ie of the form $T[z]$) are distinct from internal vertices at other vertex-disjoint $e$-$f$-paths. And, obviously, we have not employed the vertex $T$ for the substituting paths.
\end{MyProof}
%
%
\subsection{Proof of \Thm{t:MainFull}\ref{it:iiMainFull}}
\label{s:ProofiiMainFull}
\begin{MyProof}
We want to show that for $n \ge 5$, the \efg\ is $\FBound$-vertex connected. We employ the Local Menger Lemma~\ref{le:LocalMenger}. We know that the \efg\ is connected, \cite{L72}. What is left to show is that for any triangulation $T\in \fT(P)$ and edges $e$ and $f$ flippable in $T$, at least $\FBound$ vertex-disjoint $T[e]$-$T[f]$-paths exist in the \efg. $\fLink{T}$ has at least $\FBoundMinus$ vertex-disjoint $e$-$f$-paths (\Lm{le:fLink} and Menger's Theorem~\ref{t:Menger}).
Therefore, there are at least $\FBoundMinus$ $T$-avoiding vertex-disjoint $T[e]$-$T[f]$-paths (\Lm{l:fLinkToFlipGraph}). The extra path $(T[e],T,T[f])$ yields the theorem.$\!$
\end{MyProof}
%
%
\section{Covering the Edge Flip Graph with Polytopes}
\label{s:fCoveringFlipGraph}
Recall  from \Sec{s:Associahedron} that for a convex $k$-gon there is a $(k-3)$-polytope whose 1-skeleton is isomorphic to the \efg\ of triangulations of the $k$-gon, an associahedron, which we denote by ${\cal A}_{k-3}$ (the index reflecting its dimension)\footnote{To be precise, ${\cal A}_{k-3}$ is some representative realization of the associadedron}; ${\cal A}_1$ is an edge, ${\cal A}_2$ is a convex pentagon, etc. If we consider all triangulations of the $k$-gon with a given diagonal present, we get a facet of ${\cal A}_{k-3}$; all facets of ${\cal A}_{k-3}$ can be obtained in this way. In general, the $d$-faces of ${\cal A}_{k-3}$ represent the triangulations where a certain set of $k-3-d$ diagonals is present, \ie a subdivision.

\begin{figure}[htb]
\centerline{
\placefig{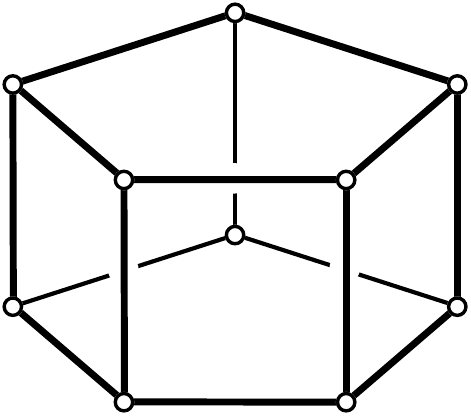}{0.14\textwidth} 
\hspace{3em} 
\placefig{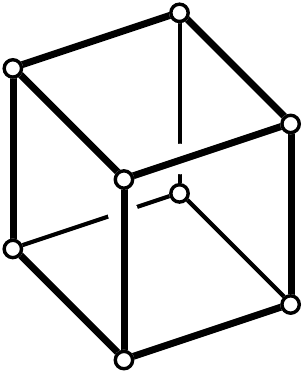}{0.09\textwidth}
\hspace{3em}
\placefig{6AssociahedronGraph}{0.18\textwidth}}
\caption{3-dimensional products of associahedra for subdivisions of slack \three: ${\cal A}_{1,2}$ (a pentagonal prism), ${\cal A}_{1,1,1}$ (a cube), and ${\cal A}_3$. }
\label{fi:Dim3Products}
\end{figure}
Suppose now that we have a subdivision $S$ of $P$ with nontriangular \regions $r_1,r_2,\ldots,r_m$, with $r_i$ of slack $d_i$. Then the set, $\fTref{S}$, triangulations refining $S$, with its \efg\ is represented by the product (see \cite{Zie95})
\[
{\cal A}_{d_1,d_2,\ldots,d_m} := {\cal A}_{d_1} \times {\cal A}_{d_2} \times \cdots \times {\cal A}_{d_m} ~,
\]
a $d$-dimensional polytope for $d:=d_1+d_2+\ldots+d_m = \Slack{S}$.

It is now easy to see that any edge of the \efg\ finds itself in the 1-skeleton of such a $\FBound$-dimensional polytope contained in the \efg\ (see also the discussion in \Sec{s:FlipComplex}).
\begin{theorem} 
\label{th:EdgeInProdAssoc}
For every edge $\{T,T[e]\}$ of the \efg\, there is an induced subgraph of the \efg\ which contains the edge $\{T,T[e]\}$ and is isomorphic to the 1-skeleton of a product ${\cal A}_{d_1,d_2,\ldots,d_m}$ of associahedra, where $d := d_1+d_2+\cdots+d_m \ge \FBound$. Therefore, the \efg\ (vertices and edges) can be covered by $1$-skeletons of $\FBound$-dimensional products of associahedra contained in the \efg.
\end{theorem}
\begin{MyProof} Let $S$ a $\preceq$-maximal coarsening of the subdivision $T_{-e}$ ($T$ with $e$ removed). Then $\Slack{S}\ge \FBound$ holds (Coarsening Lemma~\ref{l:FullCoarsening}). The subgraph induced by $\fTref{S}$ gives the claimed product of associahedra.
\end{MyProof} 
One can strengthen this and show that any pair of incident edges in the \efg\ is part of a subgraph isomorphic to the 1-skeleton of some $\lceil\frac{n}{2}-2\rceil$-dimensional polytope (a glueing of products of associahedra). This has been discussed in \cite{WW20}, but we decided to skip that part in this version.
%
%
\section{Partial Subdivisions -- Slack and Order}
\label{se:PartTriangulations}
In \Secs{se:PartTriangulations}-\ref{se:PartLink} we move on to proving \Thm{t:MainPart}, the $(n-3)$-vertex connectivity for the bistellar flip graph of partial triangulations. As indicated in the introduction, the proof will follow a similar line as for the edge flip graph, using subdivisions and links, but several new aspects and challenges will appear.
\begin{convention*}
From now on, in \Secs{se:PartTriangulations}-\ref{s:ImplRegularityPreserv}, we will mostly use \emph{triangulation} short for ``partial triangulation,'' but we return to using ``full triangulation'' and ``full subdivision.''
\end{convention*}
We define \Emph{partial subdivisions}, which form a poset in which the triangulations of $P$ are the minimal elements -- our definition is a specialization, to the plane and general position, of the established notion of a polyhedral subdivision, \cite{LRS10}. These partial subdivisions are plane graphs, possibly with isolated points. Hence, it may be useful to point out a subtlety in the definition of regions of a plane graph (\Def{d:Plane}): We defined them as the bounded connected components in the complement of the union the edges (as line segments in $\RR^2$), not taking the isolated points into account. That is, regions can contain isolated points of the graph, and isolated points will not keep them from being convex.
\begin{definition}
[partial subdivision]
\label{d:PartSubdivision}
A \Emph{partial subdivision} $S$ of $P$ is a graph with $\Pts{S} \subseteq P$ and $\EdsHull(P) \subseteq \Eds{S}$ (hence $\ext{P} \subseteq \Pts{S}$), and with all of its regions convex. 
\medskip

Similar to triangulations (\Def{d:Triang}), we define $\EdsInn{S} := \Eds{S} \setminus \EdsHull$ (the \Emph{inner edges} of $S$) and $\PtsInn{S}: = \Pts{S} \cap \inn{P}$ (the \Emph{inner points} of $S$). Moreover, we let $\PtsBy{S}$ be the points in $\Pts{S}$ which are isolated in $S$, the \Emph{bystanders} of $S$, and we let $\PtsInv{S} := \PtsInn{S} \setminus \PtsBy{S}$, the \Emph{involved points} of $S$.

For a region $r$ of $S$, let $\Pts{r} := \cl{r} \cap \Pts{S}$ ($\cl{r}$ the closure of $r$), \ie these are the vertices of the convex polygon $r$ and the bystanders in this region. 

$\Striv = \Striv(P) := (P,\EdsHull)$ is called the \Emph{trivial subdivision of} $P$.
\end{definition}
Observe that a partial subdivision $S$ is a full subdivision of $P$ iff $\Pts{S} = P$ and $\PtsBy{S} = \emptyset$. Also, a partial subdivision $S$ is a full subdivision of $\Pts{S} \setminus \PtsBy{S}$.

\begin{figure}[htb]
\centerline{
\placefig{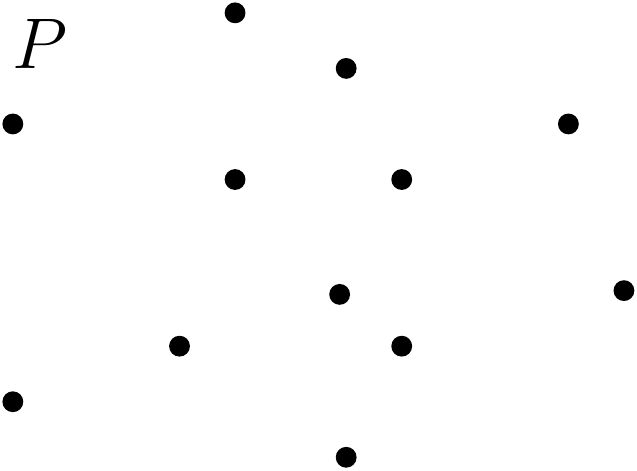}{0.2\textwidth}
\hspace{3em}
\placefig{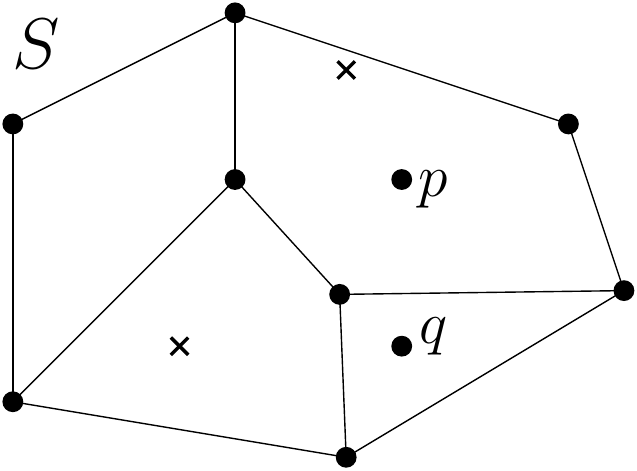}{0.2\textwidth}
}
\caption{A set $P$ and a partial subdivisions $S$ of $P$. Crosses indicate skipped points. Points $p$ and $q$ are bystanders. The region containing $p$ has slack $6-3=3$ (\Def{d:PartSlack} below).}
\end{figure}
\begin{convention*}
From now on, in \Secs{se:PartTriangulations}-\ref{s:ImplRegularityPreserv}, we will mostly use \emph{subdivision} for ``partial subdivision.''
\end{convention*}
As it should have become clear by now, $\Pts{S}$ is essential in the definition of a subdivision $S$, it is \emph{not} simply the set of endpoints of edges  in $S$, there are also bystanders. For example, for $T \in \pT(P)$, all graphs $(P',\Eds{T})$ with $\Pts{T} \subseteq P' \subseteq P$ are \psubdivisions of $P$, all different. $\Pts{S}$ partitions into boundary points, involved points, and  bystanders, \ie $\Pts{S} = \ext{P} \,\dot{\cup}\, \PtsInv{S} \,\dot{\cup}\, \PtsBy{S}$. Moreover, there are the skipped points, $P \setminus \Pts{S} = \inn{P} \setminus \PtsInn{S}$.
\medskip

A first important example of a \psubdivision is obtained from a \ptriangulation $T$ and an element $x$ flippable in $T$, \ie $\{T,T[x]\}$ is an edge of the \bfg:
\[
\pFlip{T}{x} : = (\Pts{T} \cup \Pts{T[x]}, \Eds{T} \cap \Eds{T[x]})
\]
If $x=e$ is a flippable edge, then $\pFlip{T}{e}$ has one convex quadrilateral \region $Q$; all other \regions are triangular. We obtain $T$ and $T[e]$ from $\pFlip{T}{e}$ by adding one or the other of the 2 diagonals of $Q$ to $\pFlip{T}{e}$. If $x=p$ is a flippable point, then $\pFlip{T}{p}$ is almost a \ptriangulation, all \regions are triangular, except that $p \in \Pts{\pFlip{T}{p}}$ is a bystander. We obtain $T$ and $T[p]$ by either removing this point from $\pFlip{T}{p}$ or by adding the three edges from $p$ to the points of the triangular region in which $p$ lies. The \psubdivision $\pFlip{T}{x}$ is close to a \ptriangulation and, in a 
sense, represents the flip between $T$ and $T[x]$. To formalize and generalize this we generalize the notion of slack from full to partial subdivisions.
\begin{definition}
[slack of partial subdivision]
\label{d:PartSlack}
Given a subdivision $S$ of $P$, we call a \region of  $S$ \Emph{active} if it is not triangular or if it contains at least one point in $\Pts{S}$ (necessarily a bystander) in its interior.

For $r \in \Reg{S}$, the \Emph{slack}, $\Slack{r}$, of $r$ is $|\Pts{r}|-3$. The \Emph{slack of} $S$, $\Slack{S}$, is the sum of slacks of its regions.
\end{definition}
Note that a region is active iff it has nonzero slack.
\begin{observation} 
\label{o:PartSlackMissingEdges}
For a \psubdivision $S$ with $s$ bystanders we have 
\[\Slack{S} = 3(|\Pts{S}|-s) - 3 - \extNb - |\Eds{S}| + s = 3|\Pts{S}| - 3 - \extNb - |\Eds{S}| - 2s~.
\]
\end{observation}
\begin{MyProof} The slack of a \region $r$ equals the number of edges it takes to triangulate $r$ (ignoring bystanders) plus the number of bystanders in $r$. Thus, $\Slack{S}$ is the number of edges it takes to triangulate $(\Pts{S} \setminus \PtsBy{S}, \Eds{S})$ (a full subdivision of $\Pts{S} \setminus \PtsBy{S}$) plus $|\PtsBy{S}|$. Now the claim follows from \Lm{le:EdgeCount} (or \Obs{o:SlackMissingEdges}).
\end{MyProof}
\begin{observation} 
Let $S$ be a subdivision.
\begin{EnumRom}
\item
$\Slack{S}=0$ iff $S$ is a \ptriangulation iff $S$ has no active \region.
\item
$\Slack{S}=1$ iff $S$ has exactly one active \region of slack \one; this \region is either a convex quadrilateral, or a triangular \region with one bystander in its interior.
\item
$\Slack{S}=2$ iff $S$ has either (a) exactly two active regions, both of slack \one, or (b)  exactly one active \region of slack \two, where this \region is either a convex pentagon, or a convex quadrilateral with one bystander in its interior, or a triangular \region with two bystanders in its interior.
\end{EnumRom}
\end{observation}

\begin{figure}[htb]
\centerline{
\placefig{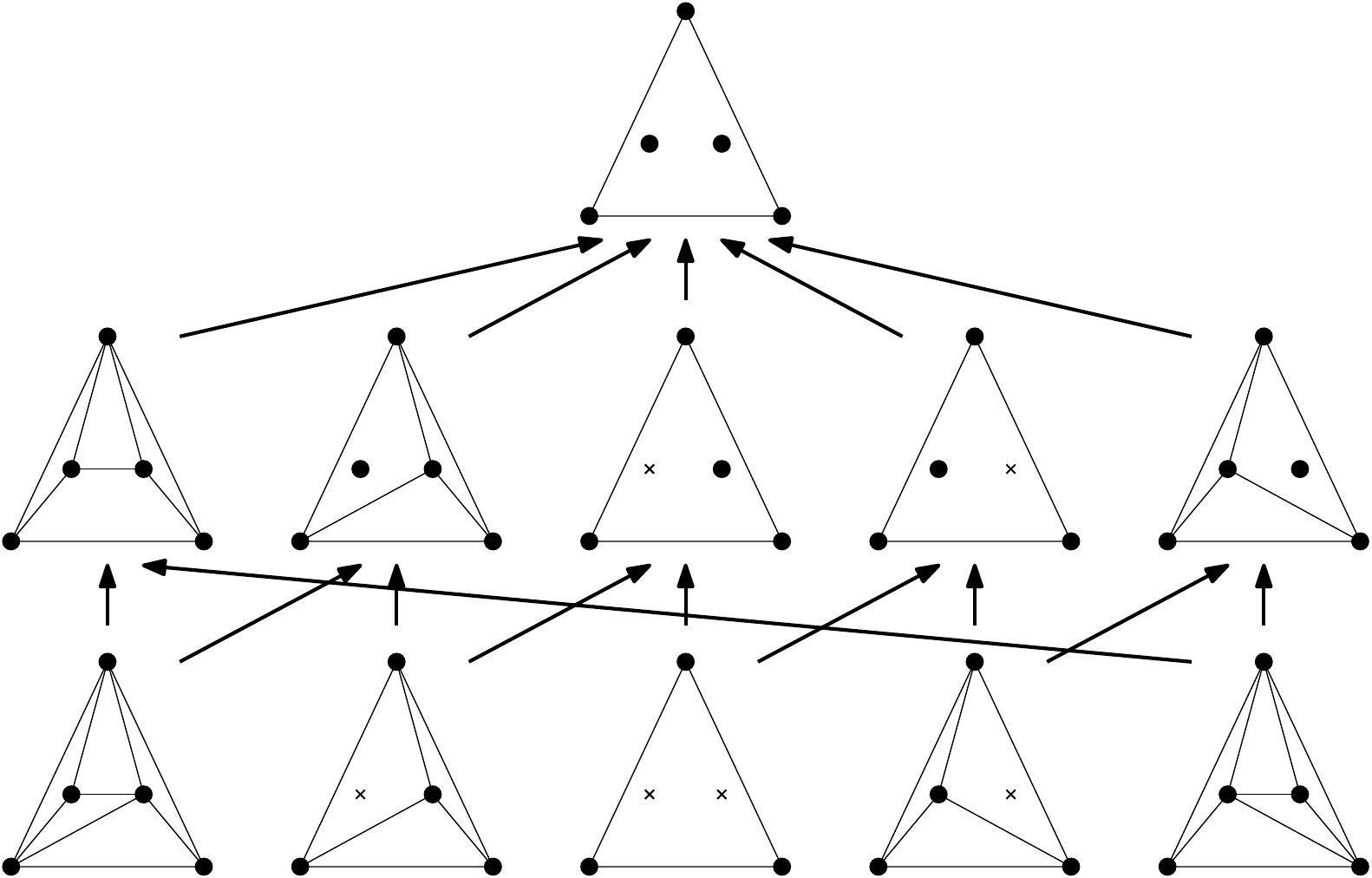}{0.55\textwidth}
}
\caption{Hasse diagram of the partial order $\preceq$ for a set of \five points.}
\label{fi:5Example3TrianXPO}
\end{figure}
\begin{figure}[htb]
\centerline{
\begin{minipage}[c]{0.19\textwidth}
\placefig{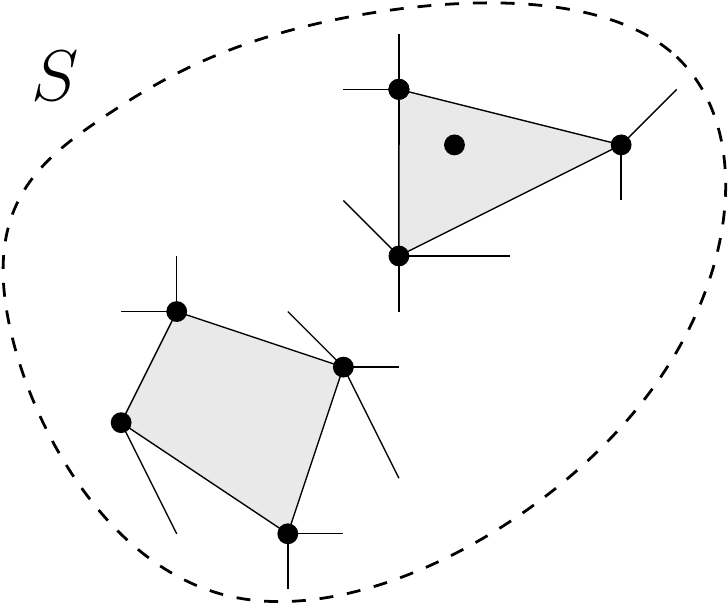}{\textwidth}
\end{minipage}
\hspace{3em}
\begin{minipage}[c]{0.45\textwidth}
\placefig{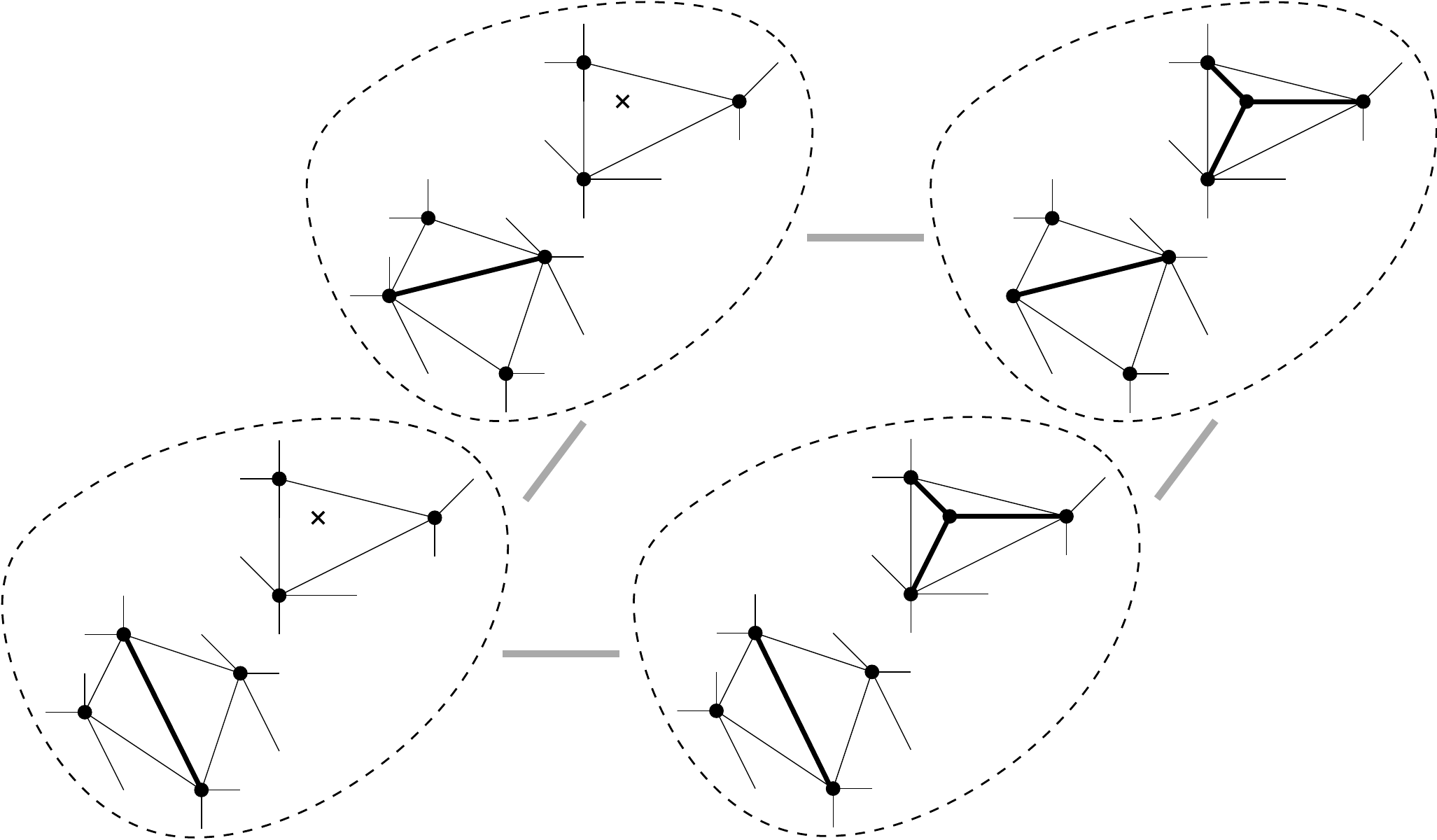}{\textwidth}
\end{minipage}
}
\caption{A \psubdivision $S$ with two active \regions of slack \one each. $\pTref{S}$ spans a $4$-cycle in the \bfg.}
\label{f:4CycleBist}
\end{figure}

\begin{definition}[coarsening, refinement] For \psubdivisions $S_1$ and $S_2$ of $P$,  $S_2$ \Emph{coarsens} $S_1$, in symbols $S_2 \succeq S_1$, if $\Pts{S_2} \supseteq \Pts{S_1}$, and $\Eds{S_2} \subseteq \Eds{S_1}$.
We also say that $S_1$ \Emph{refines} $S_2$, ($S_1 \preceq S_2$).
\end{definition}
The example in \Fig{fi:5Example3TrianXPO} hides some of the intricacies of the partial order $\preceq$; \eg in general, it is not true that all paths from a \ptriangulation to $\Striv$ have the same length $n-3$. $\Striv$ is the unique coarsest ($\preceq$-maximal) element (quite contrary to the poset of full subdivisions, where there were several $\preceq$-maximal full subdivisions). The \ptriangulations (\ie \psubdivisions of slack $0$) are the minimal elements.
\begin{definition}
[set of refining partial triangulations] 
\label{d:RefTrgs}
For a \psubdivision $S$ of $P$ we let $\pTref{ S } := \{ T \in \pT(P) \,|\, T \preceq S\}$.
\end{definition}
Note that $\pTref{ \Striv } = \pT(P)$ and for $x$ flippable in $T$, $\pTref{\pFlip{T}{x}} = \{T,T[x]\}$.
\begin{observation}
\label{o:SmallSlack2}
(i)
Any \psubdivision $S$ of slack \one of $P$ equals $\pFlip{T}{x}$ for some \ptriangulation $T \preceq S$ and some $x$ flippable in $T$.
(ii)
Let $S$ be a \psubdivision of slack \two of $P$.
If there are exactly 2 active \regions in $S$ (of slack \one each), then $\pTref{S}$ has cardinality \four, spanning a $4$-cycle in the \bfg\ of $P$ (\Fig{f:4CycleBist}). If there is exactly one active \region in $S$ (of slack \two), then $\pTref{S}$ has cardinality \five, spanning a $5$-cycle (see \Fig{fi:TheFiveExamples}). 
\end{observation}

\begin{figure}[htb]
\centerline{
\placefig{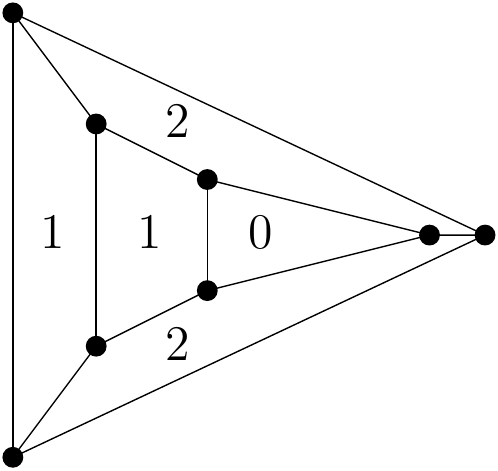}{0.16\textwidth}
}
\caption{\Eight points, with a \psubdivision of slack \six, a refinement of $\Striv$ of slack $8-3 = 5$.}
\label{f:LargeSlack}
\end{figure}

\begin{lemma}
\label{l:RefSlack2}
Any proper refinement $S$ of a subdivision $S'$ of slack \two has slack at most \one.
\end{lemma}
\begin{MyProof} For a refinement of $S'$ we add $m$ edges, thereby involving $s'$ bystanders, and we remove $s''$ bystanders (some of these parameters may be $0$, but not all, since the refinement is assumed to be proper). We have $\Slack{S} = \Slack{S'} - (m - 2s' + s'')$ (easy consequence of \Obs{o:PartSlackMissingEdges}) and we want to show $m - 2s' + s''> 0$.

Since $\Slack{S'}=2$, $S'$ has at most two bystanders and thus $s' \le 2$. If $s'=0$, then $m - 2s' + s''> 0$ holds, since some of the three parameters have to be positive. If $s'=1$, we observe that we need at least \three edges to involve a bystander and $m-2s' \ge 3 -2\cdot 1=1$. If $s'=2$, we  need  at least \five edges to involve two bystanders and $m-2s' \ge 5 -2 \cdot 2=1$. \end{MyProof}
For $D \ge 3$, a proper refinement of a subdivision of slack $D$ can have slack $D$ or even higher (\Fig{f:LargeSlack}). The proof fails, since we can involve \three bystanders with \six edges.

\begin{figure}[htb]
\centerline{
\placefig{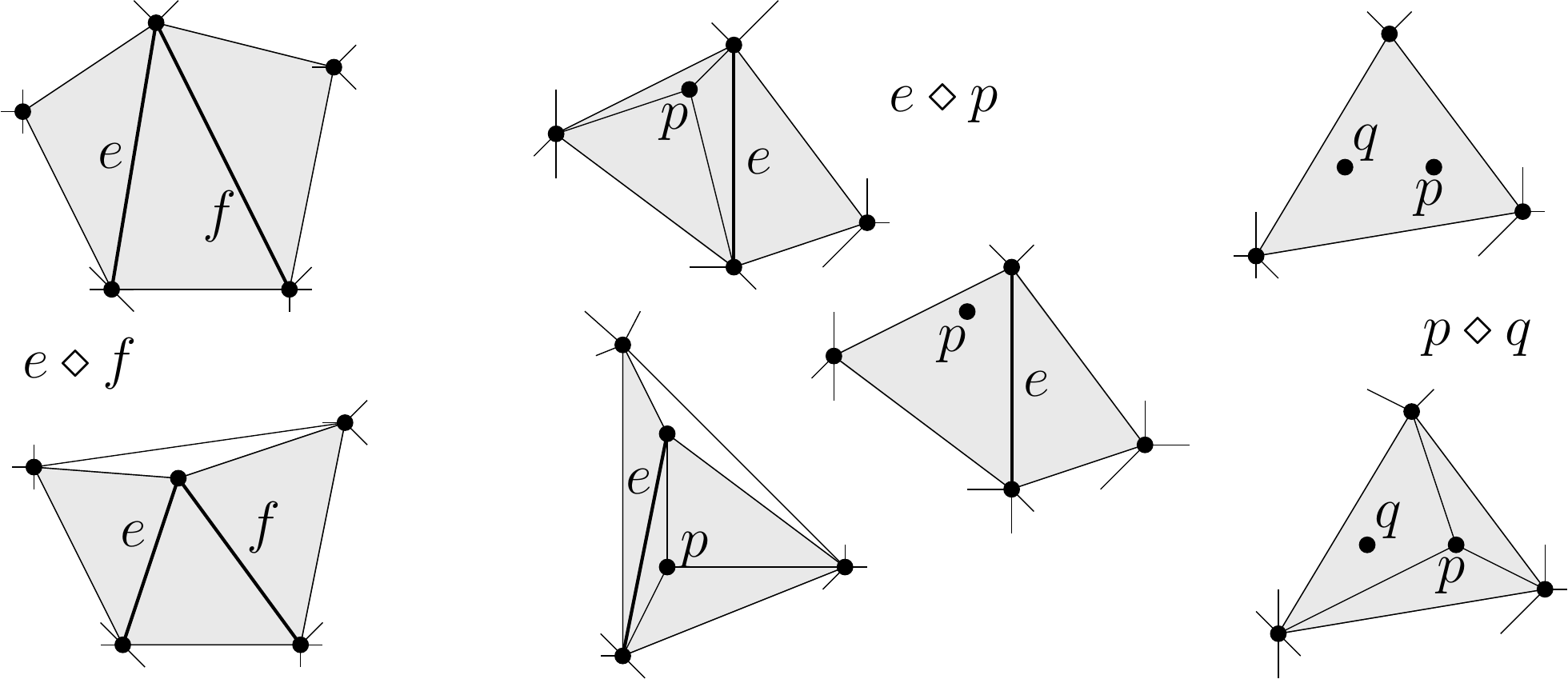}{0.53\textwidth}
\hspace{8ex}
\placefig{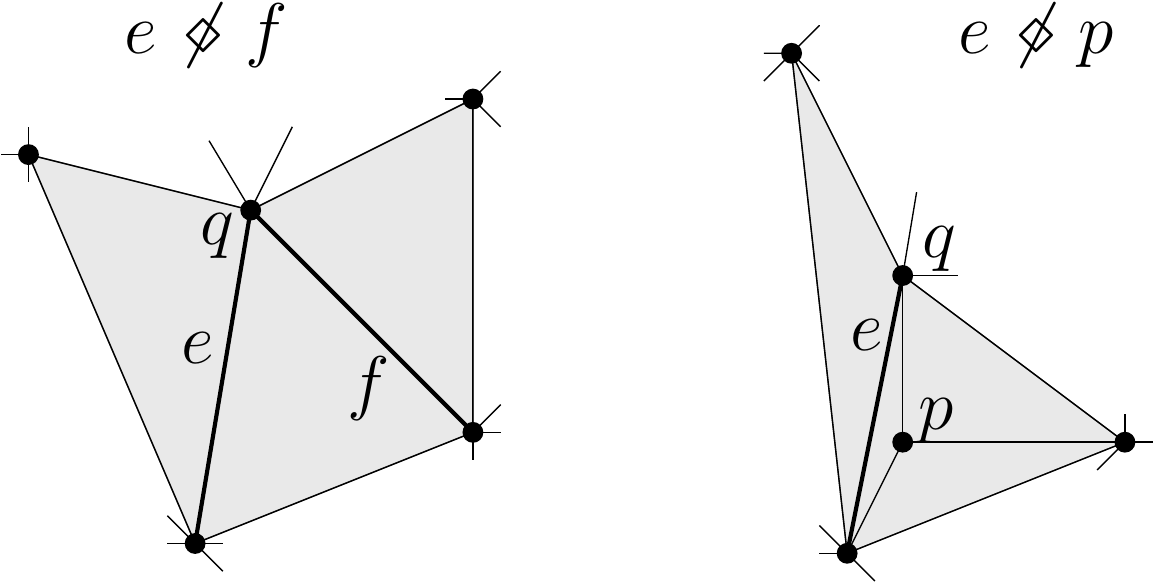}{0.31\textwidth}
}
\caption{Compatible elements (with overlapping incident regions, all contained in a \five-cycle, see \Fig{fi:TheFiveExamples}, and incompatible elements (two rightmost, where $q$ is assumed to have degree at least \five). Shaded areas are unions of incident regions of flippable elements (not the active \region in $\pFlip{T}{x,y}$!).}
\label{fi:CompUncompElements}
\end{figure}

Intuitively, as briefly alluded to at the end of \Sec{s:Associahedron} (for the special case of convex position), one can think of the subdivisions as the faces of a higher-dimensional geometric structure behind the \bfg, with slack playing the role of dimension, analogous to the secondary polytope for regular triangulations. The following lemma shows that -- for slack at most \two\ -- we have the property corresponding to the fact that faces of dimension $d$ are either equal, or intersect in a common face of smaller dimension (possibly empty). This correspondence fails for slack exceeding $2$.
\begin{lemma} 
\label{le:IntersectionProp}
\begin{EnumRom}
\item
\label{it:iiIntersectionProp}
For  \psubdivisions $S_1$ and $S_2$  of slack \two, $\pTref{S_1} \cap \pTref{S_2}$ is either \textup{(a)} empty, \textup{(b)} equals $\{T\}$ for some \ptriangulation $T$, \textup{(c)}  equals $\{T,T[x]\}$ for some \ptriangulation $T$ and some flippable element  $x$, or \textup{(d)} $S_1 = S_2$.
\item
\label{it:iiiIntersectionProp} Let $x$ and $y$ be two distinct flippable elements in \ptriangulation $T$. If there is a \psubdivision $S$ of slack \two with $\{T[x],T,T[y]\} \subseteq \pTref{S}$, then this $S$ is unique.
\end{EnumRom}
\end{lemma}
\begin{MyProof}
If $\pTref{S_1} \cap \pTref{S_2}$ contains some triangulation, then we easily see that $S_1 \meet S_2 := (\Pts{S_1} \cap \Pts{S_2}, \Eds{S_1} \cup \Eds{S_2})$ is a \psubdivision, and $\pTref{S_1 \meet S_2}=\pTref{S_1} \cap \pTref{S_2}$.
\smallskip

\noindent
\ItemRef{it:iiIntersectionProp} If (a) does not apply, let $S := S_1 \meet S_2$, a subdivision with $\pTref{S} = \pTref{S_1} \cap \pTref{S_2}$. If $\Slack{S}=0$ we have property (b), if $\Slack{S}=1$ we have property (c). 
In the remaining case $\Slack{S} \ge 2$, $S$ is a refinement of $S_1$ and of $S_2$. \Lm{l:RefSlack2} tells us that $S$ cannot be a proper refinement of $S_1$, hence $S=S_1$; similarly, $S=S_2$, hence $S_1=S_2$.
\smallskip

\noindent
\ItemRef{it:iiiIntersectionProp}
Suppose $S_1$ and $S_2$ are \psubdivisions of slack \two with $\{T[x],T,T[y]\} \subseteq \pTref{S_1} \cap \pTref{S_2}$. Since options (a-c) above cannot apply, we are left with $S_1=S_2$.
\end{MyProof}
Two edges incident to a vertex of a polytope may span a $2$-face, or not; same here, which gives rise to the following definition:
\begin{definition}
[compatible elements] 
\label{def:compatibleelements}
Two distinct flippable elements $x,y \in \PtsInn{T} \cup \EdsInn{T}$ are called \Emph{compatible in} $T$, in symbols $x \comp y$, if there is a \psubdivision $\pFlip{T}{x,y} \succeq T$ of slack \two, s.t.\ $\{T[x],T,T[y]\} \subseteq \pTref{\pFlip{T}{x,y}}$. (Note that $\pFlip{T}{x,y}$ is unique, by \Lm{le:IntersectionProp}\ItemRef{it:iiiIntersectionProp}.) Otherwise, $x$ and $y$ are called \Emph{incompatible in} $T$, in symbols $x \notcomp y$.
\end{definition}
This needs some time to digest. In particular, if two flippable edges $e$ and $f$ share a common endpoint of degree \four, then they are compatible  (\Fig{fi:CompUncompElements} bottom left), quite contrary to the situation for full triangulations as treated in \Sec{s:Interplay} (see \Def{d:FlipEdgeRel}). The configurations of 2 flippable but incompatible elements are shown in \Fig{fi:CompUncompElements} (two rightmost): (a) Two flippable edges $e$ and $f$ whose removal creates a nonconvex pentagon and whose common endpoint $q$ has degree at least \five. (b) A flippable  edge $e$ and a flippable point $p$ of degree \three whose removal creates a nonconvex quadrilateral \region whose reflex point $q$ has degree at least \five in the \ptriangulation.

What is essential for us is that whenever $x$ and $y$ are compatible in a \ptriangulation $T$, then there is a cycle of length \four or \five containing $(T[x],T,T[y])$, and therefore, apart from the path $(T[x],T,T[y])$, there exists a $T$-avoiding $T[x]$-$T[y]$-path of length \two or \three.
\begin{observation}
\label{ob:CompProps}
Let $T \in \pT(P)$.
(i)
A skipped point $p \in \inn{P} \setminus \ptsInn{T}$ is compatible with every flippable element of $T$.
(ii)
Any two flippable points $p,q \in \inn{P}$ are compatible.
\end{observation}
%
%
\section{Coarsening Partial Subdivisions}
\label{se:PartCoarse}
\begin{figure}[htb]
\centerline{
\placefig{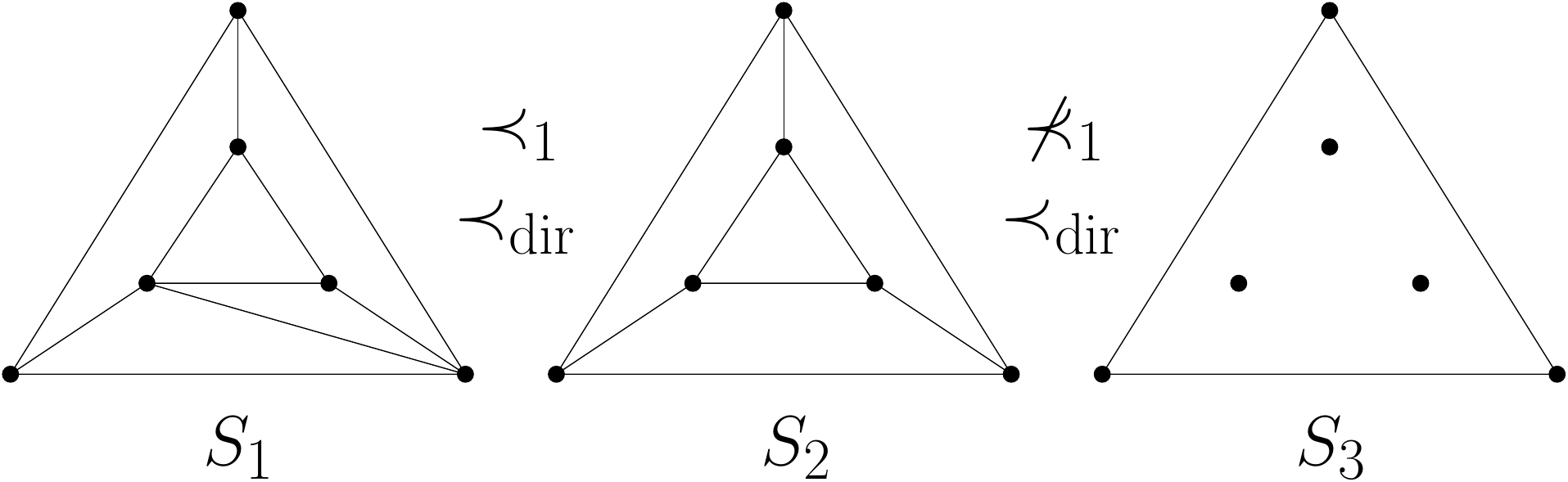}{0.45\textwidth}
}
\caption{$\Slack{S_1} = 2$,  $\Slack{S_2} = 3$, $\Slack{S_3}=3$. Note that $S_2 \precDir S_3$ but $S_2 \not\precPerf S_3$, and that $S_1 \preceq S_3$ with $\Slack{S_3} = \Slack{S_1} + 1$ but $S_1 \not\precPerf S_3$.}
\label{fi:CounterexampleCover}
\end{figure}

As in \Secs{s:FullSubdivisions} and \ref{s:FullBound2} for full triangulations, the existence of many coarsenings is essential for the vertex-connectivity of the \bfg. However, note right away that going via $\preceq$-maximal subdivisions -- as for full subdivisions -- will not work: Here, for partial subdivisions, there is a unique $\preceq$-maximal element, the trivial subdivision. Moreover, note that for full subdivisions (as employed in \Sec{s:FullSubdivisions}), if $S_1 \preceq S_2$, then $(S_1,S_2)$ is an edge in the Hasse-diagram of the partial order $\preceq$ iff $\Slack{S_2} = \Slack{S_1}+1$.  
For partial subdivisions, this is not the case (\Fig{fi:CounterexampleCover}).

\begin{definition}
[direct, perfect coarsening] 
\label{d:DirectPerfectCoarsening} 
Let $S_1$ and $S_2$ be \psubdivisions.
(i)
We call $S_2$ a \Emph{direct coarsening of} $S_1$ (and $S_1$ a \Emph{direct refinement of} $S_2$), in symbols $S_1 \precDir S_2$, if $S_1 \preceq S_2$ and any \psubdivision $S$ with $S_1 \preceq S \preceq S_2$ satisfies $S \in \{S_1,S_2\}$ (equivalently, if $(S_1,S_2)$ is an edge in the Hasse diagram of $\preceq$). 
(ii) 
We call $S_2$ a \Emph{perfect coarsening of} $S_1$ ($S_1$ a \Emph{perfect refinement of} $S_2$), in symbols $S_1 \precPerf S_2$, if $S_1 \precDir S_2$ and $\Slack{S_2} = \Slack{S_1}+1$.
(iii)
$\precPerfStar$ is the reflexive transitive closure of $\precPerf$.
\end{definition}
The reflexive transitive closure of $\precDir$ is exactly $\preceq$, while $\precPerfStar \,\subseteq\, \preceq$ and, in general, the inclusion is proper.
\medskip

\begin{figure}
\centerline{
\placefig{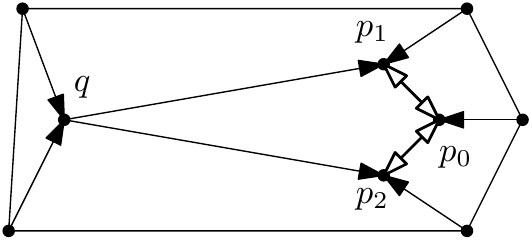}{0.3\textwidth}
}
\caption{A \psubdivision, edges are oriented to endpoints where  locked (not what we called a partial orientation, since some edges are doubly oriented). Removing the three edges incident to $p_0$ does not yield a \psubdivision, since a reflex angle occurs at $p_1$ and $p_2$. The edges incident to $\{p_0,p_1,p_2\}$ are not locked outside this set, but removing all incident edges creates a reflex angle at point $q$.}
\label{f:NonCoarseners}
\end{figure}

To motivate the upcoming definitions, let us discuss a few possibilities of coarsenings, direct coarsenings and perfect coarsenings. There are the simple operations of removing an unlocked edge, and of adding a skipped point $p \in P \setminus \Pts{S}$ as a bystander. For a \ptriangulation, we can isolate a point of degree \three. How does this generalize to \psubdivisions? Removing the edges incident to a point of degree \three does not work if some incident edge might be locked at its other endpoint (\eg $p_0$ in \Fig{f:NonCoarseners}). If, however, no edge incident to a given point $p$ (of any degree) is locked at the respective other endpoint, then we can isolate this point for a coarsening $S'$. Unless $p$ has degree \three, $S'$ is not a direct coarsening of $S$, though. If $p$ has degree at least \four, some\footnote{Actually, if $p$ has degree $d\ge 4$, at least $d-2\ge 2$ incident edges are not locked at $p$.} incident edge, say $e$, is not locked at $p$, thus not locked at all, and therefore, $S \preceq S'' \preceq S'$ for $S'':= (\Pts{S}, \Eds{S} \setminus \{e\})$. Finally, suppose we want to isolate all points in a set $U$ of points for obtaining a coarsening $S'$. For this to work, it is necessary that no edge $e$ connecting $U$ with the outside is locked at the endpoint of $e$ not in $U$. However, this is not a sufficient condition, because several edges connecting $U$ with a point not in $U$ can collectively create a reflex vertex by their removal (\eg $U=\{p_0,p_1,p_2\}$ in \Fig{f:NonCoarseners}). Moreover, for $S \precDir S'$ to hold, $U$ cannot be incident to unlocked edges, and no nonempty subset of $U$ can be suitable for such an isolation operation.

\begin{definition}
[prime, perfect \coarsener; increment]
\label{d:Coarsener}
Let $S$ be a \psubdivision and let $U  \subseteq \PtsInn{S}$.
\begin{EnumRom}
\item
$U$ is called a \Emph{\coarsener}, if 
(a) $U$ is incident to at least one edge in $S$, and
(b) removal of the set $E_U$ of all edges incident to $U$ in $S$ yields a \psubdivision.
\item
If $U$ is a \coarsener, the \Emph{increment of} $U$, $\Incr{U}$, is defined as $|E_U| - 2|U|$.
\item
$U$ is called a \Emph{prime \coarsener}, if
(a) $U$ is a \coarsener,
(b)  $U$ is a minimal \coarsener, \ie no proper subset of $U$ is a \coarsener, and
(c) all edges incident to  $U$ are locked.
\item
$U$ is called a \Emph{perfect \coarsener}, if
(a) $U$ is a prime \coarsener, and
(b)  $\Incr{U} = 1$.
\end{EnumRom}
\end{definition}

\begin{figure}[htb]
\centerline{
\placefig{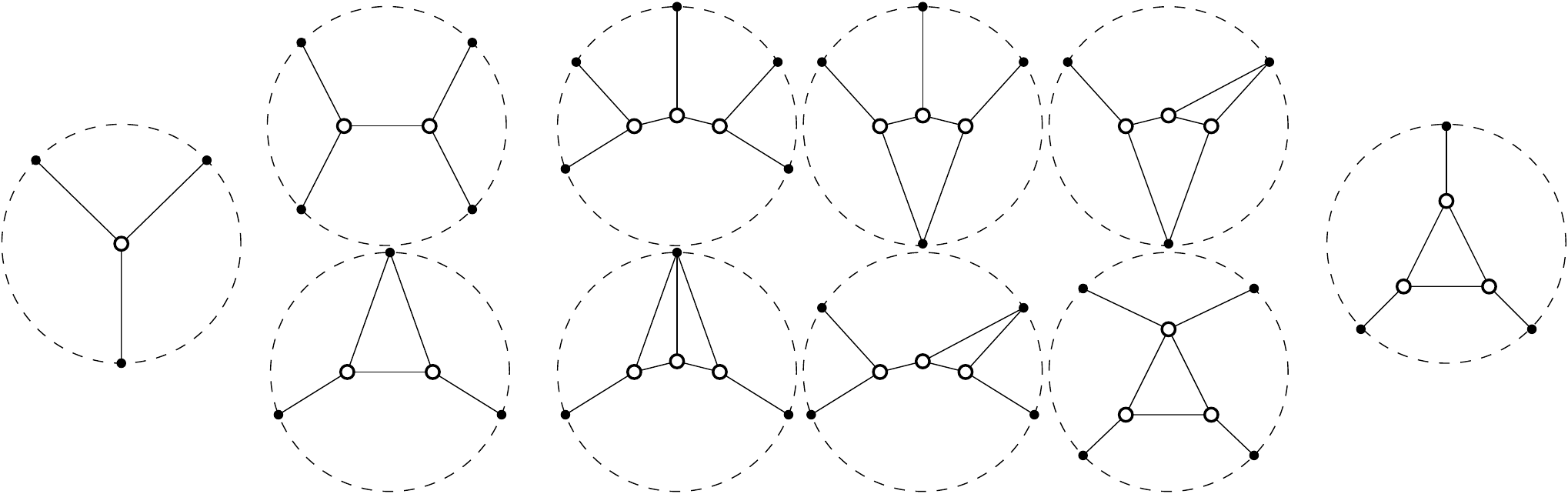}{0.75\textwidth}
}
\caption{Prime \coarseners, all perfect, except for the rightmost one (with $\incr = 0$).}
\end{figure}
The following observation, a simple consequence of \Obs{o:PartSlackMissingEdges}, explains the term ``increment''.
\begin{observation} 
\label{o:Increment}
Let $S$ be a subdivision with \coarsener $U$, and let $S'$ be the subdivision obtained from $S$ by removing all edges incident to $U$. Then $\Slack{S'} = \Slack{S} + \Incr{U}$.
\end{observation}
\begin{observation}
\label{ob:Coarsers}
\begin{EnumRom}
\item 
\label{it:iCoarsers}
Every \psubdivision $S$ with $\EdsInn{S}\neq \emptyset$ has a \coarsener (the set $\PtsInn{S}$).
\item 
\label{it:iiiCoarsers}
If $U_1$ and $U_2$ are \coarseners, then $U_1 \cap U_2$ is a \coarsener, unless there is no edge of $S$ incident to $U_1 \cap U_2$.
\item  
\label{it:ivCoarsers}
If $U_1$ and $U_2$ are prime \coarseners, then $U_1=U_2$ or $U_1 \cap U_2 = \emptyset$.
\item  
\label{it:viiiCoarsers}
If $U$ is a prime \coarsener, then the subgraph of $S$ induced by $U$ is connected.
\end{EnumRom}
\end{observation}
The following observation lists all ways of obtaining direct and perfect coarsenings. 
\begin{observation}
\label{o:direct-perfect-coarsening}
Let $S=(V,E)$ and $S'$ be \psubdivisions. 
\begin{EnumRom}
\item
$S'$ is a direct coarsening of $S$ iff it is obtained from $S$ by one of the following.
\begin{description}
\item \Emph{Adding a single point}. For $p \in P \setminus V$, $S'=(V \cup \{p\},E)$ (with $\Slack{S'} = \Slack{S}+1$).
\item \Emph{Removing a single unlocked edge}. For $e \in E$, not locked by either of its two endpoints,  $S'=(V,E \setminus \{e\})$ (with $\Slack{S'} = \Slack{S}+1$).
\item \Emph{Isolating a prime \coarsener}. For $U$ a prime \coarsener in $S$, $S'$ is obtained from $S$ by removal of the set, $E_U$, of all edges incident to points in $U$, \ie  $S'=(V,E \setminus E_U)$ (with $\Slack{S'} = \Slack{S} + \Incr{U}$).
\end{description}
\item
$S'$ is a perfect coarsening of $S$ iff it is obtained from $S$ by adding a single point, removing a single unlocked edge, or by isolating a perfect \coarsener.
\end{EnumRom}
\end{observation}
We are prepared for the right formulation and proof of the Coarsening Lemma.
\begin{lemma}
[Coarsening Lemma for partial subdivisions] 
\label{le:CoarsenPartSubd}
Every \psubdivision of slack $D$ has at least $n-3-D$ perfect coarsenings (\ie direct coarsenings of slack $D+1$).
\end{lemma}
\begin{MyProof}
We start with the case $D=0$, \ie we have a \ptriangulation $T$ and we want to show that there are at least $n-3$ direct coarsenings of slack \one. Let $N:= |\Pts{T}|$. We orient inner locked edges to their locking endpoints (recall that in a \ptriangulation there is at most one such endpoint for each inner edge). Let $C_i$, $i \in \NNnull$, be the number of points $p\in \PtsInn{T}$ with indegree $i$. The number of unoriented, thus unlocked edges is at least $N-3-C_3$ (\Lm{le:Unorient}). 

There are $n -N$ \psubdivisions obtained from $T$ by adding a single point, there are at least $N-3-C_3$ \psubdivisions obtained from $T$ by removing a single unlocked edge, and there are $C_3$ direct coarsenings obtained from $T$ by isolating an inner point of degree \three. Adding up these numbers gives at least $n-3$ perfect coarsenings of $T$.
\smallskip

\noindent
We let $S$ be a \psubdivision of slack $D\!\ge\! 1$ assuming the assertion holds for slack less than $D$.

\Case{1}{There is a bystander $p_0 \in \PtsInn{S}$.} Then $(\Pts{S} \setminus \{p_0\}, \Eds{S})$ is a \psubdivision of slack $D-1$ of $P \setminus \{p_0\}$ with at least $(n-1)-3-(D-1)=n-3-D$ perfect coarsenings of slack $D$. For each such perfect  coarsening $S'$, the subdivision $(\Pts{S'} \cup \{p_0\}, \Eds{S'})$ is a direct coarsening of $S$ of slack $D+1$, thus a perfect coarsening.

\Case{2}{There is no bystander in $S$.} Again we employ a partial orientation of $S$. The choice of the orientation is somewhat more intricate and we will proceed in three phases (\Fig{fig:threephases}). We keep the invariant that the unoriented inner edges are exactly the unlocked inner edges.
\smallskip

In a \emph{first phase}, we orient all locked inner edges to \Emph{all} of their locking endpoints, \ie we temporarily allow edges to be directed to both ends (to be corrected in the third phase); edges directed to both endpoints are called \Emph{mutual edges}. We can give the following interpretation to an edge directed from $p$ to $q$: If we decide to isolate $p$ (\ie remove all incident edges of $p$) for a coarsening of $S$, then $q$ becomes a reflex point of some \region and we have to isolate $q$ as well (\ie every \coarsener containing $p$ must contain $q$ as well). In particular, if $\{p,q\}$ is a mutual edge, then either both or none of the points $p$ and $q$ will be isolated. In fact, if we consider the graph $G$ with $\Pts{G} :=\PtsInn{S}$ and $\Eds{G}$ the mutual edges in the current orientation, then in any coarsening of $S$ either all points in a connected component of $G$ are isolated, or none. 

A connected component $K$ of $G$ is called a \Emph{candidate component}, (a) if all edges connecting $K$ with points outside are directed towards $K$, (b) no point in $K$ is incident to an unoriented edge, (c) all points in $K$ have indegree \three, and (d) the mutual edges in $K$ do not form any cycle (\ie they have to form a spanning tree of $K$). It follows that if $K$ has $k$ points then the number of edges is $3k - (k-1) = 2k+1$. The term ``candidate'' refers to the fact that removing all edges incident to $K$ \Emph{seems} like a direct coarsening step with incrementing the slack by \one (\Obs{o:PartSlackMissingEdges}); however, while individual edges connecting $K$ to the rest of the graph are not locked at their endpoints outside $K$, some of these edges collectively may actually create a reflex vertex in this way (see $K$ and $q$ in \FigLR{fig:threephases}{left}). So $K$ is only a candidate for a perfect \coarsener. 
\smallskip

\begin{figure}[htb]
\centerline{
\placefig{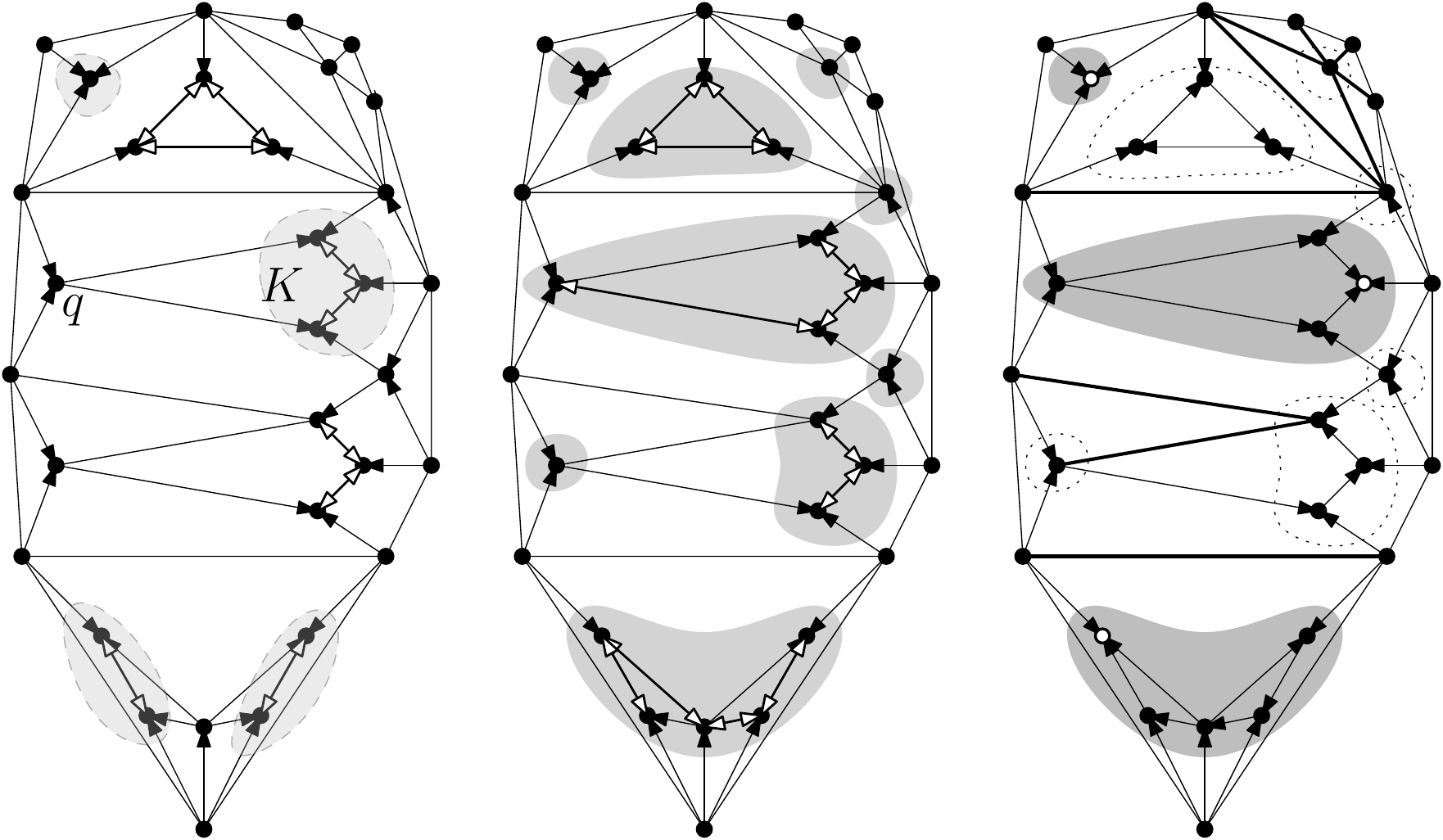}{0.75\textwidth}
}
\caption{Orientation after phase 1, with candidate components shaded (left); after phase 2 (middle), with the connected components of $G^*$; after phase 3 (right), with unoriented edges bold (each of these can be removed for a coarsening of slack \one larger), and with the candidate components with a leader shaded (perfect \coarseners).}
\label{fig:threephases}
\end{figure}
We start the \emph{second phase} of orienting edges further. In the spirit of our remarks about candidate components of $G$, suppose $q$ is an inner point outside a candidate $K$ of $G$ (thus all edges connecting $q$ to $K$ are directed from $q$ to $K$), such that removing the edges connecting $q$ to $K$ creates a reflex angle at $q$. Then we orient one (and only one) of the edges connecting $q$ to $K$, say $\{p,q\}$, also to $q$ (thereby making this edge mutual).\footnote{The reader might be worried that $q$ now joins the candidate component while possibly not having indegree \three as required in a candidate component. Fine, this just means that the enlarged component is not a candidate component, \ie we have lost a candidate component.}  
We call all the edges connecting $K$ to $q$, except for $\{p,q\}$, the \Emph{witnesses of the extra new orientation of} $\{p,q\}$ \Emph{from} $p$ \Emph{to} $q$. We successively proceed orienting edges, with the graph $G$ of mutual edges evolving in this way (and candidate components growing or disappearing).\footnote{The reader will correctly observe that our approach is very conservative towards prime \coarseners, but by what we observed and by what will follow, since we are interested only in perfect \coarseners, we can afford to leave alone connected components other than the candidate components.} The process will clearly stop at some point when the second phase is completed. We freeze $G$ and denote it by $G^*$.
\smallskip

Before we start the third phase, let us make a few crucial observations: 
\begin{EnumRom}
\item
\label{it:iAfterPhaseTwo}
If $p, q$ are inner points in the same connected component of $G^*$, then any \coarsener  contains both or none (\ie if a connected component is a \coarsener, then it is prime). This holds after phase 1, and whenever we expand a connected component, it is maintained.
\item
During the second phase, an edge can be witness only once, and it is and will never be directed to the endpoint where it witnesses. Why? (a) Before it becomes a witness, it connects different connected components of $G$, after that it is and stays in a connected component of $G$. (b) Before it becomes a witness, it is not directed to the endpoint to which it witnesses an orientation, after that it is and stays in a connected component of $G$ and can therefore not get an extra direction. (An unoriented edge can never get an orientation and it can never be a witness.)
\item
If we remove, conceptually, for each incoming edge of a point $q$ the witnesses (which direct away from $q$) for the orientation of this edge to $q$, then among remaining incident edges, all the incoming edges are locked at $q$ (an incoming edge that was oriented already in the first phase to $q$ has no witness). In particular, the indegree of $q$ cannot exceed \three, and if $q$ is incident to some not ingoing edge which is not a witness for any edge incoming at $q$, then the indegree of $q$ is at most \two. (We might generate incoming edges to a point $q$ that are not consecutive around $q$.)
\item
If an unoriented edge $e$ connects two points of the same connected component of $G^*$, then both endpoints have indegree at most \two (recall that this edge $e$ cannot be a witness at its endpoints). If an edge $e$ is directed from a connected component $K$ of $G^*$ to a point outside $K$, then the tail of this edge $e$  has indegree at most \two (recall that $e$ cannot be a witness at all, since its endpoints are in different connected components if $G^*$).
\item
A candidate component $K$ of $G^*$ is a perfect \coarsener. It is a \coarsener (otherwise, we would have expanded it further), it is a prime \coarsener (see \ItemRef{it:iAfterPhaseTwo} above) and $\Incr{K} = 1$ (we have argued before that a candidate component increases the slack by exactly \one).
\end{EnumRom}
\smallskip

The \emph{third phase} will make sure that each mutual edge loses exactly one direction. Our goal is to have in every connected component $K$ of $G^*$ at most one point with indegree \three. To be more precise, only candidate components have exactly one point with indegree \three, others don't.
Consider a connected component $K$. 

\begin{EnumAlph}
\item
If the mutual edges form cycles in $K$, choose such a cycle $c$ and keep for each edge on $c$ one orientation so that we have a directed cycle, counterclockwise, say. All other mutual edges in $K$ keep the direction in decreasing distance in $G^*$ to $c$, ties broken arbitrarily. This completed, no point in $K$  has indegree \three, since there is always a mutual edge incident that decreases the distance to $c$ and the incoming direction of this edge will be removed.
\item
If $K$ has points of indegree at most \two, choose one such point $p$ with indegree at most \two, orient all mutual edges in $K$ in decreasing distance in $G^*$ to $p$, ties broken arbitrarily. Again, this completed, no point in $K$ will have indegree \three.
\item
If none of the above applies, the mutual edges of  $K$ form a spanning tree and all points in $K$ have indegree \three. Moreover, all edges connecting $K$ with points outside  are directed towards $K$ and no edge within $K$ is unoriented (violation of these properties force a point of indegree at most \two). So this is a candidate component. We choose an arbitrary point $p$ in $K$, call it the \Emph{leader of} $K$, and for all mutual edges keep the orientation of decreasing distance in $G^*$ to $p$ (ties cannot occur, mutual edges form a tree). Now the leader $p$ is the only point of $K$ with indegree \three, all other points in $K$ have indegree exactly \two.
\end{EnumAlph}

Phase 3 is completed. Let us denote the obtained partial orientation of $S$ as ${\orient{S}}^{*}$. It has identified certain connected components  of $G^*$ which have a leader of indegree \three. In fact, every point of indegree \three after phase 3 is part of a perfect \coarsener (probably of size \one). 

We can now describe a sufficient supply of perfect coarsenings of $S$. Let $N:= |\Pts{S}|$ and let $C_3$ be the number of points of indegree \three in ${\orient{S}}^{*}$. We know that there are at least $N-3-D - C_3$ unoriented inner edges (\Lm{le:Unorient}).
\begin{EnumCapRom}
\item
There are $n-N$ perfect coarsenings obtained by adding a single point $p \in P \setminus \Pts{S}$.  
\item 
There are at least $N-3-D- C_3$ perfect coarsenings obtained by removing a single unoriented inner edge in ${\orient{S}}^{*}$. 
\item 
\label{it:IIITheCoarsenings}
And there are $C_3$ perfect coarsenings obtained by isolating all points in a candidate component in $G^*$ (with a leader of indegree \three).
\end{EnumCapRom}
In this way we have identified at least $n-3-D$ perfect coarsenings.
\end{MyProof}
Here are two immediate implications which we will need later: The first in the vertex-connectivity proof in \Sec{se:PartLink} and the second for the result about covering of the \bfg\ by $(n-3)$-polytopes in \Sec{s:ImplRegularityPreserv}.
\begin{corollary}
\label{co:CoarsenPartial}
Let $T\in\pT(P)$.
\begin{EnumRom}
\item
\label{i:icCoarsenPartial}
$T$ has at least $n-3$ flippable elements.
\item
\label{i:iicCoarsenPartial}
For every $x$ flippable in $T$ there are at least $n-4$ elements  compatible with $x$.
\end{EnumRom}
\end{corollary}
Part \ItemRef{i:icCoarsenPartial} of the corollary was proved, without general position assumption, in \cite[Thm.\,2.1]{LSU99}. 
\begin{corollary}
\label{c:UpTheLadder}
For every \psubdivision $S'$ with $\Slack{S'} \le n-3$ there is a \psubdivision $S$ with $S' \precPerfStar S$ and $\Slack{S} = n-3$. 
\end{corollary}
%
%
\section{$(n-3)$-Connectivity for Partial Triangulations}
\label{se:PartLink}
To complete the proof of \Thm{t:MainPart}, the $(n-3)$-vertex connectivity of the \bfg, we need again links, now for partial triangulations, which are graphs that represent the compatibility relation among flippable elements (\Def{def:compatibleelements}). 
%
%
\subsection{Link of a partial triangulation}
\label{s:PartLink}
Recall that if $x$ is a flippable element in a \ptriangulation $T$ then $\pFlip{T}{x}$ denotes the \psubdivision with $\pTref{\pFlip{T}{x}} = \{T,T[x]\}$, and if $y$ is compatible with $x$, denoted $x\comp y$, then $\pFlip{T}{x,y}$ denotes the unique coarsening of slack \two of $T$ with $\{T[x],T,T[y]\} \subseteq \pTref{\pFlip{T}{x,y}}$ (\Def{def:compatibleelements}).
\begin{definition}
[link of partial triangulation]
For $T \in \pT(P)$, the \Emph{link of} $T$, denoted $\pLink{T}$, is the edge-weighted graph with vertices $\ElmtsFlip{T}:= \{x \in \PtsInn{T} \cup \EdsInn{T}  \mid x \mbox{~flippable~in~} T\}$ and edge set $\{\{x,y\}\in{ \ElmtsFlip{T} \choose 2} \mid x \comp y\}$. The \Emph{weight of} an edge $\{x,y\}$ is $|\pTref{\pFlip{T}{x,y}}| - 2$ (which is $2$ or $3$).
\end{definition}
(See \Sec{s:Approach} for some intuition for this definition.) We will see that it is enough to prove $(n-4)$-vertex connectivity of all links.  The following lemma implies, via \Lm{l:C4FreeConn}, that the vertex connectivity of links is determined by the minimum vertex degree. 
\begin{lemma} 
\label{le:ForceCompPair2}
The complement of $\pLink{T}$ has no cycle of length \four, \ie if $(x_0, x_1, x_2, x_3)$ are flippable elements in $T$, then there exists $i\in \{0,1,2,3\}$ such that $x_i \comp x_{i+1\!\bmod\!4}$.
\end{lemma}
\begin{MyProof}
Recall that all $p \in \inn{P} \setminus \PtsInn{T}$ are flippable and compatible with every flippable element (\Obs{ob:CompProps}(i)), so we can assume $\{x_0, x_1, x_2, x_3\} \subseteq \PtsInn{T} \cup \EdsInn{T}$. Moreover, if $p, q \in \PtsInn{T}$ are two distinct points flippable in $T$, then $p \comp q$ (\Obs{ob:CompProps}(ii)). Hence, we assume that no two consecutive elements in the cyclic sequence $(x_0,x_1,x_2,x_3)$ are points; w.l.o.g.\ let $x_0=e$ and $x_2=f$ be edges.

\begin{figure}[htb]
\centerline{
\placefig{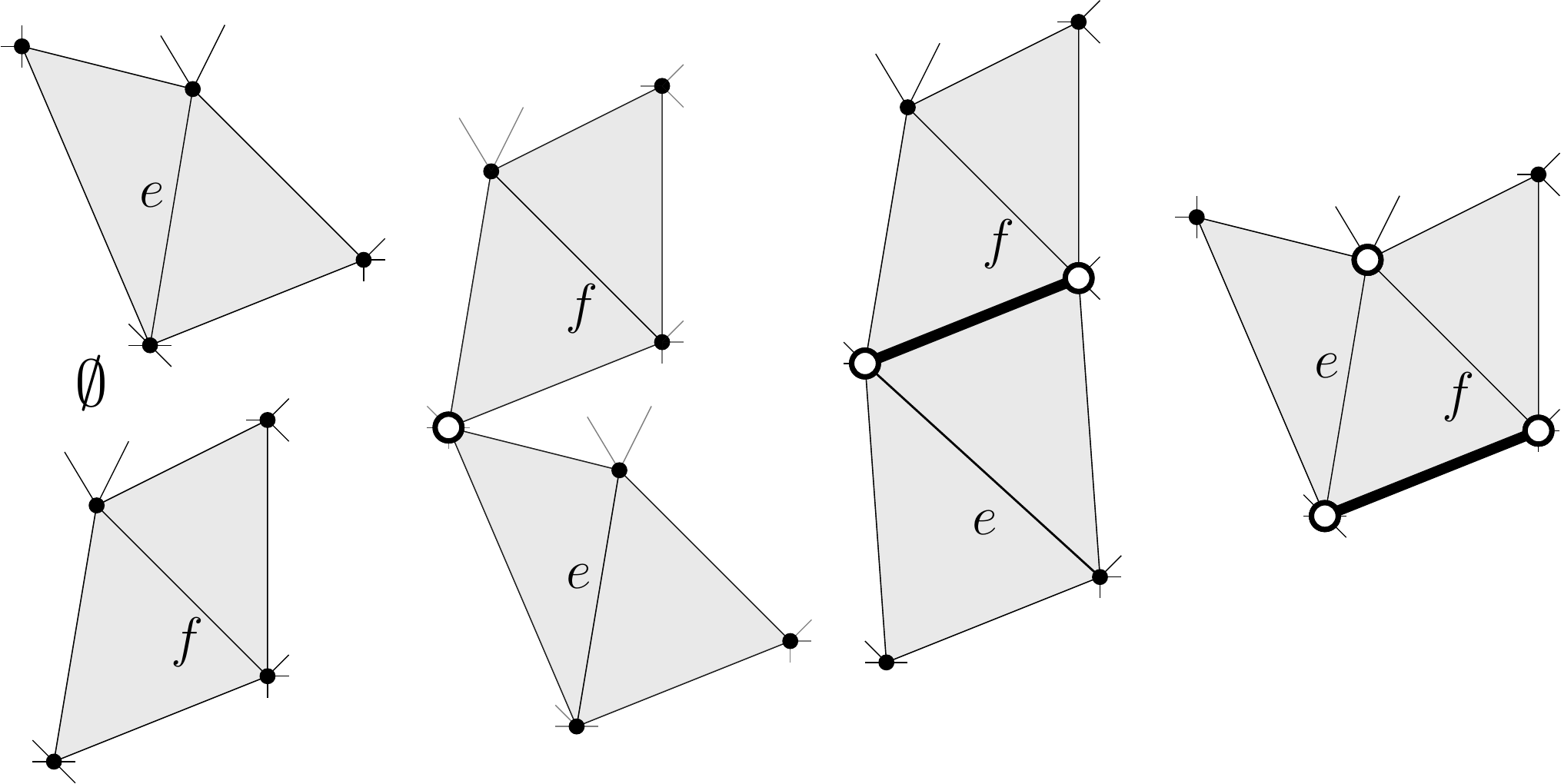}{0.5\textwidth}
}
\caption{Intersections of boundaries of territories of two flippable edges.}
\label{fi:CommonBoundary}
\end{figure}
Recall from \Def{d:Territory}, that for an inner edge $e$ in a triangulation $T$, its \emph{territory} $\Terr{e}$, is defined as the interior of the closure of the union of the two regions in $T$ incident to $e$. Obviously, $e$ is flippable in $T$ iff the quadrilateral $\TerrTwo{T}{e}$ is convex. Note that for an element $x$ to be incompatible with edge $e$, $x$ must appear on the boundary of $\Terr{e}$, and analogously elements incompatible with $f$ must appear on the boundary of $\Terr{f}$.

We show that there is at most one flippable element in the intersection of the boundaries of $\Terr{e}$ and $\Terr{f}$ (\Fig{fi:CommonBoundary}). This is obvious, if $\cl{\Terr{e}} \cap \cl{\Terr{f}}$ is empty or a single point (recall that $\cl{A}$ denotes the closure of $A \subseteq \RR^2$). If this intersection is an edge and its two endpoints, we observe that among any edge and its two incident points, at most one element can be flippable (inner degree \three points cannot be adjacent and cannot be incident to a flippable edge). This covers already all possibilties if $\Terr{e}$ and $\Terr{f}$ are disjoint (since they are convex). Finally, $\cl{\Terr{e}} \cap \cl{\Terr{f}}$ can be a triangle (see argument in the proof of \Lm{l:fLinkC4Free}), in which case the common boundary consists of the common endpoint of $e$ and $f$, clearly not flippable, and an edge with its two endpoints; again, at most one of these three can be flippable. \end{MyProof}
\begin{lemma} 
\label{le:PartLinkToFlipGraph}
Given a \ptriangulation $T$ with $x$ and $y$ flippable elements, $x \neq y$, every $x$-$y$-path of weight $w$ in $\pLink{T}$ induces a $T$-avoiding $T[x]$-$T[y]$-path of length $w$ in the \bfg. Interior vertex-disjoint $x$-$y$-paths in the link induce vertex-disjoint $T[x]$-$T[y]$-paths.
\end{lemma}

\begin{figure}[htb]
\centerline{
\placefig{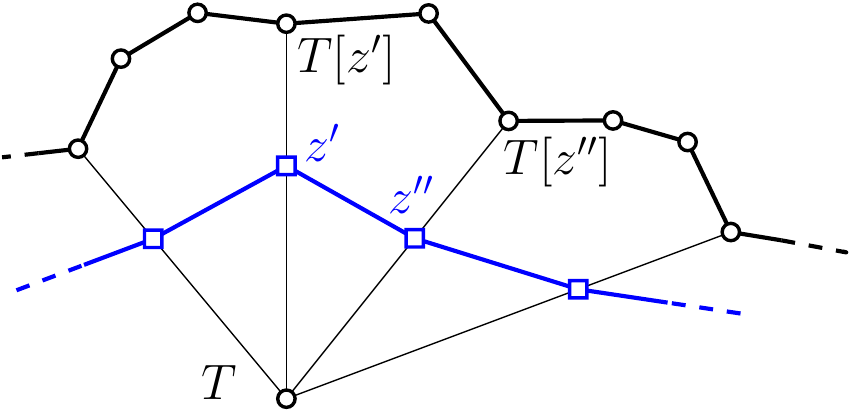}{0.35\textwidth}
}
\caption{From a path in the link to a path in the \bfg.}
\label{f:LinkPathToFGPath}
\end{figure}
\begin{MyProof} Given an $x$-$y$-path, we replace every edge $\{z',z''\}$ on this path by $(T[z'],\ldots,T[z''])$ (of length \two or \three) which draws its (\one or \two) internal vertices from $\pTref{\pFlip{T}{z',z''}} \setminus \{T[z'],T,T[z'']\}$ (\Fig{f:LinkPathToFGPath}); these vertices must have distance \two from $T$ in the  flip graph, while $T[z']$ and $T[z'']$ have distance \one. In the resulting $T[x]$-$T[y]$-path, all internal vertices adjacent to  $T$ (\ie of the form $T[z]$) are distinct from internal vertices at other paths by assumption on the initial paths in the link. For vertices at distance \two, suppose $T_1 \in \pTref{\pFlip{T}{z_1',z_1''}}$ coincides with $T_2 \in \pTref{\pFlip{T}{z_2',z_2''}}$, both at distance \two from $T$. Since  $\Slack{\pFlip{T}{z_1',z_1''}}=\Slack{\pFlip{T}{z_2',z_2''}}=2$, we have that $\pTref{\pFlip{T}{z_1',z_1''}} \cap \pTref{\pFlip{T}{z_2',z_2''}}$ either (a) equals $\{T\}$, (b) equals $\{T,T[z]\}$ for some $z$, or (c) $\pFlip{T}{z_1',z_1''} = \pFlip{T}{z_2',z_2''}$ (\Lm{le:IntersectionProp}). In (a-b) $\pFlip{T}{z_1',z_1''}$ and  $\pFlip{T}{z_2',z_2''}$ cannot possibly share a vertex at distance \two from $T$. Thus (c) holds. $\pFlip{T}{z_1',z_1''} = \pFlip{T}{z_2',z_2''}$ implies $\{z_1',z_1''\} = \{z_2',z_2''\}$.
\end{MyProof}
\begin{lemma}
\label{le:pLink}
For $T \in \pT(P)$, the link $\pLink{T}$ is $(n-4)$-vertex connected.
\end{lemma}
\begin{MyProof}
Let $x$ be a vertex of $\pLink{T}$. $\pFlip{T}{x}$, a subdivision of slack \one, has at least $n-4$ perfect coarsenings of slack $2$ (\Lm{le:CoarsenPartSubd}). Each such coarsening equals $\pFlip{T}{x,y}$ for some $y \comp x$, \ie $y$ is a neighbor of $x$ in $\pLink{T}$. Distinct coarsenings yield distinct compatible elements $y$ (since $\{T[x],T,T[y]\} \subseteq  \pTref{\pFlip{T}{x,y}}$ and $\pTref{\pFlip{T}{x,y}}$ spans a cycle, $T[y]$ is determined as the other neighbor of $T$ on this cycle). That is, the minimum vertex degree in $\fLink{T}$ is at least $n-4$. $\pLink{T}$ has no cycle of length \four in its complement (\Lm{le:ForceCompPair2}). The lemma follows  by \Lm{l:C4FreeConn}.
\end{MyProof}
%
%
\subsection{Proof of \Thm{t:MainPart}}
\begin{MyProof} We know that the \bfg\ is connected, \cite[Sec.\,3.4.1]{LRS10}, and it has at least $n-2$ vertices, since it is nonempty and every vertex has degree at least $n-3$ (\Cor{co:CoarsenPartial}\ItemRef{i:icCoarsenPartial}). Hence, for $(n-3)$-vertex connectivity, by the Local Menger Lemma~\ref{le:LocalMenger} it is left to show that for any $T\in\pT(P)$ and flippable elements $x$ and $y$, there are at least $n-3$ vertex-disjoint $T[x]$-$T[y]$-paths in the \bfg. Since $\pLink{T}$ is $(n-4)$-vertex connected (\Lm{le:pLink}),  $\pLink{T}$ has at least $n-4$ vertex-disjoint $x$-$y$-paths (Menger's Theorem~\ref{t:Menger}). Therefore, there are at least $n-4$ vertex-disjoint $T[x]$-$T[y]$-paths disjoint from $T$ (\Lm{le:PartLinkToFlipGraph}). Together with the path $(T[x],T,T[y])$, the claim is established.
\end{MyProof}
%
%
\section{Regular Subdivisions by Successive Perfect Refinements}
\label{s:SuccPerfRef}
%

\begin{figure}[htb]
\centerline{
\placefig{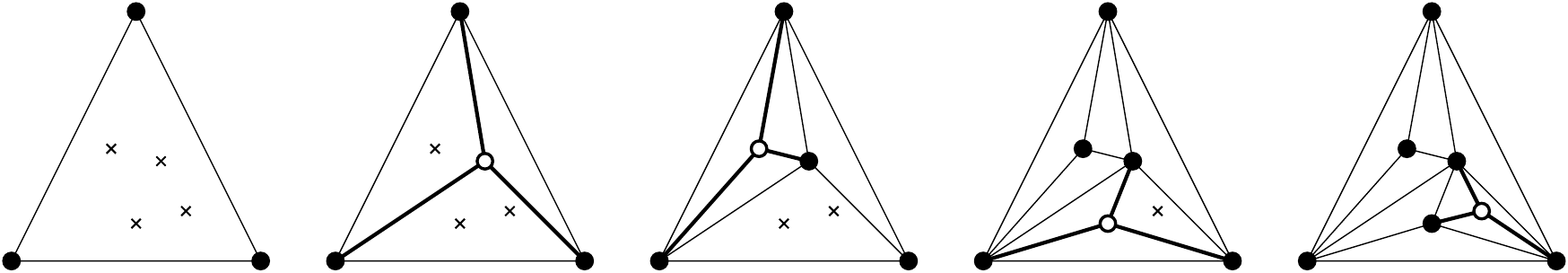}{0.7\textwidth}
}
\caption{Stacked triangulations (which are always regular).}
\label{f:Stacked}
\end{figure}

\begin{figure}[htb]
\centerline{
\placefig{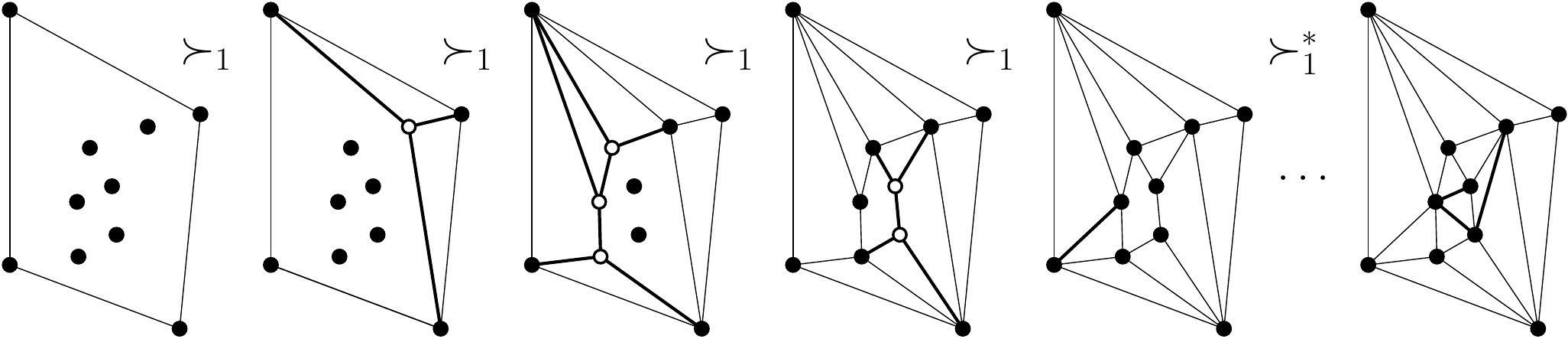}{0.9\textwidth}
}
\caption{Successive perfect refinements of a trivial subdivison (all \psubdivisions regular).}
\label{f:PerfectlyRefined}
\end{figure}

\noindent
Suppose $h= 3$ and consider stacked triangulations of $P$, \ie we start with the triangulation $(\ext{P},\EdsHull)$, and then we successively add points in $\inn{P}$ by connecting a new point to the three vertices of the triangle where it lands in (\Fig{f:Stacked}). It is easily seen that this yields regular triangulations. The result of this section is the following sufficient condition for the regularity of a \psubdivision (\Def{d:HeightFuncRegular} below), which can be seen as a generalization of the regularity of stacked triangulations (\Fig{f:PerfectlyRefined}). The condition is not necessary, see \Sec{s:Mother}.
\begin{theorem} 
\label{t:LowClosureRegular}
If $S \precPerfStar \Striv$ for a \psubdivision $S$, then $S$ is a regular subdivision.
\end{theorem}

In other words, all \psubdivisions, in particular, all \ptriangulations in the $\precPerf$-lower closure of $\Striv$ are regular. This condition will eventually allow us to show the covering of the \bfg\ by graphs of $(n-3)$-polytopes. The proof of \Thm{t:LowClosureRegular} stretches out over several definitions and lemmas with a conclusion in \Sec{s:ProofLowClosureRegular}. Before we give a brief outline of this proof shortly in \Sec{s:Outline}, we first introduce some notions.
%
%
\subsection{Height functions, liftings, and regular subdivisions}
%
\begin{definition}
[linear, compliant, realizing height function; regular subdivision]
\label{d:HeightFuncRegular}
~A \Emph{height function on} $A \subseteq \RR^2$ is a vector $\omega \in \RR^A$, $p \mapsto \omega_p$. For $p = (x_p, y_p) \in A$, we let $p^{(\omega)}:=(x_p, y_p, \omega_p)$, and for $B \subseteq A$, we set $B^{(\omega)} := \{p^{(\omega)} \mid p \in B\}$. We say that $\omega$ is \Emph{linear on} $B \subseteq A$, if there exist $a$, $b$, and $c$ in $\RR$ such that $\omega_p = a x_p + b y_p + c$ for all $p \in B$, \ie if $B^{(\omega)}$ is coplanar. 
\smallskip

\noindent
Let $S$ be a \psubdivision.
\begin{EnumRom}
\item
A height function $\omega$ on $\Pts{S}$ is \Emph{linear on} $S$ if it is linear on $\Pts{S}$. $\Lambda(S)$ denotes the set of linear height functions on $S$ and for $A \subseteq \Pts{S}$, $\Lambda_A(S)$ denotes the set of height functions on $S$ linear on $A$.
\item
A height function $\omega$ on $\Pts{S}$ \Emph{complies with} $S$, (or is $S$\Emph{-compliant}), if for every \region $r$ of $S$, $\omega$ is linear on $\Pts{r}$ (including bystanders). Let $\Gamma(S)$ be the set of $S$-compliant height functions.
\item
A height function $\omega$ on $\Pts{S}$ \Emph{realizes} $S$, if $S$ is the projection of the lower convex hull of $\Pts{S}^{(\omega)}$, with all points of $\Pts{S}$ (also the bystanders) appearing on this lower convex hull. 
\item
\label{subdef:regularsubdivision}
$S$ is called \Emph{regular} if there is a height function realizing $S$.
\end{EnumRom}
\end{definition}
Compliant height functions constitute a relaxation of realizing height functions (and of linear height functions): Every realizing height function (and every linear height function) is compliant. All height functions on a triangulation $T$ are compliant, \ie $\Gamma(T) = \RR^{\Pts{T}}$. For the trivial subdivision $\Striv$, the compliant height functions are exactly the linear height functions, \ie $\Gamma(\Striv) = \Lambda(\Striv)$. While compliant height functions exist for all \ptriangulations and the trivial subdivision, this is a non-trivial property for general \psubdivisions.
\begin{lemma}
\label{le:LinCompliant}
Let $S$ be a \psubdivision.
\begin{EnumRom}
\item 
\label{it:iLinCompliant}
$\Lambda(S)$ is a linear subspace of $\RR^{\Pts{S}}$ of dimension $\dim \Lambda(S) =3$. More generally, for every $B \subseteq \Pts{S}$ with $|B|\geq 3$, $\Lambda_{B}(S)$ is a linear subspace of $\RR^{\Pts{S}}$ of dimension $|\Pts{S}| - (|B| - 3)$.
\item
\label{it:iiLinCompliant}
$\Gamma(S) = \bigcap_{r \in \Reg{S}} \Lambda_{\Pts{r}}(S)$ ($\Reg{S}$ the set of \regions of $S$).
\item 
\label{it:iiiLinCompliant}
$\Gamma(S)$ is a linear subspace of $\RR^{\Pts{S}}$ with $\Gamma(S) \supseteq \Lambda(S)$ and  $\dim \Gamma(S) \ge |\Pts{S}|-\Slack{S}$.
\end{EnumRom}
\end{lemma}
\begin{MyProof} 
\ItemRef{it:iLinCompliant} is obvious and \ItemRef{it:iiLinCompliant} holds directly by definition. 
\smallskip

\noindent
Now, with $\Slack{r} = |\Pts{r}| - 3$ and $\Slack{S} = \sum_{r \in \Reg{S}} (|\Pts{r}| - 3)$,  assertion \ItemRef{it:iiiLinCompliant} is an immediate consequence of \ItemRef{it:iLinCompliant}, \ItemRef{it:iiLinCompliant}, and the fact that intersecting linear subspaces of co-dimension $d_1$ and $d_2$ yields a subspace of co-dimension at most $d_1+d_2$: 
\[
|\Pts{S}| - \dim \Gamma(S) 
\stackrel{\mbox{\footnotesize \ItemRef{it:iiLinCompliant}}} {\le}
\sum_{{r \in \Reg{S}}} \left(|\Pts{S}| - \dim \Lambda_{\Pts{r}}(S)\right) 
\stackrel{\mbox{\footnotesize \ItemRef{it:iLinCompliant}}}{=} 
\sum_{{r \in \Reg{S}}} \left(|\Pts{r}| - 3\right) = \Slack{S}
\]
\end{MyProof}
We see that if $\Slack{S} < |\Pts{S}|-3$, there are always compliant height functions not in $\Lambda(S)$. In order to extract among those a realizing height function we consider mountains and valleys in the lifting given by a compliant height function.
\begin{definition}[$\omega$-lifting; $\omega$-labeling]
Let $\omega \in \Gamma(S)$. The $\omega$-\Emph{lifting of} $S$ is the unique piecewise linear function $f$ on the convex hull of $\Pts{S}$, that is linear on every \region $r$ of $S$, and $\ind{f}{\Pts{S}}  = \omega$. 

We call $e \in \EdsInn{S}$ a \Emph{mountain}, a \Emph{valley}, or \Emph{flat} in the $\omega$-lifting, depending on whether the directional derivative of function $f$ decreases, increases, or remains constant, respectively, as one traverses the $f$-lifted edge from one side to the other (at a mountain, the function is locally strictly concave, at a valley it is locally strictly convex). The $\omega$-\Emph{labeling of} $S$ assigns $\oplus$, $\ominus$, and $0$ to each inner edge of $S$, depending on whether the lifted edge is a mountain, a valley, or flat, respectively.
\end{definition}

\begin{figure}[htb]
\centerline{
\placefig{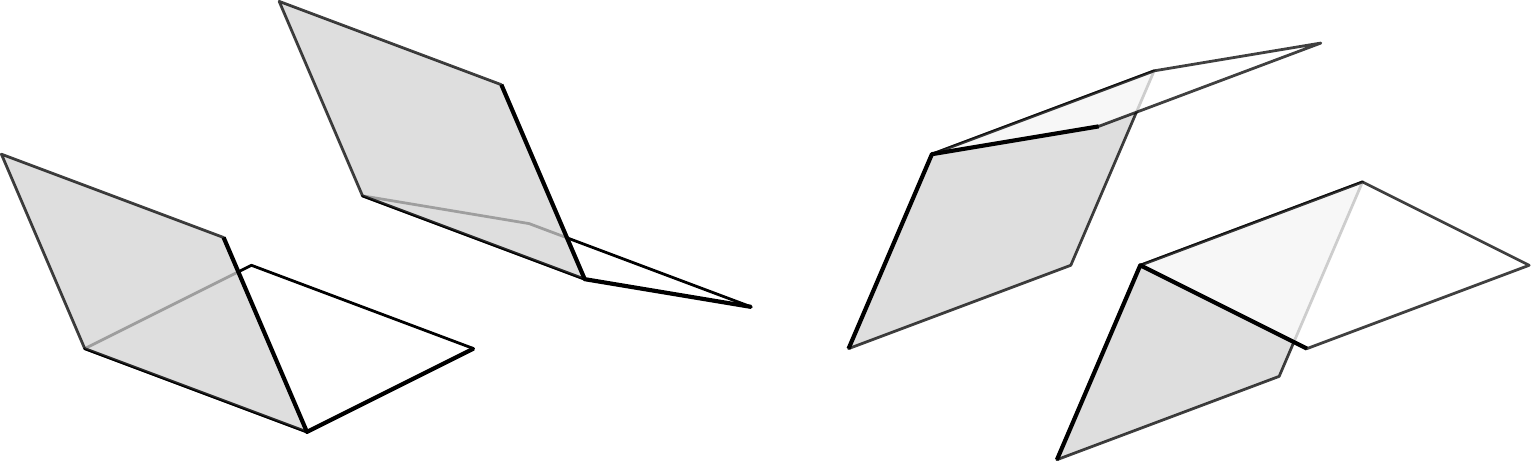}{0.45\textwidth}
}
\caption{Valleys (left) and mountains (right).}
\end{figure}
\begin{observation} 
Let $\omega \in \Gamma(S)$. 
\begin{EnumRom}
\item
$\omega$ is linear on $S$ iff the $\omega$-labeling of $S$ is constant $0$. 
\item
$\omega$ realizes $S$ iff the $\omega$-labeling of $S$ is constant $\ominus$.
\end{EnumRom}
\end{observation}
%
%
\subsection{Mother of examples}
\label{s:Mother}
\begin{figure}[htb]
\centerline{
\placefig{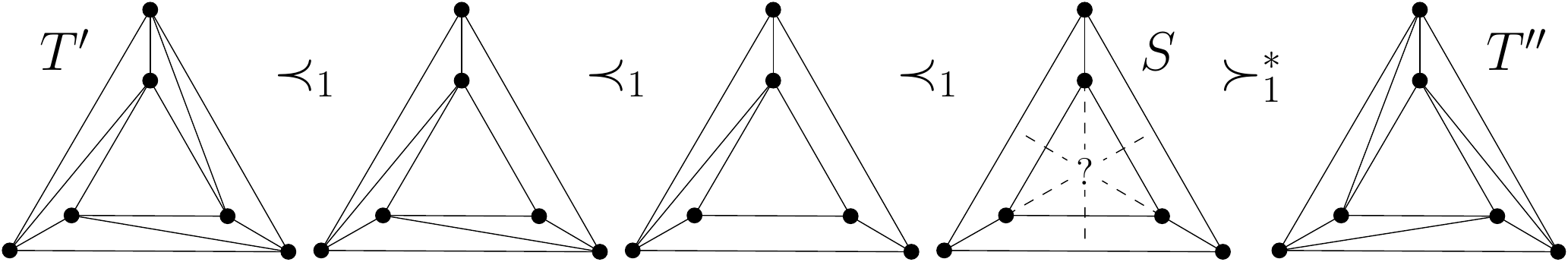}{0.8\textwidth}
}
\caption{Exactly one of $S$, $T'$, and $T''$ is regular. Which one depends on how the dashed lines meet.}
\label{f:Mother}
\end{figure}

In order to understand the subtleties of whether a \psubdivision is regular or not, we should briefly discuss the mother of examples, see \cite{LRS10}. For this consider the configuration in \Fig{f:Mother}. Whether or not the displayed \psubdivisions are regular or not depends on how exactly the three dashed lines (as indicated in $S$) meet. 
\begin{EnumAlph}
\item 
\label{i:iLinesMeet}
If the three dashes lines meet in a common point,  then $S$ is a regular subdivision, but none of  $T'$ and $T''$ is regular.
\item
\label{i:iiLinesMeet} If the three dashes lines do not meet in a common point, then $S$ is not a regular subdivision, but one of $T'$ and $T''$ is regular, the other one not.
\end{EnumAlph}
%

\noindent
The example allows us to clarify a few points.
\begin{EnumNo}
\item The condition in \Thm{t:LowClosureRegular} for regularity is not necessary (consider Case \ItemRef{i:iiLinesMeet} with $T'$ regular, and note $T' \not\precPerfStar \Striv$). This is inherently so, since the condition in \Thm{t:LowClosureRegular} depends only on the order type (or oriented matroid) of the point set. In fact, note that a perturbation of the point set does not change the order type of the set, but it affects how the dashed lines meet and, therefore, whether \psubdivisions are regular or not.
\item The condition in \Thm{t:LowClosureRegular} \emph{cannot} be generalized to: \emph{If $S \precPerfStar S'$ and $S'$ is regular, then $S$ is regular.} In fact, adding a single edge in a \psubdivision may switch from regular to non-regular (Case \ItemRef{i:iLinesMeet}). The right generalization will be given in the Regularity Preservation Lemma~\ref{le:RegularToRegular} below.
\end{EnumNo}
%
%
\subsection{Outline of proof of \Thm{t:LowClosureRegular}}
\label{s:Outline}
It is easy to see that if $p$ is an inner point of degree \three in a \ptriangulation $T$, then for any height function $\omega$ (which, as we observed, is $T$-compliant), the $\omega$-labeling assigns the same value to the three edges incident to $p$. We will generalize this observation for an $S$-compliant height function $\omega$ in two ways:
\begin{EnumCapAlph}
\item
\label{i:iOutline}
If $p \in \PtsInn{S}$ and not all incident edges are $0$-labeled, then the $\oplus$-labeled and $\ominus$-labeled edges incident to $p$ cannot be separated by a line through $p$ (\Lm{l:CompliantValid}). 
(In particular, this forces the $\omega$-labeling to be constant on the edges incident to an inner point of degree \three in any \psubdivision.)
\item
\label{i:iiOutline}
If $K$ is a perfect \coarsener of $S$, then the $\omega$-labeling assigns the same label to all the edges $E_K$ incident to a perfect coarsener $K$ (\Lm{le:PerfCoarserLabelingConstant}).
\end{EnumCapAlph}
Here is another simple observation about an inner point $p$ of degree \three in a \ptriangulation $T$. Removing the three edges incident to $p$ (while keeping $p$ as a bystander) yields a \psubdivision $S$ with $T \precPerf S$. Obviously, if $S$ is a regular subdivision, then $T$ is a regular triangulation (this was behind our observation about stacked triangulations at the beginning of this section): Given a height function $\omega$ realizing $S$, we can always perturb $p$ downwards (decrease $\omega_p$ by a sufficiently small value $\varepsilon$), obtaining a height function that realizes $T$. Again, this allows an appropriate generalization:
\begin{EnumCapAlph}
\setcounter{enumi}{2}
\item
\label{i:iiiOutline}
Suppose $S$ is a regular subdivision with $\dim \Gamma(S) = |\Pts{S}| - \Slack{S}$. Then every perfect refinement $S'$ of $S$, \ie $S' \precPerf S$, is regular, and, moreover, $\dim \Gamma(S') = |\Pts{S'}| - \Slack{S'}$ (Regularity Preservation Lemma~\ref{le:RegularToRegular}).
\end{EnumCapAlph}
Since $\Striv$ is regular and $\dim \Gamma(\Striv) = 3 = |\Pts{\Striv}| - \Slack{\Striv}$, this immediately yields an inductive argument for \Thm{t:LowClosureRegular}. For the proof of \ItemRef{i:iiiOutline}, we consider the perfect \coarsener $K$ of $S'$ whose isolation leads to $S$, and a height function $\omega_1$ that realizes $S$. First, we show that $\dim \Gamma(S') > \dim \Gamma(S)$, and, therefore, a height function $\omega' \in \Gamma(S') \setminus \Gamma(S)$ exists. According to \ItemRef{i:iiOutline}, the $\omega'$-labeling assigns the same label to all edges incident to $K$, and since $\omega' \not\in \Gamma(S)$, this label cannot be $0$. Hence, either $\omega'$ or $-\omega'$ assigns constant $\ominus$, and it can be used for a controlled perturbation $\omega_1 + \varepsilon \omega'$ which realizes $S'$.

We will now carefully work out these steps.
%
%
\subsection{Valid $\{\oplus,\ominus,0\}$-edge labelings}
\begin{figure}[htb]
\centerline{
\placefig{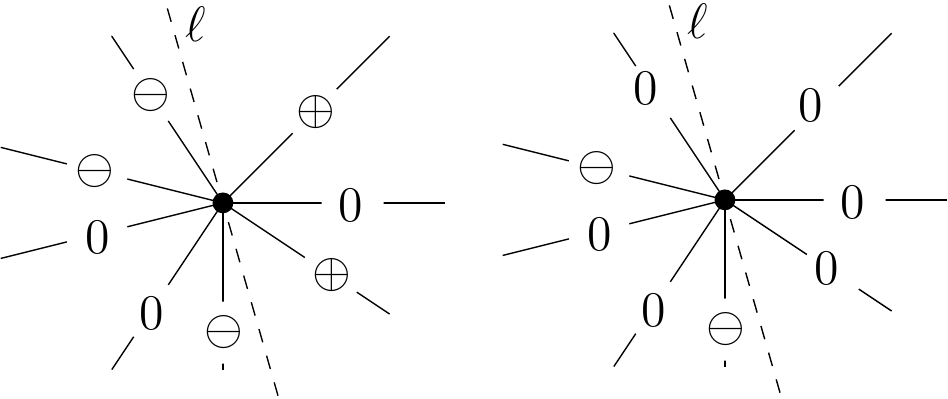}{0.4\textwidth}
}
\caption{Patterns prohibited in valid labelings.}
\label{f:InvalidPatterns}
\end{figure}

%
\begin{definition}
\label{d:ValidLab}
Let $S$ be a \psubdivision. Given a labeling $\alpha: \EdsInn{S} \rightarrow \{\oplus,\ominus,0\}$, we call an inner point $p \in \PtsInn{S}$ $\alpha$-\Emph{pointed}, if $\alpha$ is not constant $0$ on the edges incident to $p$, and if there is a line through $p$ that has all $\oplus$-labeled edges incident to $p$ strictly on one side and all $\ominus$-labeled edges incident to $p$ strictly on the other side of this line. (We do not require that both $\oplus$- and $\ominus$-labeled edges incident to $p$ exist.)

We call $\alpha$ a \Emph{valid labeling of} $\EdsInn{S}$ if no point in $\PtsInn{S}$ is $\alpha$-pointed (\Fig{f:InvalidPatterns}).
\end{definition}

For example, for an inner point  of degree \three in a \psubdivision, a valid labeling must assign the same label to its three incident edges. We can now prove \ItemRef{i:iOutline} above.
\begin{lemma} 
\label{l:CompliantValid}
Let $\omega$ be a height function compliant with \psubdivision $S$. Then the $\omega$-labeling of $S$ is a valid labeling of $\EdsInn{S}$.
\end{lemma}
\begin{MyProof} 
Let $p \in \PtsInn{S}$ and suppose there is a line $\ell$  through $p$ that has all $\oplus$-labeled edges incident to $p$ on one side, and all $\ominus$-labeled edges incident to $p$ on the other side. Sweep a vertical plane $h$ parallel to $\ell$ in $\RR^3$ over $p$ and observe its intersection with the $\omega$-lifting $f$. On the side of the $\oplus$-labeled edges, this intersection must be a locally concave function, on the side of the $\ominus$-labeled edges a locally convex function. Consequently, it has to be locally linear at the point when $h$ contains $p$ and $\ell$. Now it follows that $f$ must be locally linear around $p$ and all edges incident to $p$ must be flat.
\end{MyProof}

\begin{lemma} 
\label{le:PlusMinusLabeling} Let $K$ be a perfect \coarsener in a \psubdivision $S$. In every valid $\{\oplus,\ominus,0\}$-labeling of $\EdsInn{S}$, the edges $E_K$ incident to $K$ all get the same label.
\end{lemma}
\begin{MyProof}
We plan to prove the following.
\begin{clm}
With reference to the orientation process in the proof of \Lm{le:CoarsenPartSubd}, after the second phase, in any valid labeling, the edges incident to a candidate component of $G^*$ get the same label.
\end{clm}
It is not obvious from the proof of \Lm{le:CoarsenPartSubd} that every perfect \coarsener is identified by the three phase process. In order to close this gap (from the claim to the assertion of the lemma), isolate $K$ in $S$ obtaining a \psubdivision $S'$ with a region $r$ containing the points in $K$. Let $V_r$ be the vertices of this regions, \ie $V_r = \Pts{r} \setminus \PtsBy{S'}$. Now consider the subgraph $S_r$ of $S$ induced by $V_r \cup K$. $K$ is a perfect \coarsener of $S_r$ whose isolation yields the trivial subdivision of $V_r \cup K$. It is the only \coarsener of $S-r$ and $\Slack{S_r} = |\Pts{S_r}|-4$. Therefore, the procedure in the proof of \Lm{le:CoarsenPartSubd} must identify $K$ as a candidate component after the second phase.

We establish the claim by showing the following invariant in the process during the second phase:
\begin{EnumAlph}
\item 
\label{it:iPlusMinusLabeling}
For every candidate component $K$, the edges $E_K$ incident to $K$ obtain the same label in any valid labeling.
\item 
\label{it:iiPlusMinusLabeling}
An edge gaining a new orientation in the second phase and its witnesses obtain the same label in any valid labeling.
\end{EnumAlph}

At the end of the first phase, a point with indegree \three in $\orient{S}$ has actually degree \three in $S$, and therefore any valid labeling must give the same label to all incident edges. Since in a candidate component $K$, all points have indegree \three in $\orient{S}$ (\ie at this point, have degree \three in $S$) and since a connected component is connected [sic!], it easily follows that all edges incident to a candidate component must have the same label in any valid labeling.

During the second phase, a newly oriented edge and its witnesses are part of the edges incident to a candidate component. Hence, given \ItemRef{it:iPlusMinusLabeling}, invariant \ItemRef{it:iiPlusMinusLabeling} is maintained after an orientation step of phase 2. We are left to show that \ItemRef{it:iPlusMinusLabeling} is preserved. Consider a point $p$ of a candidate component. It must have indegree \three, all incident edges are either ingoing or witnesses for an ingoing edge (otherwise, indegree \three is impossible); we know that each bundle of an ingoing edge and its witnesses have the same label, and such a bundle can be separated from the other incident edges by a line through the given point $p$ (this is why the edge was oriented in phase 2). Hence, due to a simple consideration, any valid labeling must assign the same label to all incident edges. (The simple consideration: Suppose a single bundle is labeled $\oplus$, then this bundle can be separated from the other two bundles by a line, contradiction. Suppose exactly two bundles are labeled $\oplus$, then the remaining bundle can be separated from these two $\oplus$-labeled bundles by a line, which is a contradiction. Hence, if any incident edge is labeled $\oplus$, then all incident edges must be labeled $\oplus$. Similarly, for $\ominus$.)
This completes the proof of the claim, and thus of the lemma.
\end{MyProof}
Now, with \Lm{l:CompliantValid} we immediately get property \ItemRef{i:iiOutline}.
\begin{lemma} 
\label{le:PerfCoarserLabelingConstant}
If $\omega$ is an $S$-compliant height function, then the $\omega$-labeling is constant on any set of edges incident to a perfect \coarsener of $S$ .
\end{lemma}
%
%
\subsection{The Regularity Preservation Lemma}
%
\begin{lemma}[Regularity Preservation]
\label{le:RegularToRegular}
Let $S_1$ be a regular subdivision with $\dim \Gamma(S_1) = |\Pts{S_1}| - \Slack{S_1}$. If $S_0 \precPerf S_1$, then $S_0$ is regular and $\dim \Gamma(S_0) = |\Pts{S_0}| - \Slack{S_0}$.
\end{lemma}
\begin{MyProof} Let $\omega_1 \in \RR^{\Pts{S_1}}$ be a height function that realizes $S_1$.

\Case{1}{$S_1$ is obtained from $S_0$ by adding a single point $p \in \inn{P} \setminus \PtsInn{S_{0}}$.} Then $\restr{\omega_1}{\Pts{S_0}}$ realizes $S_0$ and $S_0$ is regular. We have $\Gamma(S_0) = \{ \restr{\omega}{\Pts{S_0}} \,\mid\, \omega \in \Gamma(S_1) \}$. For $\omega \in \Gamma(S_1)$, the value of $\omega_p$ ($p$ the added point) is determined by $\restr{\omega}{\Pts{S_0}}$, \ie $\dim \Gamma(S_1) = \dim \Gamma(S_0)$. Therefore,
\[
\dim \Gamma(S_0) = |\Pts{S_1}| - \Slack{S_1} = (|\Pts{S_0}| + 1)-(\Slack{S_0} + 1) = |\Pts{S_0}| - \Slack{S_0} ~.
\]

\Case{2}{$S_1$ is obtained from $S_0$ by removing a single unlocked edge or by isolating a perfect \coarsener in $S_0$.} We have $\Pts{S_1} = \Pts{S_0}$. The set $E^* := \Eds{S_0} \setminus \Eds{S_1}$ is either a single unlocked edge or the set $E_K$ of edges incident to a perfect \coarsener $K$ in $S_0$. Let $r^*$ be the \region in $S_1$ generated by the removal of the edges in $E^*$.

We have $\Gamma(S_0) \supseteq \Gamma(S_1)$ and (with \Lm{le:LinCompliant}\ItemRef{it:iiiLinCompliant} and $\Slack{S_0} = \Slack{S_1} - 1$)
\[
\dim \Gamma(S_0) \ge |\Pts{S_0}| - \Slack{S_0} = |\Pts{S_1}| - \Slack{S_1} + 1 = \dim \Gamma(S_1) + 1~.
\]
Therefore, there must exist $\omega' \in \Gamma(S_0) \setminus \Gamma(S_1)$, a height function that is not flat on $r^*$ and there is an edge in $E^*$ that is not flat in the $\omega'$-lifting of $S_0$. In that case, all edges in $E^*$ are mountains or all are valleys (trivially true if $|E^*|=1$, otherwise by Lemma~\ref{le:PerfCoarserLabelingConstant}). Let us suppose that the $\omega'$-labeling is constant $\ominus$ on $E^*$ (if not switch to $-\omega'$). Now, for any sufficiently small positive $\varepsilon \in \RR$, the height function $\omega_0 := \omega_1 + \varepsilon \omega'$ is compliant with $S_0$ and all inner edges in $S_0$ are valleys in the $\omega_0$-lifting of $S_0$ ($\varepsilon$ has to be small enough such that all valleys in the $\omega_1$-lifting remain valleys in $\omega_0$; this is a familar operation, see \cite[Lemma~2.3.16]{LRS10}). This establishes that $S_0$ is regular.

We are left to show that $\dim \Gamma(S_0) = |\Pts{S_0}| - \Slack{S_0}$, or, equivalently, $\dim \Gamma(S_0) = \dim \Gamma(S_1) +1$. This holds, if for any two $\omega'$, $\omega''$ in $\Gamma(S_0) \setminus \Gamma(S_1)$ there exists $a \in \RR$ and $\omega \in \Gamma(S_1)$ such that $\omega' = a \omega'' + \omega$. Suppose that all edges in $E^*$ are valleys in $\omega'$, and all edges in $E^*$ are mountains in $\omega''$ (switch signs, if necessary). Now consider $\omega_t := (1-t) \omega' + t \omega''$, $t \in [0,1]$. There must be a value $t$ in $(0,1)$ where some edge in $E^*$ is flat in the $\omega_t$-lifting, but then all edges have to be flat and $\omega_t \in \Gamma(S_1)$ (\Lm{le:PerfCoarserLabelingConstant}). We have shown $\omega' = -\frac{t}{1-t}\omega'' + \frac{1}{1-t} \omega_t$ with $\omega_t \in \Gamma(S_1)$.
\end{MyProof}
%
%
\subsection{Proof of \Thm{t:LowClosureRegular}}
\label{s:ProofLowClosureRegular}
%
\begin{MyProof}
If $S \precPerfStar \Striv$ then there is a sequence
\[
S=S_0 \precPerf S_1 \precPerf \cdots \precPerf S_\ell = \Striv ~.
\]
$\Striv$ is regular, $\Slack{\Striv} = n-3$ and $\Gamma(\Striv) = \Lambda(\Striv)$, of dimension $3 = n - \Slack{\Striv} = |\pts{\Striv}| - \Slack{\Striv}$. Along \Lm{le:RegularToRegular} we have an inductive argument that $S_0 = S$ is regular.
\end{MyProof}
\bigskip

\begin{figure}[htb]
\centerline{
\placefig{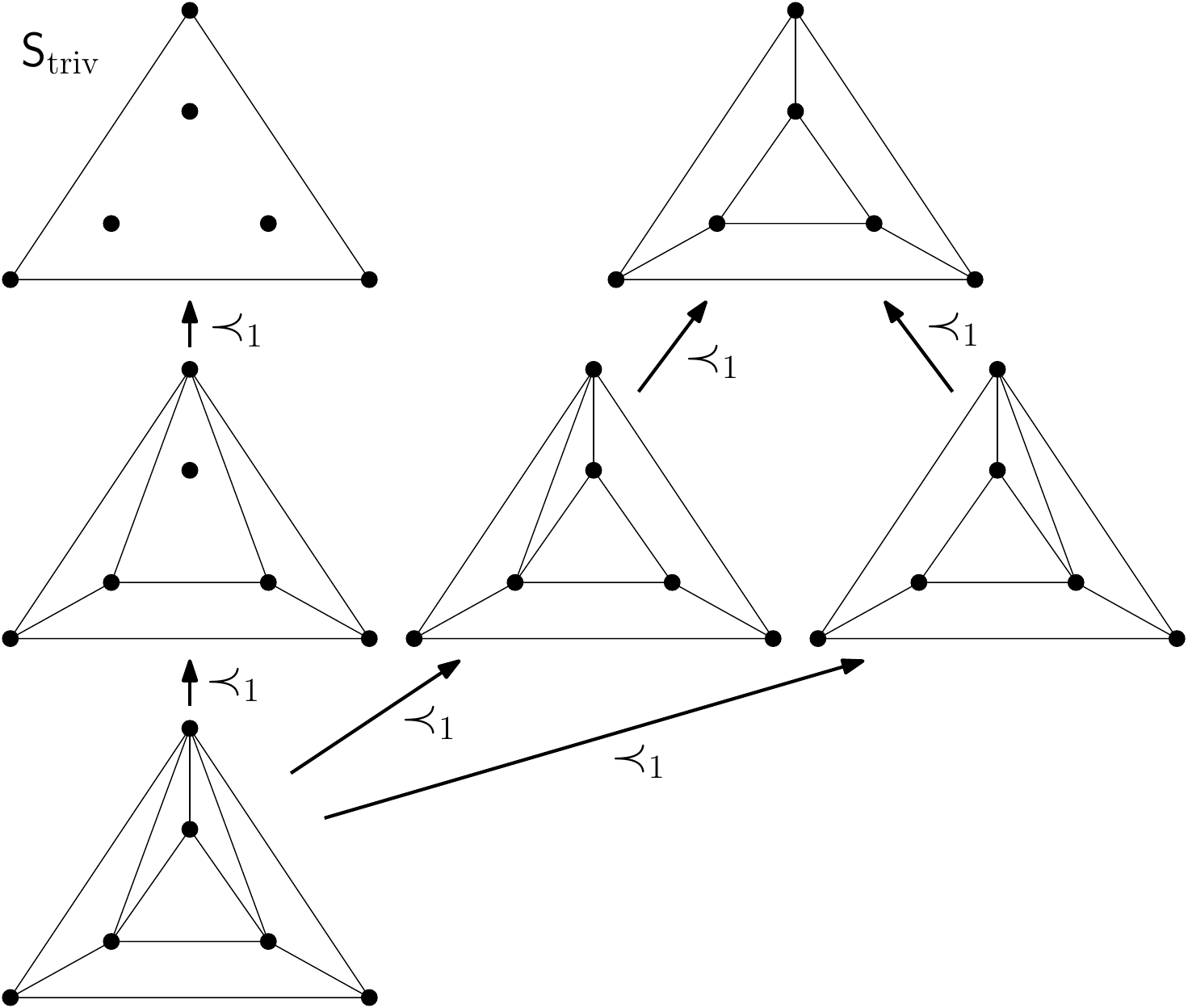}{0.45\textwidth}
}
\caption{Successive perfect coarsenings may lead to the trivial subdivisons (and thus imply regularity) or to another \psubdivision of slack $n-3$. }
\label{f:PrefCoars}
\end{figure}

We immediately get that successive perfect refinements of a \psubdivision $S^*$ fill the regions of $S^*$ with locally regular subdivisions.
\begin{definition}[restriction of \psubdivision] 
\label{d:Restriction}
Let $S \preceq S'$ be \psubdivisions, and let $r \in \Reg{S'}$. Then the \Emph{restriction of} $S$ \Emph{to} $r$, $\ind{S}{r}$, is the subgraph of $S$ induced by $\Pts{S} \cap \cl{r}$ ($\cl{r}$ the closure of $r$).
\end{definition}
\begin{corollary}
\label{c:LocallyRegular}
Let $S \precPerfStar S'$. For $r \in \Reg{S'}$, we have $\ind{S}{r} \precPerfStar \Striv(\Pts{r})$ and $\ind{S}{r}$  is a regular subdivision of $\Pts{r}$.
\end{corollary}
Let us conclude this section with the remark, that successive perfect coarsening starting from a \psubdivision is a non-deterministic process, that may -- even for the same subdivision -- lead to $\Striv$ or not (\Fig{f:PrefCoars}).

%
%
\section{Implications of Regularity Preservation}
\label{s:ImplRegularityPreserv}
%
%
\subsection{Covering the bistellar flip graph with polytopes}
\label{s:Covering}
%
\begin{theorem} 
The edge set of the \bfg\ of $P$ can be covered by subgraphs isomorphic to 1-skeletons of  $(n-3)$-polytopes (which are products of secondary polytopes).
\end{theorem}
\begin{MyProof} Given an edge $\{T,T[x]\}$ of the \bfg, let $S$ be a \psubdivision with $\pFlip{T}{x} \precPerfStar S$ and $\Slack{S} = n-3$ (\Cor{c:UpTheLadder}). For every \region $r$ of $S$, the \psubdivisions $\ind{T}{r}$, $\ind{T[x]}{r}$, and $\ind{\pFlip{T}{x}}{r}$ are regular \psubdivisions of $\Pts{r}$; this holds, since $T \precPerf \pFlip{T}{x}$ and $T[x] \precPerf \pFlip{T}{x}$ (\Lm{l:RefSlack2}), thus $T \precPerfStar S$ and $T[x] \precPerfStar S$. 

Now consider the product of polytopes (see \cite{Zie95})
\[
\prod_{r \in \Reg{S}} \SecPoly{\Pts{r}}~,
\]
where  $\SecPoly{A}$ denotes the secondary polytope of $A \subseteq P$ \cite{LRS10}, see also \Sec{s:SecondaryPolytope}, \Thm{t:SecondaryPolytope}. The dimension of this product is $\sum_r (|\Pts{r}|-3) = \Slack{S} = n-3$. Its faces correspond to the refinements $S'$ of $S$ such that for each \region $r$ of $S$, $\ind{S'}{r}$ is regular, \ie this includes $T$ and $T[x]$ (as vertices), and $\pFlip{T}{x}$ (as edge) (\Cor{c:LocallyRegular}). This completes the argument.
\end{MyProof}

\begin{figure}[htb]
\centerline{
\begin{minipage}[c]{0.18\textwidth}
\placefig{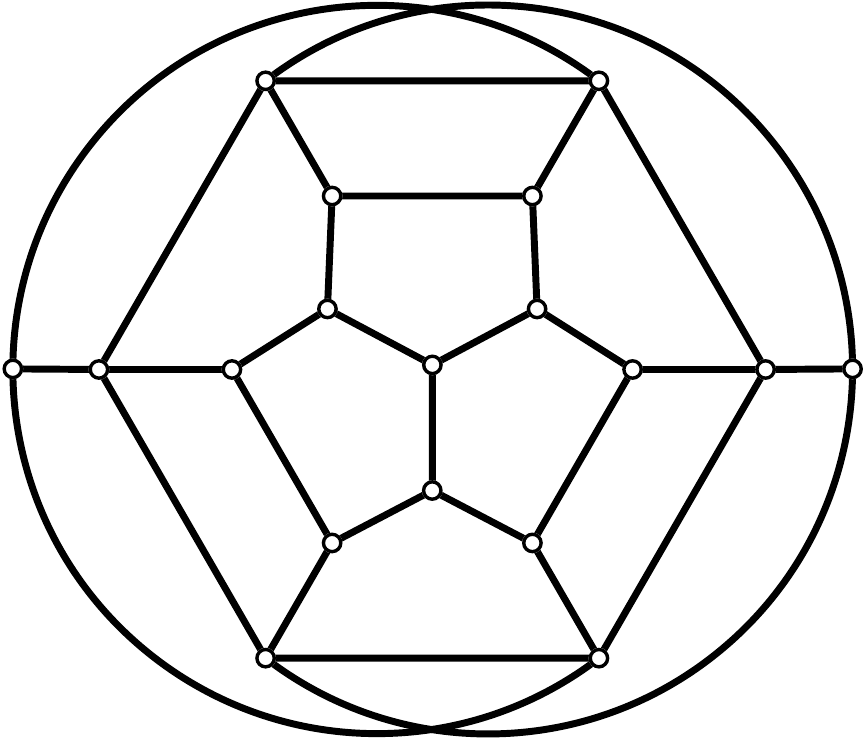}{\textwidth} 
\end{minipage}
\begin{minipage}[c]{1em}
$=$
\end{minipage}
\hspace{-0.5em}
\begin{minipage}[c]{0.18\textwidth}
\placefig{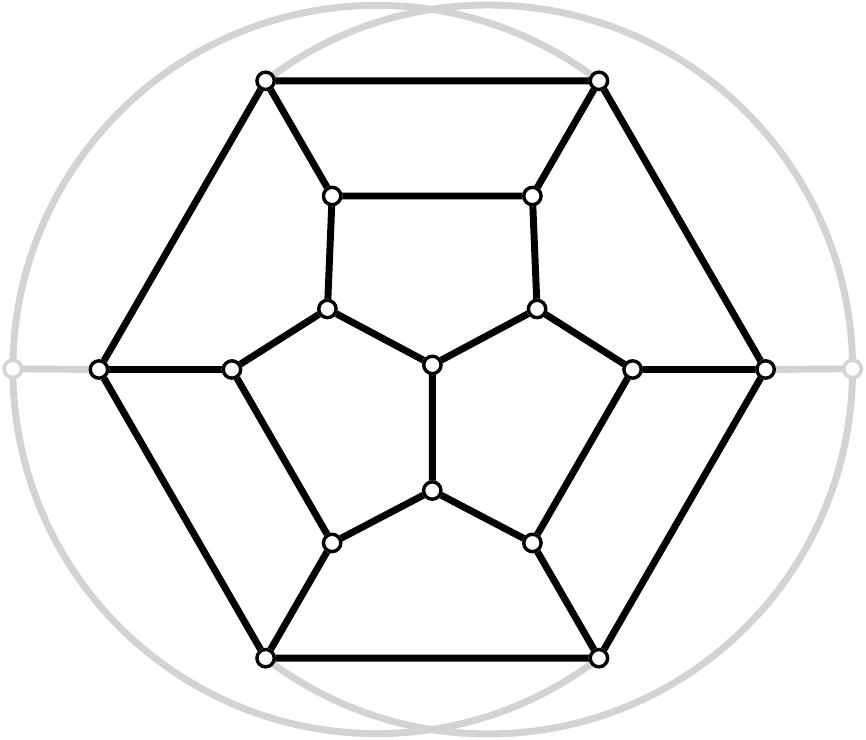}{\textwidth}
\end{minipage}
\begin{minipage}[c]{1em}
$\cup$
\end{minipage}
\hspace{-0.7em} 
\begin{minipage}[c]{0.18\textwidth}
\placefig{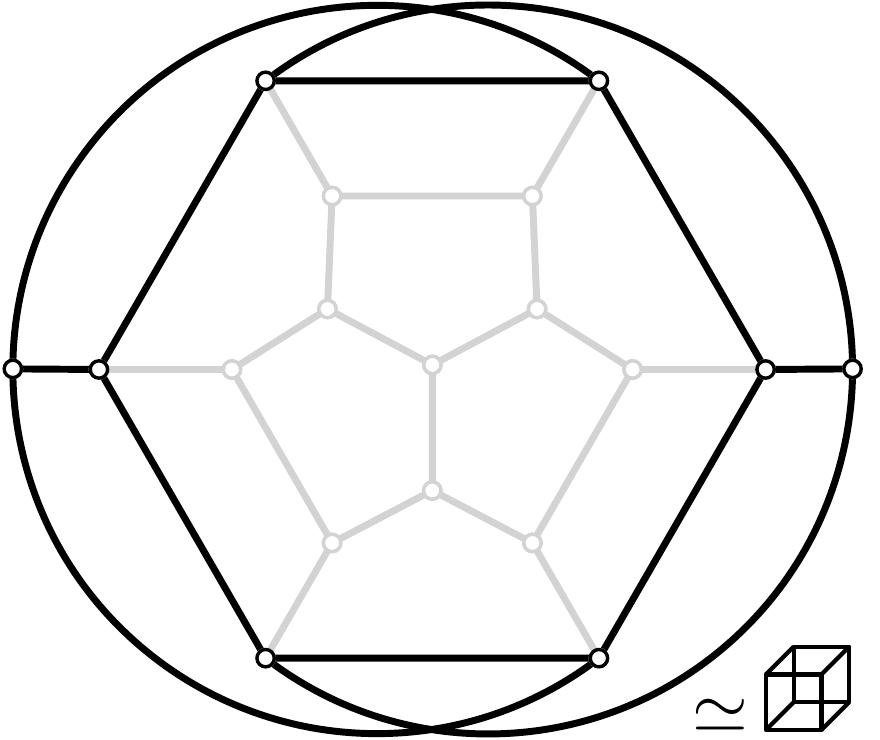}{\textwidth}
\end{minipage}
}
\caption{The \bfg\ of the mother-of-examples configuration as the union of the graphs of two $3$-polytopes.}
\label{f:UnionOfPolytopes}
\end{figure}
%
%
\subsection{Sets with all triangulations regular}
We give characterizations of point sets for which all \ptriangulations are regular (as, \eg it is the case for point sets in convex position). In particular, we show that this can be easily read off the height of the partial order $\preceq$. In a first step we prove that property to be equivalent to requiring that all \psubdivisions are regular.
\begin{lemma}
\label{le:TrRegVsSubdReg}
All \psubdivisions are regular iff all \ptriangulations are regular.
\end{lemma}
\begin{MyProof} The direction ($\Rightarrow$) is obvious.
\smallskip

\noindent
For ($\Leftarrow$) it suffices to show that every non-regular \psubdivision $S$ with $\Slack{S} > 0$ has a direct refinement which is not regular.

\Case{1}{$S$ has a bystander $p \in \PtsBy{S}$.} Clearly, the direct refinement $(\Pts{S} \setminus \{p\}, \Eds{S})$ is not regular iff $S$ is not regular.

\Case{2}{$S$ has no bystander.} Since $\slack{S}>0$, $S$ must have an active region $r^*$ which is a $k$-gon for $k\ge4$. Choose two crossing diagonals $e_0$ and $e_1$ in $r^*$ and consider the \psubdivisions $S_i := (\Pts{S}, \Eds{S} \cup \{e_i\})$, $i=0,1$. We want to show that if $S$ is not regular, then at least one of $S_0$ and $S_1$ is not regular. So let us suppose that, for $i=0,1$, $\omega_i \in \RR^{\Pts{S}}$ is a height function realizing $S_i$ (as a regular \psubdivision) and, for $t \in [0,1]$, consider the convex combination $\omega_t := (1-t) \omega_0 + t \omega_1$.

We say that a height function $\omega$ \Emph{respects} region $r$ in \psubdivision $S$, if $\omega$ is linear on $\Pts{r}$ and all points in $\Pts{S}^{(\omega)} \setminus \Pts{r}^{(\omega)}$ lie strictly above the plane spanned by $\Pts{r}^{(\omega)}$. Clearly, $\omega$ realizes $S$ iff it respects all regions $r \in \Reg{S}$. Moreover, if two height functions respect a region, then all convex combinations do.

It follows, that $\omega_t$ respects all regions  in $\Reg{S}$ except for $r^*$, since these are regions both in $S_0$ and $S_1$. We have that $e_0^{(\omega_0)}$ lies below $e_1^{(\omega_0)}$ (as segments in the lifting in $\RR^3$), while $e_1^{(\omega_1)}$ lies below $e_0^{(\omega_1)}$ and, therefore, there must be a $t \in (0,1)$, where $e_0^{(\omega_t)}$ and $e_1^{(\omega_t)}$ intersect (in the lifting in $\RR^3$). For that value of $t$, $\omega_t$ is linear on $\Pts{r^*}$. Moreover, all edges in $\ind{S}{r}$ are valleys in the $\omega_t$-lifting, since these are valleys both in the $\omega_0$-lifting and the $\omega_1$-lifting (and that property is preserved for all convex combinations of $\omega_0$ and $\omega_1$). Hence, $\omega_t$ realizes $S$ and we have a contradiction.
\end{MyProof}
We recall the definition of the height of an element in a partial order, and of the height of the partial order.
\begin{definition}[height] 
\label{d:Height}
The \Emph{height, $\Height{S}$, of} a \psubdivision $S$ (in the partial order $\preceq$) is recursively defined: (a) If $S$ is a \ptriangulation, then $\Height{S} : = 0$, and (b) if $S$ is not a \ptriangulation, then $\Height{S} := 1+ \max_{S' \precDir S} \Height{S'}$. (Equivalently, $\Height{S}$ is the size of the longest $\preceq$-chain ending in $S$ minus \one.) We let $\HeightMax = \HeightMax(P)$ be the maximum height of any \psubdivision of $P$ (\ie $\HeightMax = \Height{\Striv}$).
\end{definition}
\begin{theorem} 
\label{t:AllRegular} The following six conditions are equivalent.

\noindent
\begin{minipage}[t]{0.5\textwidth}
\begin{EnumRom}
\item
\label{i:iAllRegular}
All \ptriangulations are regular.
\item
\label{i:iiAllRegular}
All \psubdivisions are regular. 
\item
\label{i:ivAllRegular}
$\HeightMax = n-3$.
\end{EnumRom}
\end{minipage}
\begin{minipage}[t]{0.5\textwidth}
\begin{EnumRom}
\setcounter{enumi}{3}
\item
\label{i:ivAllRegular}
$\precDir = \precPerf$.
\item
\label{i:vAllRegular}
$\Height{} = \slack$.
\end{EnumRom}
\end{minipage}
\end{theorem}
\begin{MyProof} For \ItemRef{i:iAllRegular} $\Leftrightarrow$ \ItemRef{i:iiAllRegular} see \Lm{le:TrRegVsSubdReg}.
For the rest we show the implication cycles  
\begin{center}
\emph{all \psubdivisions are regular $\stackrel{\mathrm{(a)}}{\Rightarrow}$ $\HeightMax = n-3$ $\stackrel{\mathrm{(b)}}{\Rightarrow}$ $\precDir = \precPerf$ $\stackrel{\mathrm{(c)}}{\Rightarrow}$ all \psubdivisions are regular}
\end{center}
and 
\begin{center}
\emph{$\HeightMax = n-3$ $\stackrel{\mathrm{(b)}}{\Rightarrow}$ $\precDir = \precPerf$ $\stackrel{\mathrm{(d)}}{\Rightarrow}$ $\Height{} = \slack$ $\stackrel{\mathrm{(e)}}{\Rightarrow}$ $\HeightMax = n-3$}.
\end{center}

(a) \emph{All \psubdivisions are regular $\Rightarrow $ $\HeightMax = n-3$.}~~ This is well known and discussed, \eg in ``Twelve proofs of non-regularity'' in \cite[Sec.\,7.1.2, (6)]{LRS10}: On the one hand, if all subdivisions are regular, they all correspond to faces of the secondary polytope, an $(n-3)$-polytope where every chain of proper faces (excluding the empty face and the polytope itself) has size at most $(n-3)$. On the other hand, if $\HeightMax > n-3$, that gives a chain of size exceeding $n-3$ of non-trivial subdivisions.\footnote{Note here: There is a maximum chain of size $1+ \HeightMax$ in the $\preceq$-partial order. If we remove the trivial subdivision (not corresponding to a proper face), we get still a chain of length $\HeightMax$.}
\medskip

\noindent
(b) \emph{$\HeightMax = n-3$ $\Rightarrow$ $\precDir = \precPerf$.}~~ Consider a maximal chain
$S_0 \precDir S_1 \precDir \cdots \precDir S_m$; because of maximality, $S_0$ is a \ptriangulation (of slack $0$), and $S_m = \Striv$ (of slack $n-3$). We know that $\Slack{S_i} \le \Slack{S_{i-1}} + 1$ (\Lm{le:CoarsenerBounds} below) with equality iff $S_{i-1} \precPerf S_i$. It follows that $m \ge n-3$. Moreover, if $m=n-3$ (which is given if $\HeightMax = n-3$), then $S_{i-1} \precPerf S_i$ for all $i=1,2,\ldots,n-3$. Since every pair $S' \precDir S$ is part of a maximal chain, the claim follows.
\medskip

\noindent
(c) \emph{$\precDir = \precPerf$ $\Rightarrow$ all \psubdivisions are regular.}~~ Every \psubdivision $S$ has a chain of direct coarsenings to $\Striv$. If every direct coarsening is a perfect coarsening, this shows $S \precPerfStar \Striv$ and therefore $S$ is regular (\Thm{t:LowClosureRegular}).
\medskip

\noindent
(d) \emph{$\precDir = \precPerf$ $\Rightarrow$ $\Height{} = \slack$.}~~ For proving $\Height{S} = \Slack{S}$, we can proceed by induction on the height of $S$, where the induction basis holds without assumptions. With the assumption of $\precDir = \precPerf$ and with the induction hypothesis
\[
\Height{S} = 1+ \max_{S' \precDir S} \Height{S'} 
= 1+ \max_{S' \precPerf S} \Height{S'}
= 1+ \max_{S' \precPerf S} \Slack{S'}
= 1+ \max_{S' \precPerf S} (\Slack{S}-1)
= \Slack{S}
\]
For the last equality, note that for a subdivision that is not a triangulation, a perfect refinement can always be obtained by simply adding an edge, or by involving a bystander as a point of degree \three. If there is no edge to add and if there is no bystander, we have a triangulation.
\medskip

\noindent
(e) \emph{$\Height{} = \slack{}$ $\Rightarrow$ $\HeightMax = n-3$.}~~ If $\Height{} = \slack{}$, then $\HeightMax = \Height{\Striv} = \Slack{\Striv} = n-3$.
\end{MyProof}
%
%
\subsection{More properties of coarseners}
We derive two more properties of prime and perfect \coarseners, \Lm{le:CoarsenerBounds}, which we used  in the proof of \Thm{t:AllRegular} above, and \Lm{le:TreeCoarsenerPerfect}, which we will need in \Sec{se:Certificates} below.
\begin{lemma}
\label{le:CoarsenerBounds}
$\Incr{U} \le 1$ for every prime \coarsener $U$ in a \psubdivision $S$.
\end{lemma}
\begin{MyProof} 
Let $r$ be the region in $S'$ obtained by removing from $S$ the edges $E_U$ incident to $U$. The subgraph of $S$ induced by $U$ is connected (\Obs{ob:Coarsers}\ItemRef{it:viiiCoarsers}), that is, all points of $U$ have to lie in the same \region of $S'$. We consider the restriction $\ind{S}{r}$ (\Def{d:Restriction}), a \psubdivision of $\Pts{\ind{S}{r}} = \Pts{S} \,\cap\, \cl{r}$. Isolating $U$ in $\ind{S}{r}$ yields $\ind{S'}{r}$, the trivial subdivision of $\Pts{\ind{S}{r}}$. Therefore, 
\[
\Slack{\ind{S}{r}} + \incr{U} \stackrel{\mbox{\footnotesize \Obs{o:Increment}}}{=} \Slack{\ind{S'}{r}} = |\Pts{\ind{S}{r}}| - 3,
\]
that is, $\incr{U}= |\Pts{\ind{S}{r}}| - 3 - \Slack{\ind{S}{r}}$. On the one hand, $U$ is the only coarsener of $\ind{S}{r}$ (since $U$ is prime and it exhausts all inner points in $\Pts{\ind{S}{r}}$). On the other hand, the Coarsening Lemma~\ref{le:CoarsenPartSubd} guarantees $|\Pts{\ind{S}{r}}| - 3 - \Slack{\ind{S}{r}}$, \ie  $\incr{U}$ (perfect) coarsenings of $\ind{S}{r}$. Therefore, $\incr{U} \le 1$.
\end{MyProof}

\begin{lemma}
\label{le:TreeCoarsenerPerfect}
A prime \coarsener $U$ inducing a tree in its \psubdivision $S$ is perfect.
\end{lemma}
\begin{MyProof} Let $k:= |U|$ and let $\ell$ be the number of edges in $\EdsInn{S}$ that are incident to exactly one point in $U$. We have $|E_U| = (k-1)+\ell$. Since every point in $U$ has degree at least \three, we have $|E_U| \ge \frac{3k + \ell}{2}$. Hence, $(k-1)+\ell \ge \frac{3k + \ell}{2}$, \ie $\ell \ge k+2$. Now $\incr{U} = |E_U| - 2|U| = (k-1)+\ell - 2k = \ell - k - 1 \ge 1$, \ie by \Lm{le:CoarsenerBounds}, $\incr{U} = 1$.\footnote{Since, $\incr{U} \le 1$, this also shows that if $U$ induces a tree in $S$, then all points in $U$  have degree exactly \three.}
\end{MyProof}
%
%
\subsection{Large minimal sets with non-regular triangulations --  Proof of \Thm{t:MinNonRegExample}}
\label{se:Certificates}
\begin{observation} 
\label{o:Hereditary}
If $P'\subseteq P$ and $P'$ has non-regular triangulations, then $P$ has non-regular triangulations.
\end{observation}
Given a set $P$ with non-regular triangulations, is there always a small subset $P'$ of $P$ that witnesses this fact? This is a question asked by F.\ Santos \cite{SanPersCom}. An equivalent reformulation is: How large can minimal sets $P$ with non-regular triangulations be? (Here, ``minimal'' means that every proper subset of $P$ has only regular triangulations.) Santos \cite{SanPersCom} describes such a minimal set of \eight points and conjectures this to be the largest example of such a minimal set. We will show that there exist such minimal sets of arbitrarily large (even) size.

\begin{figure}[htb]
\centerline{
\placefig{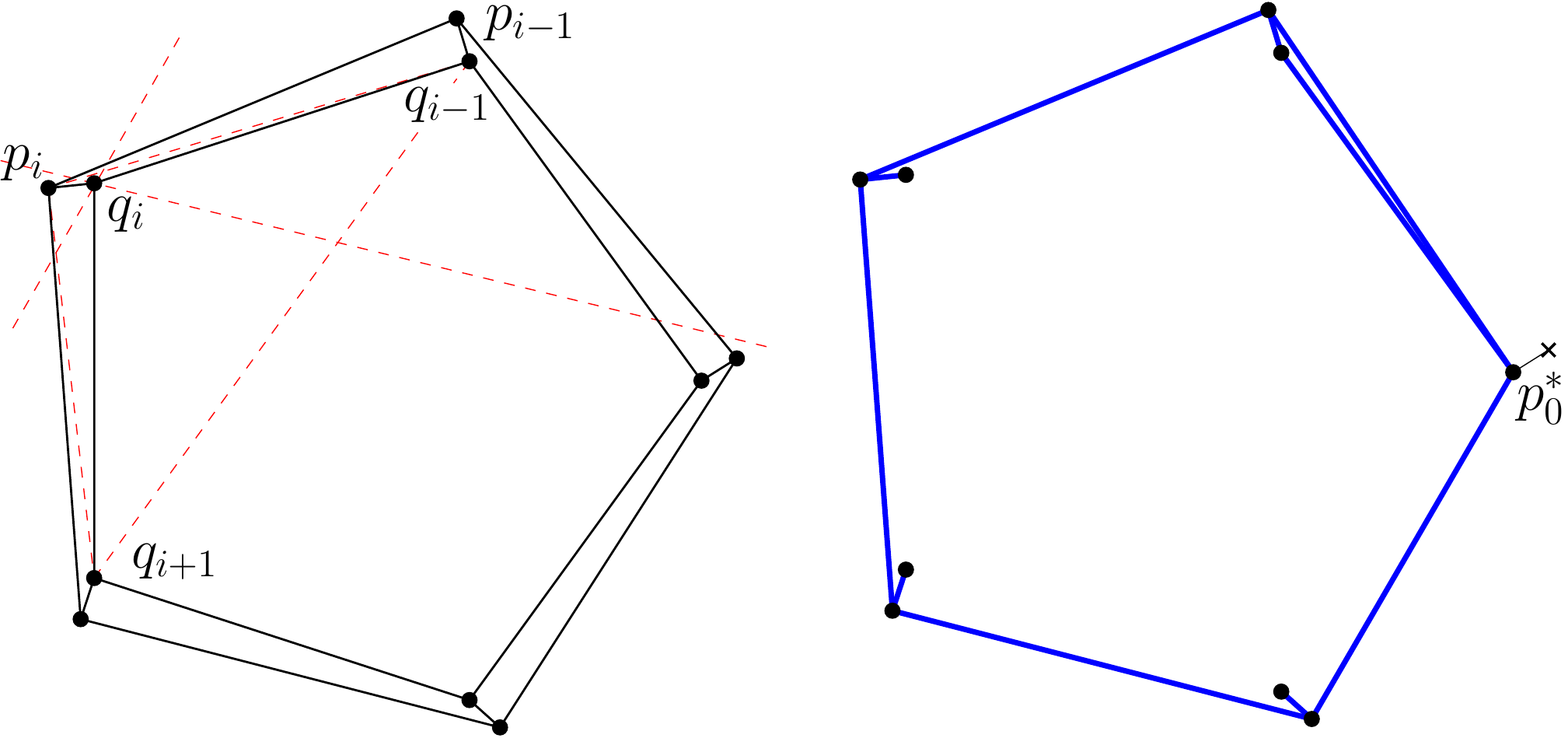}{0.67\textwidth}
}
\caption{A twisted double-gon of \ten points with \psubdivision $S^\Box$ (left). Subset $P^*$ (case $p_0^* = q_0$) of a twisted double-gon (right).}
\label{fi:MinimalNonRegular}
\end{figure}
\begin{definition}[twisted double-gon]
\label{d:Windmill}
A set $P$ of $n$ points in general position, $n=2k$ even, is called a \emph{twisted double-gon} if the following holds.
\begin{EnumCapRom}
\item
$|\ext{P}| = k$. 
\end{EnumCapRom}
Let $p_0,p_1,\ldots,p_{k-1}$ be a counter-clockwise numbering of $\ext{P}$ along the boundary of the convex hull of $P$.
\begin{EnumCapRom}
\setcounter{enumi}{1}
\item
\label{it:iWindmill}
The set $Q:=\inn{P}$ of inner points is in convex position.
\end{EnumCapRom}
There is a numbering $q_0,q_1,\ldots,q_{k-1}$ of $Q$ following the order along the boundary of the convex hull of $Q$ such that for all $i$, $0 \le i \le k-1$, it holds:
\begin{EnumCapRom}
\setcounter{enumi}{2}
\item
\label{it:iiWindmill}
$q_i$ is extreme in $P \setminus \{p_i\}$.
\item
\label{it:iiiWindmill}
$q_i$ is extreme in $P \setminus \{p_{i-1},q_{i-1}\}$.
\item
\label{it:ivWindmill}
$q_i$ lies in the triangle $q_{i-1}p_iq_{i+1}$.
\end{EnumCapRom}
\end{definition}
\Fig{fi:MinimalNonRegular} indicates that such twisted double-gons exist\footnote{These are not double-circles, see \eg \cite{AHN04}; actually, double-circles have only regular triangulations.} for all even $n\ge6$. For $n=6$, this is the mother-of-examples configuration (\Sec{s:Mother}). Here are a few simple observations.
\begin{observation} 
\label{o:TwistCond}
Let $P$ be a twisted double-gon, with notation as in \Def{d:Windmill}.
\begin{EnumRom}
\item
\label{i:iTwistCond}
The graph $S^\Box:=(P, \EdsHull \cup \{\{q_i,p_i\}, \{q_i,q_{i+1\!\bmod\!k}\} \mid i=0,1,\ldots,k-1\}$ is a \psubdivision (uses \ItemRef{it:iWindmill} and \ItemRef{it:ivWindmill}). Its slack is $n-3$.
\item
\label{i:iiTwistCond}
If $q_i$ is involved in a \psubdivision of a subset of $P$ where it is not extreme, it is connected to $p_i$ (by \ItemRef{it:iiWindmill}) and to at least one of $\{p_{i-1},q_{i-1}\}$ (by \ItemRef{it:iiiWindmill}).
\end{EnumRom}
\end{observation}
Here comes the lemma that shows that twisted double-gons constitute examples showing \Thm{t:MinNonRegExample}.
\begin{lemma} 
\label{l:MinNonRegExample}
If $P$ is a twisted double-gon, then
\begin{EnumRom}
\item
\label{it:iMinNonRegExample}
$P$ has a non-regular \ptriangulation, and
\item
\label{it:iiMinNonRegExample}
any proper subset of $P$ has only regular triangulations.
\end{EnumRom}
\end{lemma}
\begin{MyProof}
\ItemRef{it:iMinNonRegExample}
Since $\Slack{S^\Box} = n-3$, we have $\Height{S} \ge n-3$ (by \Lm{le:CoarsenerBounds}), therefore $\Height{\Striv} > n-3 = \Slack{\Striv}$ and $\Height{} \neq \slack$. \Thm{t:AllRegular} implies that $P$ has non-regular triangulations. Note that we do not claim that $S^\Box$ is a non-regular subdivision; this depends on the concrete coordinates of the point set $P$.
\smallskip

\noindent
\ItemRef{it:iiMinNonRegExample} We remove $p_0$ or $q_0$ from $P$, we denote the resulting set by $P^*$ with $p_0^*$ the point among $p_0$ and $q_0$ remaining in $P^*$, see \FigLR{fi:MinimalNonRegular}{right}. If we can show that all \ptriangulations of $P^*$ are regular, then the proof is complete (by symmetry and \Obs{o:Hereditary}).

It is enough to show that any prime coarsener $U$ of any \psubdivision $S$ of $P^*$ is perfect (by \Lm{le:CoarsenerBounds} this is equivalent to $\incr{U} = |E_U|-2|U| \ge 1$), since then $\precDir = \precPerf$  and \Thm{t:AllRegular} can step in. So let us consider such a prime coarsener $U$, let $E^\mathrm{in}_U$ be the edges in $S$ connecting two points in $U$, and let $E^\mathrm{out}_U$ be the set of edges in $S$ connecting a point in $U$ to a point in $\Pts{S} \setminus U$. We have $E_U = E^\mathrm{in}_U \cup E^\mathrm{out}_U$.
\begin{EnumAlph}
\item 
The subgraph of $S$ induced by $U$ has to be connected (since $U$ is prime, \Obs{ob:Coarsers}\ItemRef{it:viiiCoarsers}), hence $|E^\mathrm{in}_U| \ge |U|-1$.
\item
If the subgraph of $S$ induced by $U$ is a tree, then, by \Lm{le:TreeCoarsenerPerfect}, $U$ is a perfect coarsener.
\end{EnumAlph}
So let us assume that $U$ does not span a tree, which implies $|E^\mathrm{in}_U| \ge |U|$.
\begin{EnumAlph}
\setcounter{enumi}{2}
\item
Every point in $U$ has to connect to at least one point in $\Pts{S} \setminus U$. This holds, since $U \subseteq Q\setminus \{q_0\}$ and $q_i \in U$, $i \ge 1$, has to connect to $p_i$ (\Obs{o:TwistCond}\ItemRef{i:iiTwistCond}). Hence, $|E^\mathrm{out}_U| \ge |U|$.
\end{EnumAlph}
At this point we have already shown that $|E_U| = |E^\mathrm{in}_U| + |E^\mathrm{out}_U| \ge 2|U|$. Therefore, $\incr{U} \ge 0$, and $U$ is perfect unless $|E^\mathrm{in}_U| = |U|$ and $|E^\mathrm{out}_U| = |U|$.
\begin{EnumAlph}
\setcounter{enumi}{3}
\item Let $j := \min\{i\in \NN \mid i \ge 1, q_i \in U\}$. If $j=1$, \ie $q_1 \in U$, then $q_1$ connects to $p_1$ and $p_0^*$. Hence, $|E^\mathrm{out}_U| \ge  |U|+1$. If $j \ge 2$, then $q_j$ has to connect to $p_j$ and $p_{j-1}$ (by (by \Obs{o:TwistCond}\ItemRef{i:iiTwistCond}), since $q_{j-1}$ is not available). Hence, $|E^\mathrm{out}_U| \ge  |U|+1$. 
\end{EnumAlph}
We can conclude that $U$ has to be perfect.
\end{MyProof}
%
%
\section{Discussion and Open Problems}
\label{s:Discuss}
We conclude by briefly discussing some questions that naturally arise in connection with this work.
%
%
\subsection{The min-degree bound for small sets}
We have put some effort into getting rid of the ``$P$ large enough''-condition in the min-degree bound of \Thm{t:MainFull}\ref{it:iMainFull} for the edge flip graph.\footnote{In fact, the min-degree bound was our initial result, and the other results have emerged in the quest to get rid of the  ``$P$ large enough''-condition.} We have, however, not tried to find counter-examples by checking the Graz order type data base, \cite{AAK02}. Such an endeavour is not trivial, since an ad-hoc approach would have to produce and check vertex connectivity for reasonably large edge flip graphs, even when the underlying sets have no more than $11$ points.
%
%
\subsection{General vs. special position}
\begin{figure}[htb]
\centerline{
\placefig{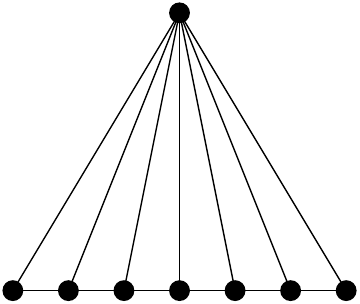}{0.15\textwidth}
\hspace{3em}
\placefig{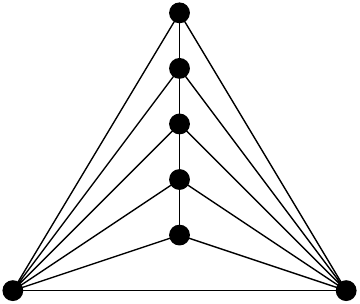}{0.15\textwidth}
}
\caption{Point sets with a unique full triangulation.}
\label{f:Degenerate}
\end{figure}

When abandoning the general position assumption, \ie allowing collinearities of \three or more points can occur,\footnote{Actually, even repeated points have to be considered.} the situation changes significantly. The edge flip graph may degenerate to a single vertex even for large point sets (\Fig{f:Degenerate}).

\begin{figure}[htb]
\centerline{
\placefig{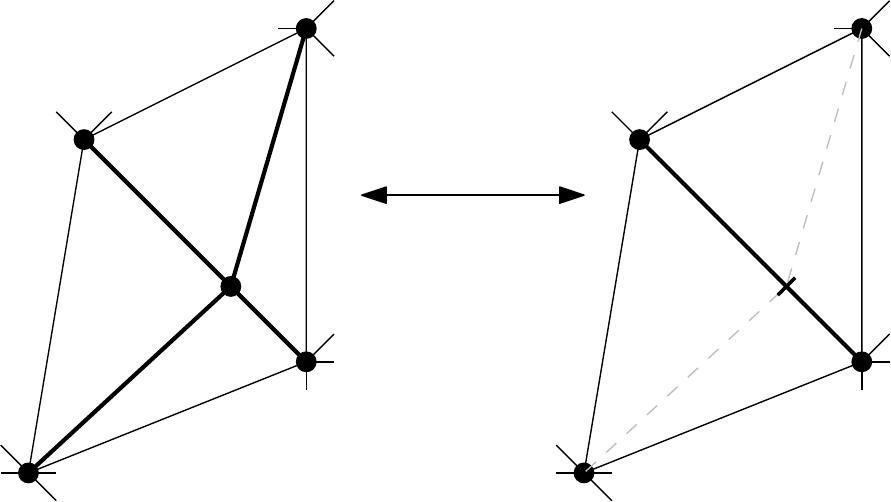}{0.3\textwidth}
}
\caption{A flip in the presence of collinearities.}
\label{f:DegFlip}
\end{figure}

For bistellar flips, new flips as shown in \Fig{f:DegFlip} come into play. \Thm{t:PartNbFlips} (\cite{LSU99}, see also \cite[Thm.\,3.4.9]{LRS10}) still holds in this situation and $n-3$ bistellar flips are guaranteed. We are currently investigating with N.\,Grellier to what extent our methods generalize from general to special position towards $(n-3)$-vertex connectivity of the bistellar flip graph, which is by no means obvious.

%
%
\subsection{Stably regular triangulations}
It might be interesting to understand, which regular triangulations are captured by \Thm{t:LowClosureRegular}, \ie partial triangulations in the  $\precPerfStar$-lower closure of $\Striv$. From the discussion in \Sec{s:Mother} (and \Fig{f:Mother}) we recall one reason why this condition for regularity cannot possibly be necessary:  It depends only on the order type (or oriented matroid) of $P$, while regularity depends on the coordinates of the point set. So let us define a partial triangulation $T$ of $P$ \Emph{stably regular}, if it is regular on any realization of the order type of $P$.\footnote{A careful formulation of ``regular on any realization of the order type" is: Whenever there is a bijective mapping $p \mapsto p'$ from $P$ to a set $P'$ which preserves orientation of triples, then the partial triangulation $T'$ of $P'$ obtained by $\Pts{T'} = \{p' \mid p \in \Pts{T}\}$ and $\Eds{T'} = \{\{p',q'\} \mid \{p,q\} \in \Eds{T}\}$ is also regular.}\,\footnote{If we can move the points in $P$ freely while preserving $T$ as a triangulation, then we can turn any triangulation into a regular one. This follows basically from Steinitz' Theorem (see \cite[Thm.\,4.1]{Zie95}).} We believe that \Thm{t:LowClosureRegular} goes as far as one can go with an order type based condition.
\begin{conjecture} 
A triangulation $T \in \pT(P)$ is stably regular iff $T \precPerfStar \Striv$.
\end{conjecture}
One might ask whether the subdivisions in the $\precPerfStar$-down set of $\Striv$ correspond to the face lattice of a polytope. In general, this is not the case, it is not even an abstract polytope, (see \cite[Chap.\,2A]{MS02} for abstract polytopes): We can choose $\Slack{S}$ as a rank function, have $\Striv$ as maximal/greatest element, and add an empty face as minimal/least element of rank $-1$. All flags (\ie maximal chains) have the same size. But we fail on the so-called \Emph{diamond axiom}: ``If $S$ is a subface of $S'$ with the ranks differing by \two, then there are exactly \two faces strictly between $S$ and $S'$.'' Fig.\,\ref{f:PrefCoars} shows an example where this axiom fails: $S$ and $\Striv$ in this figure differ by \two in slack, but there is only one subdivision strictly in between the two (in the $\precPerf$-order).
%
%
\subsection{Minimal sets with non-regular triangulations}
In \Sec{se:Certificates} we have identified arbitrarily large minimal sets with non-regular triangulations. Given that such sets of arbitrary size exist, it is not clear how to check efficiently whether a set allows only regular triangulations. A first step would be to understand all minimal sets with non-regular triangulations. It is possible that twisted double-gons as defined in \Def{d:Windmill} are the only such sets.
%
%
\subsection{The fundamental group}
Consider the set of closed walks in an edge or bistellar flip graph, starting in a fixed triangulation $T^*$. Consider a cycle $c=(T_0,T_1,\ldots, T_{k-1})$, $k \ge 2$, in the flip graph, \ie  $T_i$ and $T_{i + 1\!\bmod\!k}$ are adjacent triangulations for $0 \le i < k$. An \emph{insertion of $c$} adds $c$ before an occurrence of $T_0$ in a closed walk, and a \emph{deletion of $c$} removes $c$ if it occurs before $T_0$ in a closed walk. Let us consider two closed walks equivalent, if they can be obtained from each other by insertion and deletion of cycles of length $2$, or of cycles spanned by the refinements of a subdivision of slack \two (these are $4$- or $5$-cycles\footnote{One can show that these are exactly the $4$- and $5$-cycles, both for the edge and the bistellar flip graph.}). Lubiw \etal\ \cite{LMW19} show that in the \emph{edge} flip graph all closed walks containing a triangulation $T^*$ are equivalent to the trivial walk $(T^*)$; in other words, the fundamental group of the complex spanned by slack $2$ subdivisions (which correspond to the $2$-faces of the flip complex) is trivial. This is the decisive step in proving the orbit conjecture of \cite{Bose:Flipping-edge-labelled-triangulations-2018} (see discussion in \Sec{s:FlipComplex}). It is natural to ask, whether the analogous result holds for the bistellar flip graph, or, more generally, whether there is a bistellar analogue of the flip complex. 
%
%
\subsection{Higher dimensions}
Flip graphs can be defined on triangulations in higher dimensions, \ie tetrahedralizations in $\RR^3$, \etc, see \cite{LRS10}, but even for sets of points in convex and general position in $3$-space it is open, whether these graphs are always connected. As mentioned before, because of the Local Menger Lemma~\ref{le:LocalMenger}, we were able to show high vertex connectivity without ever providing any evidence that the graph is connected. Basically, we showed high vertex connectivity of the connected components of the flip graphs, and Lawson gave us the connectedness on top of it. Still, it might be interesting to see whether one can show, \eg $(n-4)$-vertex connectivity for bistellar flip graphs of sets in convex and general position in $\RR^3$. A first step is supplied: There are always at least $n-4$ flips for such sets \cite[Prop.\,3.6.19]{LRS10}.
%
%
\subsection{Mixing rate of the random flip process}
Last, but not least, a natural next question is to show expansion properties of the flip graphs, ideally yielding rapid mixing of the process of flipping random edges in the flip graph. This is known for points in convex positions, still with a big gap between the upper bound of $O(n^5)$ and the lower bound of $\Omega(n^{3/2})$ (Mcshine \& Tetali 1998, \cite{MT98}, and Molloy \etal\ 2001, \cite{MRS01}). There are related results for lattice point sets by Caputo \etal\ 2013, \cite{CMSS13}.
%
%
\bibliographystyle{acm}
\bibliography{FlipGraph-ArXiv-v1}{}
\end{document}